\documentclass[5p,preprint,twocolumn,sort&compress]{elsarticle}

\makeatletter
\def\ps@pprintTitle{%
 \let\@oddhead\@empty% chktex 21
 \let\@evenhead\@empty% chktex 21
 \def\@oddfoot{\footnotesize\itshape Preprint\hfill\today}% chktex 21 chktex 1
 \let\@evenfoot\@oddfoot} % chktex 21
\makeatother

\usepackage{lineno}
\modulolinenumbers[5]

\journal{}

\usepackage[T1]{fontenc}
\usepackage[utf8]{inputenc}

\usepackage{url}
\usepackage{xcolor}

\usepackage{setspace}

\usepackage{graphicx}
\usepackage{wrapfig}
\usepackage{subcaption}

\usepackage{amsmath}
\usepackage{amssymb}
\usepackage{latexsym} 

\usepackage{pdfpages}

\usepackage{float}
\usepackage{dblfloatfix}
\usepackage{tabto}

\usepackage{nameref}
\usepackage{hyperref}
\hypersetup{%
    pdfauthor={Tom Mahler},%
    pdftitle={A New Methodology for Information Security Risk Assessment for Medical Devices and Its Evaluation},%
    pdfsubject={Preprint},%
    pdfkeywords={Cyber Security; Information Security; Risk Assessment; Medical Devices; Medical Imaging Devices; CAPEC; Utility Elicitation},% chktex 13
    pdfproducer={},%
    pdfcreator={},%
    hidelinks,%
    hyperindex,%
    breaklinks,%
}

\newcommand{\attackref}[2]{\textcolor{#1}{\ref{#2}. \nameref{#2}}}
\newcommand{\hrefblue}[2]{\textcolor{blue}{\href{#1}{#2}}}
\newcommand{\emailref}[1]{\hrefblue{mailto:#1}{#1}}
\newcommand{\afdref}[2]{\hyperref[#1]{\ref{#1}. #2}}

\usepackage[nameinlink, capitalise]{cleveref}

\usepackage[super]{nth}
\usepackage[shortcuts]{extdash}

\usepackage{paralist}
\usepackage{enumitem}

\usepackage{csquotes}

\usepackage[flushleft]{threeparttable}
\usepackage{supertabular,booktabs}

\usepackage{array}
\newcolumntype{P}[1]{>{\centering\arraybackslash}p{#1}}
\newcolumntype{M}[1]{>{\centering\arraybackslash}m{#1}}
\newcolumntype{B}[1]{>{\centering\arraybackslash}b{#1}}
\usepackage{multirow}
\usepackage[export]{adjustbox}
\usepackage{rotating}
\newcommand{\rot}[1]{\begin{sideways}{#1}\end{sideways}}

\usepackage{tikz}
\usepackage{tikz-qtree}
\usepackage{flowchart}
\usetikzlibrary{external, shapes.geometric, arrows.meta, positioning, trees, snakes, shapes, backgrounds, decorations.markings}

\usepackage{pgfplots}
\pgfplotsset{compat=1.14}

\tikzstyle{process} = [
    component,
    minimum width=5cm,text width=5cm,
    font=\large,
]
\tikzstyle{arrow} = [
    decoration={markings,mark=at position 1 with {\arrow[scale=2,>=stealth]{>}}},
    postaction={decorate},
]
\tikzstyle{annotation} = [
    font=\fontsize{35.83}{35.83}\selectfont\bfseries,
]
\tikzstyle{component} = [
    rectangle,rounded corners=2ex,
    minimum width=4cm,minimum height=2cm,inner sep=1pt,text width=4cm,
    text centered,font=\LARGE,execute at begin node=,
    draw=black,
]
\tikzstyle{subcomponent} = [
    predproc,
    minimum width=4cm,minimum height=2cm,inner sep=1pt,text width=3.5cm,
    text centered,font=\LARGE,execute at begin node=,
    draw=black,
]
\tikzstyle{terminator} = [
    rectangle,rounded corners=7ex,
    minimum width=4cm,minimum height=2.1cm,inner sep=1pt,text width=4cm,
    text centered,font=\LARGE,execute at begin node=,
    draw=black,fill=gray!20!white,
]
\tikzstyle{network} = [
    cloud,cloud puffs=9,cloud ignores aspect,
    minimum width=3cm,minimum height=2cm,inner sep=1pt,text width=3cm,
    text centered,font=\LARGE,execute at begin node=,
    draw=black,fill=gray!20!white,
]
\tikzstyle{directed} = [
    decoration={markings,mark=at position 1 with {\arrow[scale=4,>=stealth]{>}}},
    postaction={decorate},
]
\tikzstyle{directedText} = [
    text centered,font=\Large,execute at begin node=,
]
\tikzstyle{bolded} = [
    arrow,
    line width=1mm,
]
\tikzstyle{emptycomponent} = [
    rectangle,rounded corners=2ex,
    minimum width=4cm,minimum height=2cm,inner sep=0pt,text width=4cm,
    text centered, 
]

\usepackage{imakeidx}
\usepackage[all]{foreign}

\usepackage[abbreviations,numberedsection=autolabel,automake]{glossaries-extra}

\makeglossaries%

\glsenableentrycount%
\glssetcategoryattribute{abbreviation}{entrycount}{0}
\glssetcategoryattribute{name}{entrycount}{1}
\glssetcategoryattribute{footname}{entrycount}{0}
\setabbreviationstyle[footname]{short-footnote}
\glssetcategoryattribute{footname}{nohyper}{true}

\newabbreviation[type=main,category=footname]{aecl}{AECL}{\textit{Atomic Energy of Canada Limited}}
\newabbreviation[type=main,category=footname]{av}{AV}{anti\=/virus}
\newabbreviation[type=main,category=footname]{ccta}{CCTA}{\textit{Central Computer and Telecommunications Agency}}
\newabbreviation[type=main,category=footname]{cramm}{CRAMM}{\textit{\cgls{ccta} Risk Analysis and Management Method}}
\newabbreviation[type=main,category=footname]{cwe}{CWE}{\textit{Common Weaknesses Enumeration}}
\newabbreviation[type=main,category=footname]{cve}{CVE}{\textit{Common Vulnerabilities and Exposures}}
\newabbreviation[type=main,category=footname]{dos}{DoS}{Denial of Service}
\newabbreviation[type=main,category=footname]{ecg}{ECG}{electrocardiography}
\newabbreviation[type=main,category=footname]{fair}{FAIR}{\textit{Factor Analysis of Information Risk}}
\newabbreviation[type=main,category=footname]{hipaa}{HIPAA}{\textit{Health Insurance Portability and Accountability Act}}
\newabbreviation[type=main,category=footname]{iec}{IEC}{\textit{International Electrotechnical Commission}}
\newabbreviation[type=main,category=footname]{isamm}{ISAMM}{\textit{Information Security Assessment \& Monitoring Method}}
\newabbreviation[type=main,category=footname]{nistsp}{NIST\=/SP}{\textit{National Institute of Standards and Technology Special Publication}}
\newabbreviation[type=main,category=footname]{nsa}{NSA}{\textit{National Security Agency}}
\newabbreviation[type=main,category=footname]{obgyn}{OB/GYN}{obstetrics and gynecology}
\newabbreviation[type=main,category=footname]{octave}{OCTAVE}{\textit{Operationally Critical Threat, Asset, and Vulnerability Evaluation}}
\newabbreviation[type=main,category=footname]{smb}{SMB}{Server Message Block}
\newabbreviation[type=main,category=footname]{url}{URL}{uniform resource locator}
\newabbreviation[type=main,category=footname]{vha}{VHA}{\textit{Veterans Health Administration}}
\newabbreviation[type=main,category=footname]{snr}{SNR}{signal to noise ratio}

\newabbreviation{dicom}{DICOM}{\textit{Digital Imaging and Communications in Medicine}}
\newabbreviation{fda}{FDA}{\textit{Food and Drug Administration}}
\newabbreviation{nhs}{NHS}{\textit{National Health Service}}
\newabbreviation{iso}{ISO}{\textit{International Organization for Standardization}}
\newabbreviation{capec}{CAPEC}{\textit{Common Attack Pattern Enumeration and Classification}}
\newabbreviation{tldr}{TLDR}{\textit{Threat identification, ontology\=/based Likelihood, severity Decomposition, and Risk integration}}
\newabbreviation{afd}{AFD}{\textit{Attack Flow Diagram}}
\newabbreviation{ifv}{IFV}{\textit{Information Flow Vector}}
\newabbreviation{ise}{ISE}{\textit{Information Security Expert}}
\newabbreviation{medle}{MEDLE}{\textit{Mean of the Experts' Direct\=/Likelihood Estimates}}
\newabbreviation{mecble}{MECBLE}{\textit{Mean of the Experts' \cgls{capec}\=/Based Likelihood Estimates}}
\newabbreviation{me}{ME}{\textit{Medical Expert}}
\newabbreviation{rme}{RME}{\textit{Radiology Medical Expert}}
\newabbreviation{mid}{MID}{medical imaging device}

\newabbreviation{irs}{IRS}{Image Reconstruction System}

\newabbreviation{dna}{DNA}{deoxyribonucleic acid}
\newabbreviation{it}{IT}{information technology}
\newabbreviation{ct}{CT}{computed tomography}
\newabbreviation{mri}{MRI}{magnetic resonance imaging}
\newabbreviation{spect}{SPECT}{single\=/photon emission computed tomography}
\newabbreviation{pet}{PET}{positron\=/emission tomography}
\newabbreviation{icd}{ICD}{implantable\=/cardioverter defibrillator}
\newabbreviation{pacs}{PACS}{picture archiving and communication system}
\newabbreviation{ciso}{CISO}{chief information security officer}
\newabbreviation{hmo}{HMO}{health maintenance organization}
\newabbreviation{mems}{MEMS}{microelectromechanical systems}
\newabbreviation{vpn}{VPN}{virtual private network}
\newabbreviation{emr}{EMR}{electronic medical record}
\newabbreviation{cis}{CIS}{contrast injection system}
\newabbreviation{rf}{RF}{radio frequency}

\newabbreviation[category=name]{ot}{OT}{operational technology}
\newabbreviation[category=name]{os}{OS}{operating system}
\newabbreviation[category=name]{cr}{CR}{computed radiography}
\newabbreviation[category=name]{cia}{CIA}{confidentiality, integrity, and availability}
\newabbreviation[category=name]{hmi}{HMI}{human\=/machine interface}
\newabbreviation[category=name]{dfd}{DFD}{data flow diagram}
\newabbreviation[category=name]{ris}{RIS}{radiological information system}
\newabbreviation[category=name]{www}{WWW}{world wide web}
\newabbreviation[category=name]{pp}{PP}{pulse programmer}
\newabbreviation[category=name]{sar}{SAR}{specific absorption rate}
\newabbreviation[category=name]{fov}{FoV}{field of view}
\newabbreviation[category=name]{mr}{MR}{magnetic resonance}
\newabbreviation[category=name]{dr}{DR}{digital radiography}
\newabbreviation[category=name]{ppm}{ppm}{parts per million}
\newabbreviation[category=name]{dsv}{DSV}{diameter spherical volume}
\newabbreviation[category=name]{vr}{VR}{virtual reality}
\newabbreviation[category=name]{lan}{LAN}{local area network}

\def\bitcoinA{%
    \leavevmode
    \vtop{\offinterlineskip%
        \setbox0=\hbox{B}%
        \setbox2=\hbox to\wd0{\hfil\hskip-.03em% chktex 41 chktex 1
        \vrule height .3ex width .15ex\hskip .08em% chktex 41
        \vrule height .3ex width .15ex\hfil}
        \vbox{\copy2\box0}\box2}}% chktex 41

\usepackage{siunitx}
\DeclareSIUnit\peak{p}
\DeclareSIUnit\kVp{\kilo\volt\peak}
\DeclareSIUnit\mAs{\milli\ampere\second}

\usepackage{amsthm}
\usepackage{thmtools}

\declaretheorem[
    name=Remark, 
    refname={remark,remarks}, 
    Refname={Remark,Remarks}, 
    style=remark,
]{remark}

\newtheoremstyle{breakindent}%
  {}%
  {}%
  {\addtolength{\leftskip}{2.5em}}%
  {-2.5em}%
  {\bfseries}%
  {.}%
  {\newline}%
  {}%

\newtheoremstyle{break}%
  {}%
  {}%
  {}%
  {}%
  {\bfseries}%
  {.}%
  {\newline}%
  {}%
\declaretheorem[
    name=Attack, 
    refname={attack,attacks}, 
    Refname={Attack,Attacks}, 
    style=break,
]{attack}

\declaretheorem[
    name=CAPEC,
    refname={CAPEC,CAPECs},
    Refname={CAPEC,CAPECs},
    style=definition,
]{CAPEC}
\declaretheorem[
    name=CAPEC NEW,
    refname={CAPEC,CAPECs},
    Refname={CAPEC,CAPECs},
    style=definition,
    numbered=no,
]{CAPEC-NEW}

\declaretheorem[
    name=Definition,
    refname={definition,definitions},
    Refname={Definition,Definitions},
    style=definition, 
    numberwithin=section,
]{definition}

\begin{document}
%-------------------------------------------------------------------------------

\begin{frontmatter}
    \title{A New Methodology for Information Security Risk Assessment for Medical Devices and Its Evaluation}
    
    \author{Tom~Mahler\corref{correspondingauthor}}
    \author{Yuval~Elovici}
    \author{Yuval~Shahar}
    
    \cortext[correspondingauthor]{Corresponding author\newline \textit{Email address:} \emailref{mahlert@post.bgu.ac.il} (Tom~Mahler)}
    \address{The Department of Software and Information Systems Engineering (SISE), Ben\=/Gurion University of the Negev, Israel}
    
    %-------------------------------------------------------------------------------
\begin{abstract}
%-------------------------------------------------------------------------------
    As technology advances towards more connected and digital environments, medical devices are becoming increasingly connected to hospital networks and to the Internet, which exposes them, and thus the patients using them, to new cybersecurity threats.
    Currently, there is a lack of a methodology dedicated to information security risk assessment for medical devices.
    
    In this study, we present the \glsxtrfull{tldr} methodology for information security risk assessment for medical devices.
    The \glsxtrshort{tldr} methodology uses the following steps: 
    \begin{inparaenum}[(1)]
        \item \emph{identifying the potentially vulnerable components} of medical devices, in this case, four different \glsxtrfullpl{mid};\
        \item \emph{identifying the potential attacks}, in this case, \emph{23 potential attacks} on \glsxtrshortpl{mid};\
        \item \emph{mapping the discovered attacks} into a known attack ontology --- in this case, the \glsxtrfullpl{capec};\
        \item \emph{estimating the likelihood} of the mapped \glsxtrshortpl{capec} in the medical domain with the assistance of a panel of senior healthcare \glsxtrfullpl{ise};
        \item \emph{computing the \glsxtrshort{capec}\=/based likelihood estimates} of each attack;
        \item \emph{decomposing} each attack into several \emph{severity aspects} and assigning them \emph{weights};
        \item assessing the \emph{magnitude of the impact} of each of the severity aspects for each attack
        with the assistance of a panel of senior \glsxtrfullpl{me};
        \item \emph{computing the composite severity} assessments for each attack; and finally,
        \item \emph{integrating} the likelihood and severity of each attack into its risk, and thus \emph{prioritizing} it.
    \end{inparaenum}
    The details of steps six to eight are beyond the scope of the current study;
    in the current study, we had replaced them by a single step that included asking the panel of \glsxtrshortpl{me} [in this case, radiologists], to \emph{assess the overall severity} for each attack and use it as its severity.
    
    We have demonstrated that using the \glsxtrshort{tldr} methodology, healthcare \glsxtrshortpl{ise} can reach a form of a consensus regarding the likelihood estimates of cyber attacks and \glsxtrshortpl{capec} while maintaining the validity of the risk assessments' absolute values.
    To show this, we asked the \glsxtrshortpl{ise} to estimate also the \emph{direct} overall likelihood for each attack individually.
    We demonstrated the value and the validity of using the \glsxtrshort{tldr} methodology with respect to the relative ranking of the 23 potential threats, by calculating the pairwise Spearman's correlation between the \glsxtrshort{capec}\=/based likelihood estimates and the direct estimates of each of our \glsxtrshortpl{ise}, and showing that the correlation is high, while the pairwise correlation between the mean of the direct estimates and each of the \glsxtrshortpl{ise}' overall direct estimates was lower.
    We also calculated the paired \(T\)\=/test statistic between the \glsxtrshort{capec}\=/based estimates and the direct \glsxtrshortpl{ise}' estimates, and showed that the null hypothesis is accepted (i.e., there was no significant difference between them). % chktex 21
    This implies that the \glsxtrshort{capec}\=/based likelihood estimates, using the \glsxtrshort{tldr} methodology, are as valid as current direct risk assessment methodologies' likelihood estimates; however, the \glsxtrshort{capec}\=/based likelihood estimates are much easier to calculate.
    
    With the \glsxtrshort{tldr} methodology, members of the healthcare industry who provide the severity assessments are likely to be more aware of the implications of these potential risks, might be able to address them more efficiently, and might be better positioned to prioritize the protection efforts for \glsxtrshortpl{mid} and for multiple other medical devices and ecosystems to which the \glsxtrshort{tldr} methodology can be applied.
    In addition, mapping all potential medical devices attacks into a small number of \glsxtrshortpl{capec}, as we had demonstrated comprehensively in the case of \glsxtrshortpl{mid}, facilitates the acquisition of their likelihoods from \glsxtrshortpl{ise} and the computation of the \glsxtrshort{capec}\=/based likelihood estimates for each attack, and thus, also, of its risk.
    \end{abstract}
    

\end{frontmatter}

\clearpage

%-------------------------------------------------------------------------------
\section{Introduction}
%-------------------------------------------------------------------------------

As part of the rapid development of medical device technology, medical devices are becoming increasingly connected to hospital networks and even to the Internet, including many devices that have become critical resources in healthcare organizations, such as \cglspl{mid}.
Consequently, medical devices today are facing new cyber security threats which include:
\begin{inparaenum}[(1)]
    \item an increasing potential attack surface due to connection to hospital networks and the Internet;
	\item a rising number of outdated and unpatched devices, for which known vulnerabilities that can be easily exploited already exist;
	\item ongoing difficulty in updating the devices' software, due to strict regulations which takes too long to comply with (certification was typically given only to the device that included the original hardware and software);
	\item lack of awareness and insufficient investment in security among medical device manufacturers, regulatory authorities, healthcare personnel, and patients; and
	\item increasing interest in the healthcare industry by cyber warfare's leading players due to the potential high revenue of this domain (e.g., ransomware of expensive or life\=/supporting devices, or black market trade of highly\=/priced patients medical records).
\end{inparaenum}

Such threats are no longer just the imagination of William Gibson's science fiction stories\footnote{A science fiction writer that notably coined the term \emph{cyberspace} in his short story \textit{Burning Chrome} (1982)~\cite{Gibson2017} and made it famous in his ground\=/breaking novel, \textit{Neuromancer} (1984).}.
In May 2017, the \nameref{appendix:WannaCry} ransomware attack~\cite{Larson17, Liptak17, Millar17} spread worldwide, infecting over 200,000 devices in more than 150 countries~\cite{Brenner17}, including tens of thousands of the UK's \cgls{nhs} hospitals' devices~\cite{BBC17} and \cglspl{mid}~\cite{UngoedThomas17}, causing them to be non\=/operational by encrypting them.
This attack caused several hospitals to turn away patients~\cite{Hern17, CarrieWong2017} and divert ambulance routes~\cite{Foxx17} (see appendix~{\S}~{\ref{appendix:WannaCry}} for more details).
A year earlier, in February 2016, the \textit{Hollywood Presbyterian Medical Center} in Los Angeles was hit by another ransomware attack~\cite{Skinner16, Strickland16}, potentially related to the \textit{Locky} ransomware~\cite{Davis16}, which seized control of the hospital's computer system and encrypted all of its files; the \textit{Guardian}~\cite{Wong16} reported that ``the hospital attackers are demanding a ransom of {\bitcoinA}9,000 Bitcoin (about \$3.6 million [at that time]) to decrypt the files'', and the hospital eventually ``paid a \$17,000 ransom'' to release the files, as reported by the \textit{Los Angeles Times}~\cite{Winton16}.
These examples demonstrate the high financial motivation of criminals to target critical healthcare infrastructure, such as \cglspl{mid}.\

Furthermore, medical devices, such as \cglspl{mid} and radiation therapy medical devices, could be used maliciously to cause physical harm to patients due to the use of powerful ionizing radiation, strong magnets, gamma radiation, etc.
To date, there have been no reports of cyber attacks that caused such physical harm to patients; however, the \textit{Therac\=/25} incident teaches us that this scenario is indeed possible.
\textit{Therac\=/25} was a radiation therapy medical device for the treatment of cancer, designed by the \cgls{aecl}; this device was involved with several incidents of radiation overdose in 1985\=/1987, which Leveson and Turner~\cite{Leveson93} reviewed.
The device's software contained dangerous vulnerabilities associated with improper safety interlock implementations, which were implemented by software instead of hardware.
These vulnerabilities enabled a race condition to occur in certain situations, resulting in patients receiving massive amounts of direct radiation, sometimes a hundred times more than the usual dose.
The incident involved at least six patients that suffered severe radiation burns and paralysis of different areas of the body (e.g., hands, legs, and vocal cords), neurogenic bowel and bladder, disorientation, coma, and even death.
As far as we know, this incident was not the result of a cyber attack \foreign{per~se}; nevertheless, it demonstrates the potential risks of these devices if they are compromised.

Today, reports on cyber attacks and adversarial events targeting medical devices such as \cglspl{mid} are still not frequent, possibly due to the lack of incentives, lack of federal safe harbor policies, lack of clear, actionable guidance, and lack of meaningful and convenient reporting mechanisms~\cite{Kramer12}.
A publication by Ayala~\cite{Ayala16} provides general information regarding the security of medical devices, briefly describing several potential attacks.
While this is an important initial step towards raising awareness about the security of medical devices, it lacks a methodology for information security risk assessment of these threats.

The \cgls{iso} defines information security risk assessment as the: ``process to comprehend the nature of risk and to determine the level of risk, [providing] the basis for risk evaluation and decisions about risk treatment [including] risk estimation\ldots''~\cite{ISO73:2009}.
Risk assessment is usually the main part of an overall risk management strategy, which also includes control measures of mitigating the potential risks~\cite{Popov2016,Rausand2013}.
The \cgls{hipaa}, the US legislation for security and privacy of medical information, requires healthcare organizations to ``conduct an accurate and thorough assessment of the potential risks and vulnerabilities to the confidentiality, integrity, and availability of electronically protected health information held by the [organization].''~\cite{HHS06a} (\cgls{hipaa}~{\S}~{164.308(a)(1)(ii)(A)}). % chktex 36
Note that \cgls{hipaa} does not require the use of a specific information security risk assessment methodology (see~{\S}~{164.316(b)(1)}~\cite{HHS06a}). % chktex 36
Consequently, many \cglspl{ciso} in \cglspl{hmo} are using common risk assessment methods, which are designed to protect the patients' \emph{information} rather than the patients' \emph{well\=/being}.

Therefore, there is a need for a new methodology for information security risk assessment for medical devices.
We believe that the application of the new information security risk assessment methodology we are proposing, the \cgls{tldr} methodology, to \cglspl{mid}, and perhaps to additional medical devices in the future, will pave the way for future research of medical devices security and lead to the development of more secure medical devices.

The main contributions of this study are:
\begin{itemize}[noitemsep]
    \item We present the essential background for our study, including the current state\=/of\=/the\=/art medical devices security and the security challenges for the healthcare industry.
	\item We present the \cgls{tldr} methodology for information security risk assessment for medical devices, which includes:
    \begin{inparaenum}[(1)]
        \item \emph{identifying the potentially vulnerable components} of medical devices using \cglspl{afd};\
        \item \emph{identifying the potential attacks} and marking them on the \cglspl{afd};\
        \item \emph{mapping the discovered attacks} into their relevant \cglspl{capec};\
        \item \emph{estimating the likelihood} of the mapped \cglspl{capec} in the medical domain, with the assistance of a panel of senior healthcare \cglspl{ise};\
        \item \emph{computing the \cgls{capec}\=/based likelihood estimates} for each attack, using the mean of the \cglspl{ise}' likelihood estimates of the \cglspl{capec} into which the attack is mapped;
        \item \emph{decomposing} each attack into several \emph{severity aspects} and assigning them \emph{weights};
        \item assessing the \emph{magnitude of the impact} of each of the severity aspects for each attack, with the assistance of a panel of \cglspl{me};\;
        \item \emph{computing the composite severity} assessments for each attack; and finally,
        \item \emph{integrating} the likelihood and severity of each attack into its risk, and thus \emph{prioritizing} it.
    \end{inparaenum}\newline
    The details of steps six to eight are beyond the scope of the current study and are a part of another study.
	\item We present the application of the \cgls{tldr} methodology to \cglspl{mid}:\
    \begin{itemize}[noitemsep]
        \item We identified the potentially vulnerable components of \cglspl{mid} by creating four \cglspl{afd} for the generic \cgls{mid}, a generic \cgls{ct}, a generic \cgls{mri}, and a generic ultrasound;
        \item we identified a total of 23 potential attacks on \cglspl{mid}: 15 known attacks and eight new attacks that we discovered using the \cglspl{afd};
        \item We mapped of all 23 discovered attacks to just eight out of 517 existing \cgls{capec}, and a new, \nth{518}, \cgls{capec} attack pattern that we suggest adding;
        \item We estimated the likelihood of the nine generic \cglspl{capec} with the assistance of our panel of four senior healthcare \cglspl{ise} and used these estimates to compute the \cgls{capec}\=/based likelihood estimates of each attack;
        \item We assessed, for all medical device attacks, the overall severity of each attack\footnote{Note that this single simple step would usually include steps~\ref{methods:Decomposing each attack into several severity aspects and assigning them weights} to~\ref{methods:Computing the composite severity assessments for each attack} of the \cgls{tldr} methodology; however, these steps and their validation are beyond the scope of the current study and are a part of another study.} with the assistance of a panel of four senior \cglspl{me} [in this case, radiologists]; and finally,
        \item We integrated the likelihood and severity of each attack into its risk and prioritized the attacks.
    \end{itemize}    
    \item We \emph{demonstrated} that by using the \cgls{tldr} methodology, \cglspl{ise} could reach a [mean] consensus on the relative ranking of the \cgls{capec}\=/based likelihood estimates of the different attacks, while also maintaining the validity of the risk assessments' absolute values.
\end{itemize}

To show this, we asked the \cglspl{ise} to estimate also the \emph{direct} overall likelihood for each attack individually.
We then defined two ``consensus'' vectors of the 23 potential attacks:
\begin{inparaenum}[(1)]
    \item the \glsxtrfull{mecble} (computed indirectly), and
    \item the \glsxtrfull{medle}.
\end{inparaenum}
We demonstrated the value and the validity of using the \cgls{tldr} methodology with respect to the relative ranking of the 23 potential threats, by calculating the pairwise Spearman's correlation between the \cgls{capec}\=/based (\cgls{mecble}) likelihood estimates and the direct estimates of each of our \cglspl{ise}, and showing that the correlation is high, while the pairwise correlation between the mean of the direct estimates (the \cgls{medle}) and each of the \cglspl{ise}' overall direct estimates was lower.
We also demonstrated that we maintain the validity of the risk assessment process, by calculating the paired \(T\)\=/test statistic between the \cgls{capec}\=/based (\cgls{mecble}) estimates and the direct \cglspl{ise}' estimates, and showing that the null hypothesis is accepted (i.e., there was no significant difference between them). % chktex 21
This implies that the \cgls{capec}\=/based (\cgls{mecble}) likelihood estimates, using the \cgls{tldr} methodology, are as valid as current direct risk assessment methodologies' likelihood estimates; however, the \cgls{capec}\=/based likelihood estimates, computed from the mapped \cglspl{capec}, are much easier to calculate.

The rest of the paper is structured as follows.
First, we provide the essential background~({\S}~{\ref{background:Background}}) for this study, on medical devices security~({\S}~{\ref{background:Medical Device Security}}), information security risk assessment methodologies ({\S}~{\ref{background:Information Security Risk Assessment Methodologies}}), \cgls{capec} mechanism of attack~({\S}~{\ref{background:CAPEC Mechanisms of Attack}}), and \cglspl{mid}~({\S}~{\ref{background:medical imaging devices}}).
We then present the \cgls{tldr} methodology for information security risk assessment for medical devices and the evaluation methods of the \cgls{tldr} methodology~({\S}~{\ref{methods:Methods}}).
We then present the results~({\S}~{\ref{results:Results}}) of the application of the \cgls{tldr} methodology to \cglspl{mid} and the validation of its correctness, with a panel of four healthcare \cglspl{ise}.
The \cgls{tldr} composite severity assessment part of the \cgls{tldr} methodology, which includes severity decomposition into severity aspects, its assessment, and its validation, is beyond the scope of the current study and will be discussed in a separate paper that we are preparing.
Finally, we summarize our study~({\S}~{\ref{sec:Summary}}), discuss the implications of our results~({\S}~{\ref{sec:Discussion}}), and conclude the study~({\S}~{\ref{sec:Conclusions}}).
The following supplementary information is provided in the appendices: a description of the \nameref{appendix:WannaCry} attack~({\S}~{\ref{appendix:WannaCry}}), and a list of abbreviations~({\S}~{\hyperref[appendix:Abbreviations]{Appendix~B}}).

%-------------------------------------------------------------------------------
\section{Background}\label{background:Background}
%-------------------------------------------------------------------------------

In this section, we present some essential background for our study.
We first discuss the current state of the art of medical devices security~({\S}~{\ref{background:Medical Device Security}}) and stress why \cglspl{mid} are particularly vulnerable.
Following that, we briefly describe currently used information security risk assessment methodologies and define how to calculate the risk of an attack~({\S}~{\ref{background:Information Security Risk Assessment Methodologies}}).
We then present the \cgls{capec} mechanism of attack~({\S}~{\ref{background:CAPEC Mechanisms of Attack}}), which we use in the \cgls{tldr} methodology in the next section~({\S}~{\ref{methods:Methods}}).
Finally, we present a background on \cglspl{mid}~({\S}~{\ref{background:medical imaging devices}}).

\subsection{Medical Device Security}\label{background:Medical Device Security}

Medical devices are in the process of evolving from analog, standalone, disconnected from the outside world, devices to sophisticated devices with advanced computing and communication capabilities, such as \cgls{lan} and wireless connectivity, and, in extreme cases, even public cloud connectivity~\cite{Camara15}.
Whereas these technology advancements have many benefits (e.g., improved patient care), they also poses numerous security vulnerabilities that could be exploited by attackers to create dangerous scenarios for patients such as disabling the device and its services (e.g., ransomware), disrupting the device's proper operation and resources (e.g., battery depletion), tampering with the private data stored and transmitted by the device, etc.
Furthermore, wired and wireless connectivity enables remote attacks that do not even require the attackers to be physically close to the targeted device or connected to hospital networks~\cite{Panescu08}.

For example, by disabling or reprogramming therapies delivered by implanted devices, such as an \cgls{icd} with wireless connectivity~\cite{Li11}, researchers were able to induce a shock to the \cgls{icd} capable of causing ventricular fibrillation (i.e., a fatal heart rhythm)~\cite{Fu09} and cause depletion of the battery, rendering the device inoperative~\cite{Hei10} and resulting in a situation in which the patient necessitated a surgical procedure to replace the device.
Such risks led former US vice president Dick Cheney to disable the wireless capabilities on his \cgls{icd} in 2007 to avoid being targeted by terrorists~\cite{Franzen13}.
Currently, nothing prevents such remote attacks from finding their way into \cglspl{mid} as well.

Today, cyber warfare's leading players are groups of cyber terrorists and hacktivists (i.e., politically\=/driven), cyber criminals (i.e., profit\=/driven), and nation\=/state supported attackers (i.e., militarily\=/driven), which do not target \cgls{it} systems exclusively but also target critical infrastructure industries, such as manufacturing, financial services, transportation, government, and healthcare.
According to the \cgls{fda}: ``cyber actors will likely increase cyber intrusions against health care systems --- to include medical devices\ldots''~\cite{FBICyberDivision2014a}.
In January 2016, the \glsxtrshort{vha}\footnote{\glsxtrlong{vha}, an entity encompassing hospitals and facilities for US war veterans, serving more than 20 million veterans.} reported~\cite{USIS16} blocking more than 76 million intrusion attempts, over 638 million malware\=/based attacks, more than 99 million malicious emails, and three infected medical devices --- in just one month~\cite{USVA16}.
Hence, it is expected that in the near future, cyber attacks targeting the healthcare industry will become more frequent, more sophisticated, and more dangerous.

\cGlspl{mid} are particularly vulnerable and attractive targets to cyber attacks since they are:
\begin{inparaenum}[(1)]
	\item connected to hospital networks and in many cases to the Internet, increasing the potential attack surface of attackers;
	\item considered critical resources in the healthcare industry, widely used and depended upon in all aspects of health care, from diagnoses of patients' medical conditions to guiding physicians during surgery;
	\item very expensive and thus many hospitals can only procure a few \cglspl{mid}, making then an even more critical resource;
	\item very complex, often consisting of an entire ecosystem of components, making the cost of a single \cgls{mid} extremely high;
	\item storing and accessing patients' private medical data, exposing them to privacy associated risks (e.g., leakage of confidential data); and
	\item accompanied by risks to patients' physical health (e.g., radiation exposure).
\end{inparaenum}

A recent paper, Ferrara~\cite{Ferrara2019} surveyed cybersecurity in medical imaging and reached similar conclusions.
In his paper, Ferrara emphasizes that healthcare organizations are not expected to be completely resistant to cyber attacks; however, they should consider carefully how to manage and mitigate risks.
In the next section, we discuss currently used information security risk assessment methodologies and why there is a need for a new information security risk assessment methodology for medical devices.

\subsection{Information Security Risk Assessment Methodologies}\label{background:Information Security Risk Assessment Methodologies}

Fenz \etal{~\cite{Fenz14}} reviewed commonly used risk assessment methods including the \cgls{nistsp} 800\=/30~\cite{NISTSP12}, the \cgls{iso}/\glsxtrshort{iec} 27005~\cite{isoiec2700508} by the \cgls{iso} and the \cgls{iec}, the \cgls{octave}~\cite{OCTAVE03}, \cgls{cramm}, by the UK \cgls{ccta}~\cite{Farquhar1991, Yazar02}, the \cgls{fair}~\cite{Freund2015}, and the \cgls{isamm}~\cite{ISAMM07}.
We found that these methods usually consists of the following steps: 
\begin{inparaenum}[(1)]
	\item \emph{definition of the potentially vulnerable assets} of the organization,
	\item \emph{identification of the potential threats} to these assets,
	\item \emph{estimation of the likelihood} of the identified threats to occur,
	\item \emph{assessment of the severity} (or \emph{impact}) of the identified threats if occurred,
	\item \emph{integration of the risk} of the identified threats using the likelihood and severity, and 
	\item \emph{prioritization of threats} based on the risk.
\end{inparaenum}

A conventional method for integrating the risk of potential threats, which we found to appear in most risk assessment methodologies such as those listed above, can be simply defined as follows:
\begin{definition}[Risk]\label{def:Risk}
	Risk of a threat (e.g., cyber attack) is a function \(Risk:Likelihood\times Severity\to Risk\), defined in \Cref{eq:Basic Risk Calculation}:
	\begin{equation}\label{eq:Basic Risk Calculation}
	    Risk = Likelihood \cdot Severity
	\end{equation}
	Where \(Likelihood\) represents the probability of the occurrence of the threat, and \(Severity\) represents the potential impact of the threat.
\end{definition}
\begin{remark}
    The \(Severity\) is usually decomposed into \cgls{cia} aspects.
\end{remark}
\begin{remark}
	Although economists and decision analysts use other methods, it is common in the information security community, in which knowledge of likelihoods and impacts is often scarce, to estimate the \(Likelihood\) probability and the \(Severity\) qualitatively (e.g., using a zero to five scale).
\end{remark}

In 2017, Stine \etal{~\cite{Stine2017}} proposed a risk scoring system specifically designed for medical devices.
Their methodology focuses on decomposing the severity of each attack into five aspects: loss or denial of view, loss or denial of control, manipulation of view, manipulation of control, and denial or manipulation of safety, and qualitatively estimating them.
The likelihood of each attack, in their methodology, was fixed to a probability of 1, which the authors justified by following the \cgls{fda} post\=/market guidance of using a naive worst\=/case estimate in the absence of adequate probability data~\cite{FDA2016}

In 2019, Yasqoob \etal{~\cite{Yasqoob2019IntegratedDevices}} proposed an integrated safety, security, and privacy risk assessment framework for medical devices.
In their methodology, the authors calculate the risk of specific vulnerabilities based on the \cgls{cwe} database of \cgls{cve}.\
They present a non\=/straightforward approach for estimating the likelihood of \cglspl{cve} as expanding the traditional \cgls{cia}, adding a fourth ``safety'' aspect.

While these methodologies represent important attempts to consider aspects that are specifically designed for medical devices with a \cgls{hmi}, which most previously mentioned risk assessment methods (that decomposed the severity to only three \cgls{cia} aspects) lacked --- they are not complete.
Stine's method lacks an estimation of likelihood.
Yasqoob's method is designed for only estimating \cglspl{cve}' risks, while attacks usually consist of several \cglspl{cve};\ thus, it is not clear how to estimate the risk of sophisticated attacks.
The results of both methods were not statistically validated.
Nevertheless, these methodologies demonstrate the unique requirements of an information security risk assessment for medical devices in the healthcare industry and strengthen our understanding that a complete and more reasonably accurate means are necessary.

\subsection[CAPEC:\ mechanisms of attack]{\Glsxtrshort{capec}: Mechanisms of Attack}\label{background:CAPEC Mechanisms of Attack}
The mechanisms of attacks we describe in this study are essentially the \emph{attack patterns}, which define how the attacks are performed.
Attack patterns can be classified using the \glsxtrfull{capec} taxonomy by the \textit{MITRE Corporation}~\cite{CAPEC}, which already includes 517 different attack patterns well\=/organized, well\=/reviewed, and confirmed by the community.

The \cgls{capec} hierarchy is divided into different levels of abstraction:
\begin{inparaenum}[(1)] 
	\item category, a collection of attack patterns based on some common characteristic;
	\item meta attack pattern, an abstract characterization of a technique used in an attack; and
	\item standard attack pattern, a specific technique used in an attack.
\end{inparaenum}
Furthermore, many of the attack patterns in \cgls{capec} also include additional useful information, such as:
\begin{enumerate}[noitemsep]
	\item\label{itm:Summary description} \emph{Summary description}: a standard description of the attack pattern that defines the weakness of the attack target and the steps performed by the attacker.
	\item\label{itm:Attack prerequisites} \emph{Attack prerequisites}: the necessary conditions for the attack to succeed.
	\item\label{itm:Attack severity} \emph{Attack severity}: range from one (i.e., very low) to five (i.e., very high).
	\item\label{itm:Likelihood of Attack Success} \emph{Likelihood of attack success}: the probability of an attack to succeed, considering target vulnerabilities, prerequisites, required skills, required resources, and the effectiveness of potential mitigation, range from one (i.e., very low) to five (i.e., very high).
	\item\label{itm:Methods of Attack} \emph{Methods of attack}: the attack vectors that identify the mechanisms used in the attack (e.g., brute force).
	\item\label{itm:Specific Knowledge or Skill Required to Conduct Attack} \emph{Specific knowledge or skill required to conduct attack}: the specific knowledge or skill required by an attacker to perform the attack, including an indication of the level of knowledge or skill required from 1 to 3.
	\item\label{itm:Potential Solutions or Mitigation} \emph{Potential solutions or mitigation}: any action or approach that may prevent the attack, reduce the probability, reduce the impact, or mitigate the effects of the attack.
\end{enumerate}

Thus, if we can map attacks on medical devices into \cgls{capec} attack patterns, we will be able to infer the relevant information from \cgls{capec} and apply it to these attacks (e.g., potential solutions or mitigation options, which will already exist within the \cgls{capec} attack pattern), or at least use such information as a good starting point.

%-------------------------------------------------------------------------------
\subsection[Medical Imaging Devices (MIDs)]{\glslink{mid}{Medical Imaging Devices (MIDs)}}\label{background:medical imaging devices}
%-------------------------------------------------------------------------------

Medical imaging is a broad field associated with creating a visual representation of the human body and internal tissues, using advanced technologies to diagnose, monitor, treat, manage, or study medical conditions~\cite{Hendee2010b}.
Different imaging technologies produce different results regarding the area of the body being inspected or treated; thus, medical imaging provides physicians with a vast amount of information about patients' medical conditions and the effectiveness of medical treatment~\cite{FDA17MedicalImaging}.
Medical imaging is used extensively throughout the healthcare system and patient lifespan, from prenatal imaging to geriatrics, encompassing personal imaging to population imaging, emergency care to chronic care, and treatment of various patients such as oncologic patients.

\cGlspl{mid} are composed of very sophisticated detectors that can measure the effects of various signals which are used to produce medical images.
The type of equipment used in each \cgls{mid} is referred to as the \emph{modality}, and the type of signal depends on the modality.
Unlike regular cameras (e.g., digital cameras), \cglspl{mid} do not usually detect visible light.
Instead, they detect other signals, such as magnetic field resonance, X\=/rays, gamma rays, ultrasonic waves, etc., which reveal the internal structures of the human body.
Each type of signal affects various tissues differently, providing diverse information about the area of the body being studied or treated; this information can be related to possible disease, injury, or the effectiveness of medical treatment.

\cGlspl{mid} can measure the effects of signals produced by the \cgls{mid} itself or signals transmitted naturally from the patient or preinjected material.
The latter can be done in most imaging exams by using substances that absorb or change the signals produced by the \cgls{mid}, or, by particular radiation\=/emitting substances that are injected into the patient's body before the scan.
Also, intravenous injections of iodinated contrast are often used for enhanced anatomical and functional assessment of various organs~\cite{FDA17CT}.
The \cgls{fda} has indicated that this exposes patients to potential risks associated with possible reactions to the agent~\cite{FDA17Xray}.

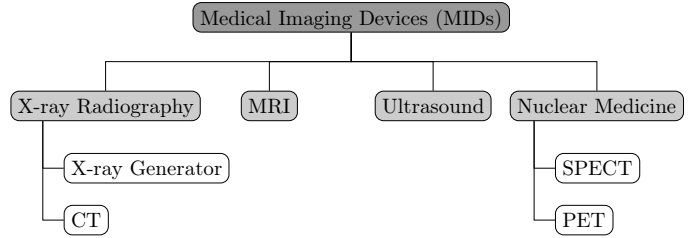
\begin{figure}[!ht]
	\centering
	\resizebox{\linewidth}{!}{\begin{tikzpicture}[%
    every node/.style={
            rectangle,rounded corners,
            minimum height=5mm,
            draw=black,
            align=center,
            text depth=0pt,
            text centered,
    },
    subdevice/.style={grow=down,xshift=-20pt,anchor=west,edge from parent path={(\tikzparentnode.south) +(-30pt,0) |- (\tikzchildnode.west)}},
    first/.style={level distance=30pt},
    second/.style={level distance=55pt},
    level 1/.style={sibling distance=2.75cm}
  ]
    \node[fill=gray!80] {\nameref{background:medical imaging devices}}
    [edge from parent fork down]
        child { node[fill=gray!40] {X\=/ray Radiography}
            child[subdevice,first] { node {\nameref{background:xray generator}}}
            child[subdevice,second] { node {\hyperref[background:ct]{CT}}}
        }
        child { node[fill=gray!40] {\hyperref[background:mri]{MRI}}}
        child { node[fill=gray!40] {\nameref{background:ultrasound}}}
        child { node[fill=gray!40] {\nameref{background:nuclear medicine}}
            child[subdevice,first] { node {\Glsxtrshort{spect}}}
            child[subdevice,second] { node {\Glsxtrshort{pet}}}};
\end{tikzpicture}\unskip}
	\caption[Taxonomy of MIDs.]{Taxonomy of \glsxtrshortpl{mid}.}\label{fig:Taxonomy of MIDs}
\end{figure}

\Cref{fig:Taxonomy of MIDs} presents the main \cglspl{mid} currently used by healthcare facilities worldwide.
Each modality has its applications and technology: X\=/rays are used in \cgls{ct}, \cgls{rf} is used in \cgls{mri}~\cite{hartwig09}, gamma rays are used in nuclear medicine, and high\=/frequency sound waves are used in ultrasound.
In the context of patient safety, a property shared by all modalities is the trade\=/off between the power of the projected energy, emitted by the device or modality, and its impact on patient safety and the quality of the image: a high X\=/ray dose, long magnetic resonance time, or high ultrasound power may improve the quality of the image; however, it may damage the tissue being scanned and harm the patient.
The cost of \cglspl{mid} and their technologies range from a few thousand dollars (e.g., ultrasound) to a few million dollars.

\paragraph{X\=/ray Generator}\label{background:xray generator}
X\=/ray generators produce radiographic images by projecting short X\=/ray pulses through the patient using electromagnetic energy sources and measuring the X\=/ray waves passing from the patient (i.e., after passing through the patient) using X\=/ray detectors.
Different tissue such as bone, soft tissue, and air inside the patient result in a heterogeneous distribution of X\=/rays that creates the image~\cite{Mahesh13}.
The X\=/ray's detector might be a \cgls{cr} cassette, where photo\=/stimulated luminescence screens capture the X\=/rays and transform them into digital form~\cite{Benseler06}, or \cgls{dr}, where images are directly transformed into digital form; this makes \cgls{dr} scans faster, accurate, and radiation dose effective (i.e., use the minimal radiation dose, above which there is no significant improvement in quality).
The unique properties of X\=/ray technology, such as projecting densities of different tissues on the image, makes X\=/ray use very popular in many medical domains, such as diagnosis of bone fractures, lung pathologies (including cancer), or dental diagnosis.
X\=/ray radiation is ionizing, with enough energy to potentially cause damage to the \cgls{dna} structure of different organic tissues.
Potential risks from exposure to such ionizing radiation include damage to tissues and skin or increased risk of cancer.
The actual radiation risks of different organs and tissues vary according to their different sensitivity to radiation exposure~\cite{RSNA17}.

\paragraph[Computed Tomography (CT)]{\Glsxtrfull{ct}}\label{background:ct}
\cGls{ct} is an essential and popular X\=/ray generator modality ({\S}~{\ref{background:xray generator}}).
\cGls{ct} images are produced by passing X\=/ray waves through the body at various angles; these X\=/ray waves are emitted from X\=/ray tubes rotating around the patient's body, producing many slices (two\=/dimensional images of a three\=/dimensional organ).
Arrays of detectors located opposite the X\=/ray tubes measure the exiting X\=/ray waves in analog form, and then the data is reconstructed into digital images using a computer.
Similar to the X\=/ray generator ({\S}~{\ref{background:xray generator}}), \cgls{ct} modalities produce ionizing radiation and thus may be dangerous~\cite{FDA17CT}.
This unique imaging technique enables the production of high\=/resolution images, making \cgls{ct} a favorite, significant, and critical resource for the diagnosis of many medical conditions, such as emergency diagnosis of a subdural hematoma, ruptured disks, aneurysms, and other pathologies.
\cGls{ct} modalities are very expensive; the cost depends on the number of detectors that the modality has (e.g., 16 slices, 64 slices, 128 slices, and even 256 slices).

\paragraph[Magnetic Resonance Imaging (MRI)]{\Glsxtrfull{mri}}\label{background:mri}
\cGls{mri} uses magnetic fields to create \cgls{rf} effects, which are measured to produce images.
During the scan, the patient is positioned inside a scanner tube which, consists of a magnet (currently a superconducting magnet is usually used), which produces strong magnetic fields (about 10,000 to 60,000 times stronger than the earth's magnetic field) around the patient.
The magnetic field aligns protons (i.e., hydrogen atoms) inside water molecules contained in different tissues of the patient.
\cGls{rf} pulses are used to excite the protons and move them from their precessional frequency.
When the protons return to their original position, they emit \cgls{rf} radiation, which is then measured by the \cgls{mri} modality.
Each tissue has a different proton resonance frequency; therefore, by measuring the \cgls{rf} emitted, the \cgls{mri} modality can reconstruct an image that represents internal tissues.
Since protons exist in water molecules, \cgls{mri} achieves the best results when diagnosing soft tissues, rich with hydrogen atoms.
Hence, it is popular for neurological imaging, as well as musculoskeletal imaging.
\cGls{mri} modalities are expensive, usually even more expensive than \cgls{ct} modalities, due to the cost of the magnet; thus, in many hospitals, only a few \cgls{mri} modalities are installed.
To improve the accuracy, the process of aligning the protons using the magnetic field and measuring the emitted \cgls{rf} radiation is repeated several times, making \cgls{mri} scans longer in duration than \cgls{ct} scans (i.e., several minutes compared to several seconds).
Because of this, \cgls{mri} scans may not be appropriate for individual patients, such as pediatric patients and patients that are unable to hold still for some time.
Moreover, the use of large magnetic fields prevents patients with pacemakers or other metal implants from getting \cgls{mri} scans.
Although \cgls{mri} does not emit ionizing radiation~\cite{Brown14}, the combination of static gradient magnetic fields and \cgls{rf} radiation during \cgls{mri} scans may generate heat and harm a patient's tissues or cause damage to the patient's \cgls{dna} structure~\cite{hartwig09}.

\paragraph{Ultrasound}\label{background:ultrasound}
Ultrasound uses a short duration of high\=/frequency sound waves, which are generated and transmitted into the patient's body.
The waves are then reflected to the device by the internal structures of the body, creating echoes.
These echoes are reconstructed into a linear array of tomographic slices of the tissues of interest, showing the motion of the tissues (e.g., fetus movements).
Ultrasound imaging is preferred in obstetrical patients to monitor the fetus during pregnancy, making ultrasound modalities common in a hospital's \cgls{obgyn}, and maternal care departments.
However, since sound waves do not travel well in air and bones, ultrasound is more limited compared to X\=/ray generators, \cgls{ct}, or \cgls{mri}.\

\paragraph{Nuclear Medicine}\label{background:nuclear medicine}
Nuclear medicine uses a chemical substance containing a radioactive isotope and measures the radioactive decay of the isotope, which is emitted in the form of gamma radiation, in order to produce the image.
The isotope is given to the patient before the exam orally, by injection, or by inhalation, and the exam begins once the agent (i.e., the isotope) has distributed itself according to the physiological status of the patient.
The most popular nuclear medicine techniques are gamma cameras: \cgls{spect} and \cgls{pet}~\cite{Mahesh13}.
These technologies are often used in neurological and cardiac imaging.
Hybrid imaging allows combining \cgls{spect} and \cgls{pet} technologies with \cgls{ct} and \cgls{mri} into one modality, such as \cgls{spect}/\cgls{ct}, \cgls{spect}/\cgls{mri}~\cite{Beyer11}, \cgls{pet}/\cgls{ct}, or \cgls{pet}/\cgls{mri}~\cite{Townsend08}.
Hybrid imaging usually produces enhanced results due to the combination of different technologies.
For example, in the cardiac domain, \cgls{spect}/\cgls{ct} imaging is used to image the heart at rest and during exercise to evaluate the heart's blood flow, which helps detect narrowing of the arteries~\cite{AHA16}.
Similarly, \cgls{pet}/\cgls{ct} imaging is used for diagnosing the effectiveness of chemotherapy in cancer or diagnosing the effectiveness of glucose metabolism.

%-------------------------------------------------------------------------------
\section[The TLDR Methodology]{The \glsxtrshort{tldr} Methodology}\label{methods:Methods}
%-------------------------------------------------------------------------------

\begin{figure}[!htb]
	\centering\captionsetup{singlelinecheck=off}
	\resizebox{\linewidth}{!}{\begin{tikzpicture}[
    align=center,
    node distance=.5cm and 0cm,
]
    \node (t) [annotation] {T};
    \node (311) [process, below=of t] {\nameref{methods:Identifying the potentially vulnerable components using AFDs} ({\S}~{\ref{methods:Identifying the potentially vulnerable components using AFDs}})};
    \node (312) [process, below=of 311] {\nameref{methods:Identifying the potential attacks and marking them on the AFDs} ({\S}~{\ref{methods:Identifying the potential attacks and marking them on the AFDs}})};
    \node (d) [annotation, right=of 312, xshift=1.75cm, yshift=-.5cm] {D};
    \node (l) [annotation, left=of 312, xshift=-2cm, yshift=-.5cm] {L};

    \draw [arrow] (311.south) -- (312.north);
    
    \node (316) [process, below right=of 312, xshift=-2cm] {\nameref{methods:Decomposing each attack into several severity aspects and assigning them weights} ({\S}~{\ref{methods:Decomposing each attack into several severity aspects and assigning them weights}})};
    \node (317) [process, below=of 316, yshift=-.5cm] {\nameref{methods:Assessing the magnitude of the impact of each of the severity aspects for each attack} ({\S}~{\ref{methods:Assessing the magnitude of the impact of each of the severity aspects for each attack}})};
    \node (318) [process, below=of 317, yshift=-.375cm] {\nameref{methods:Computing the composite severity assessments for each attack} ({\S}~{\ref{methods:Computing the composite severity assessments for each attack}})};
    
    \draw [arrow] (312.east) -| (316.north);
    \draw [arrow] (316) -- (317);
    \draw [arrow] (317) -- (318);
    
    \node (313) [process, below left=of 312, xshift=2cm] {\nameref{methods:Mapping the Discovered Attacks into Their Relevant CAPECs} ({\S}~{\ref{methods:Mapping the Discovered Attacks into Their Relevant CAPECs}})};
    \node (314) [process, below=of 313, yshift=-.5cm] {\nameref{methods:Estimating the likelihood of the generic CAPECs into which the potential attacks are mapped} ({\S}~{\ref{methods:Estimating the likelihood of the generic CAPECs into which the potential attacks are mapped}})};
    \node (315) [process, below=of 314, yshift=-.5cm] {\nameref{methods:Computing the CAPEC-based likelihood estimates for each attack} ({\S}~{\ref{methods:Computing the CAPEC-based likelihood estimates for each attack}})};
    
    \draw [arrow] (312.west) -| (313.north);
    \draw [arrow] (313) -- (314);
    \draw [arrow] (314) -- (315);
    
    \node (319) [process, below right=of 315, xshift=-2cm] {\nameref{methods:Integrating the likelihood and severity of each attack into its risk and prioritizing it} ({\S}~{\ref{methods:Integrating the likelihood and severity of each attack into its risk and prioritizing it}})};
    \node (r) [annotation, below=of 319] {R};

    \draw [arrow] (315.south) |- (319.west);
    \draw [arrow] (318.south) |- (319.east);
\end{tikzpicture}\unskip}
	\caption[The TLDR Methodology]{The \glsxtrshort{tldr} methodology:
	\begin{enumerate}[noitemsep,nolistsep]
		\item[\textbf{T}] \textbf{T}hreat identification 
		(\ref{methods:Identifying the potentially vulnerable components using AFDs},~\ref{methods:Identifying the potential attacks and marking them on the AFDs}).
		\item[\textbf{L}] ontology\=/based \textbf{L}ikelihood (\ref{methods:Mapping the Discovered Attacks into Their Relevant CAPECs} to~\ref{methods:Computing the CAPEC-based likelihood estimates for each attack}) performed by \glsxtrshortpl{ise}.
		\item[\textbf{D}] severity \textbf{D}ecomposition (\ref{methods:Decomposing each attack into several severity aspects and assigning them weights} to~\ref{methods:Computing the composite severity assessments for each attack}) performed by \glsxtrshortpl{me}\\%
		(beyond the scope of the current study).
		\item[\textbf{R}] \textbf{R}isk integration (\ref{methods:Integrating the likelihood and severity of each attack into its risk and prioritizing it}).
	\end{enumerate}}\label{fig:The TLDR Methodology}
\end{figure}
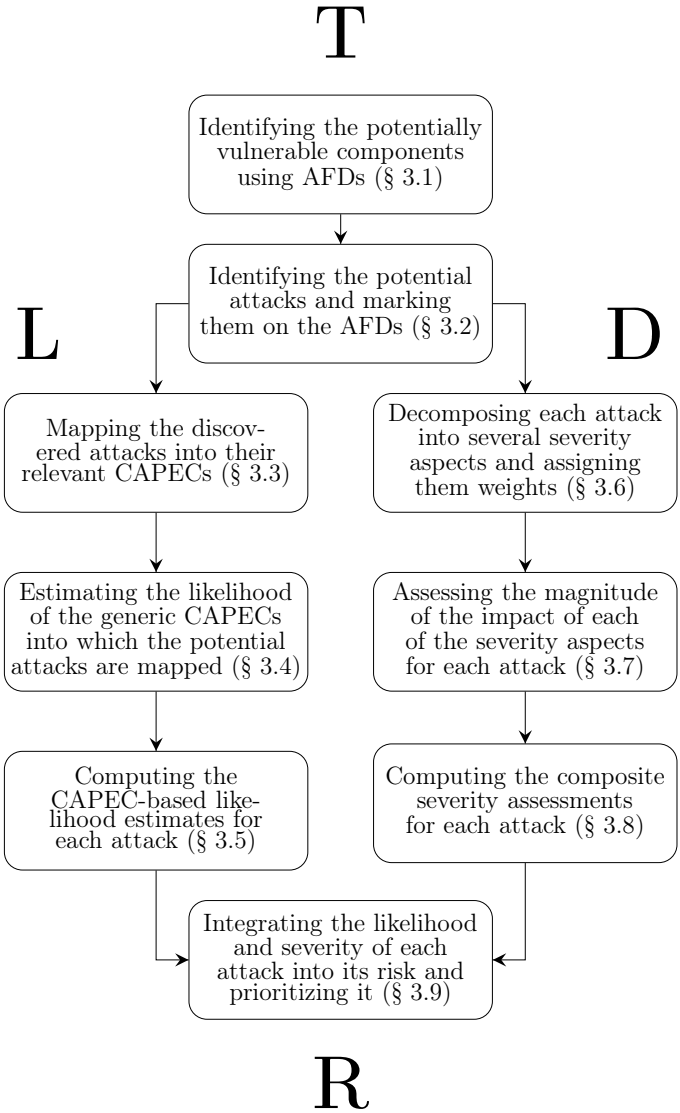

In this section, we present the \glsxtrfull{tldr}%
\footnote{\cGls{tldr} is our way of referring to the common Internet\=/based acronym Too Long; Didn't Read (TL;DR), since we hope that our methodology can facilitate the lengthy process of risk assessment.} %
methodology for information security risk assessment for medical devices.
Note that the \textbf{D} part of the \cgls{tldr} methodology ({\S}~{\ref{methods:Decomposing each attack into several severity aspects and assigning them weights}} to {\S}~{\ref{methods:Computing the composite severity assessments for each attack}}) is beyond the scope of the current study and will be discussed in a separate paper that we are preparing.

\subsection[Identifying the potentially vulnerable components using AFDs]{Identifying the Potentially Vulnerable Components Using \glsxtrshortpl{afd}}\label{methods:Identifying the potentially vulnerable components using AFDs}

The first step of the \cgls{tldr} methodology for information security risk assessment for medical devices is identifying the potentially vulnerable components using \glsxtrfullpl{afd}, which are diagrams of medical devices, consisting of their main components, and \cglspl{ifv} between these components.
On each \cgls{ifv}, we will mark in the next step~({\S}~{\ref{methods:Identifying the potential attacks and marking them on the AFDs}}) the IDs of all attacks that potentially utilize the \cgls{ifv} as part of the attack.
This helps to visualize, for each device: 
\begin{inparaenum}[(1)]
    \item its main components,
    \item the information flow between these components,
    \item all its relevant attacks, and
    \item the potential flow of these attacks.
\end{inparaenum}
This approach is based on the \cgls{dfd}, a widely used methodology for graphically representing the data flow inside complex information systems~\cite{Bruza93}.
The \cgls{afd} symbols are self\=/explanatory; nevertheless, a detailed legend of the symbols is provided in \Cref{tab:detailed legend of AFD's symbols}.

\begin{table}[!ht]
	\small%
	\centering%
	\caption[detailed legend of AFD's symbols]{Detailed Legend of \glsxtrshort{afd}'s Symbols.}\label{tab:detailed legend of AFD's symbols}%
	\begin{tabular}{@{}cm{.8\linewidth}@{}}
		\toprule%
		\textbf{Shape} & \multicolumn{1}{c}{\textbf{Description}} \\%
		\midrule%
		\begin{minipage}{.2\linewidth}
			\centering%
			\resizebox{\linewidth}{!}{\begin{tikzpicture}
    \node (component) [component] {Component};
\end{tikzpicture}\unskip}%
		\end{minipage} 
		&%
		A fundamental component that represents a process (or several processes) that performs a well\=/defined operation.\\\midrule%
		\begin{minipage}{.2\linewidth}
			\centering%
			\resizebox{\linewidth}{!}{\begin{tikzpicture}
    \node (subcomponent) [subcomponent] {Sub-component};
\end{tikzpicture}\unskip}%
		\end{minipage} &%
		A subcomponent is a component that is further expanded into an additional \glsxtrshort{afd}.\\\midrule%
		\begin{minipage}{.2\linewidth}
			\centering%
			\resizebox{\linewidth}{!}{\begin{tikzpicture}
    \node (terminator) [terminator] {Terminator};
\end{tikzpicture}\unskip}%
		\end{minipage} &%
		A terminator is a component which represents a beginning or termination of the flow.\\\midrule%
		\begin{minipage}{.2\linewidth}
			\centering%
			\resizebox{\linewidth}{!}{\begin{tikzpicture}
    \node (network) [network] {Network};
\end{tikzpicture}\unskip}%
		\end{minipage} &%
		A network represents an abstract structure of several connected components, such as the Internet or internal hospital networks.\\\midrule%
		\begin{minipage}{.2\linewidth}
			\centering%
			\resizebox{\linewidth}{!}{\begin{tikzpicture}[
    align=center,
]
    \draw[dashed] (0,0) rectangle (2,2) node[pos=.5,text width=2cm,text centered] {
        Logical\\
        Encapsulation
    };
\end{tikzpicture}\unskip}%
		\end{minipage} &%
		A logical encapsulation represents an encapsulation of several components, separating them from outer components.
		Outer components can consider the logical encapsulation as a ``black box'' with well\=/defined inputs and outputs for interaction.\\\midrule%
		\begin{minipage}{.2\linewidth}
			\centering%
			\resizebox{\linewidth}{!}{\tikzstyle{component} = [
    rectangle,rounded corners=2ex,
    minimum width=4cm,minimum height=2cm,inner sep=0cm,text width=4cm,
    text centered,font=\LARGE,execute at begin node=\setlength{\baselineskip}{1pt},
    draw=black,
]

\begin{tikzpicture}
    \node (component) [component,fill=gray!20!white] {
        Outer\\
        Component
    };
\end{tikzpicture}\unskip}%
		\end{minipage} &%
		An outer component is a component which is not directly a part of the analyzed device, usually outside a logical encapsulator.\\\midrule
		\begin{minipage}{.2\linewidth}
			\centering%
			\resizebox{\linewidth}{!}{\begin{tikzpicture}
    \node (A) [component] {Component A};
    \node (B) [component, below=of A, yshift=-1cm] {Component B};
    
    \draw [directed] (A.south) -- node[directedText, anchor=west, text width=1.5cm, align=center] {
        Directed\\
        Edge
    } (B.north);
\end{tikzpicture}\unskip}%
		\end{minipage} &%
		A directed edge is a potential \glsxtrshort{ifv} between two components.\\\midrule%
		\begin{minipage}{.2\linewidth}
			\centering%
			\resizebox{\linewidth}{!}{\begin{tikzpicture}
    \node (A) [component] {Component A};
    \node (B) [component, below=of A, yshift=-1.5cm] {Component B};

    \draw [bolded] ([xshift=-1cm]A.south) -- node[directedText, anchor=west, text width=2.75cm, align=center] {
        Bold Edge\\
        \textcolor{blue}{(1)}\\
        \textcolor{red}{(2)}
    } ([xshift=-1cm]B.north);
\end{tikzpicture}\unskip}%
		\end{minipage} &%
		A \textbf{bold edge} is a directed edge that takes part in an attack; thus, on this edge, we mark the discovered attacks' IDs (see~{\S}~{\ref{methods:Identifying the potential attacks and marking them on the AFDs}}), using \textcolor{blue}{blue} to represent IDs of existing attacks, and \textcolor{red}{red} to represent IDs of new attacks that we have discovered.\\\midrule%
		\begin{minipage}{.2\linewidth}
			\centering%
			\resizebox{\linewidth}{!}{\begin{tikzpicture}
    \node (A) [component] {Component A};
    \node (B) [component, below=of A, yshift=-1.5cm] {Component B};

    \draw [bolded] ([xshift=-1cm]A.south) -- node[directedText, anchor=west, text width=2.75cm, align=center] {
        Bold Numbers\\
        \textcolor{blue}{(1),\textbf{(3)}}\\
        \textcolor{red}{(2),\textbf{(4)}}
    } ([xshift=-1cm]B.north);
\end{tikzpicture}\unskip}%
		\end{minipage} &%
		\textbf{Bold numbers} represent IDs of newly discovered attacks (with regards to previously identified attacks in the \glsxtrshort{tldr} methodology), which are only relevant to the current medical device, marked on edge previously introduced in another \glsxtrshort{afd} (e.g., the \glsxtrshort{afd} of the generic \glsxtrshort{mid}).\\\bottomrule%
	\end{tabular}
\end{table}

In addition to creating \cglspl{afd} for specific medical devices (e.g., \cgls{ct}), it is possible to create \cglspl{afd} for generic medical devices.
For example, we have created an \cgls{afd} for the generic \cgls{mid} that consists of components that typically appear in most \cglspl{mid}.\
Defining a generic medical device is useful for many purposes, such as for
\begin{inparaenum}[(1)]
	\item describing attacks targeting the surrounding components of the medical device;
	\item consolidating basic attack scenarios which potentially target any wide variety of medical devices;
	\item describing attacks on medical devices for which we did not find unique attacks;
	\item lay the groundwork for advanced attack scenarios of specific medical devices; and
	\item provide a solid foundation for future research to expand the abstract \cgls{afd} for additional unique modalities.
\end{inparaenum}

\subsection[Identifying the potential attacks and marking them on the AFDs]{Identifying the Potential Attacks and Marking Them on the \glsxtrshortpl{afd}}\label{methods:Identifying the potential attacks and marking them on the AFDs}

The second step of the \cgls{tldr} methodology includes identifying all the potential attacks and marking them on the \cglspl{afd}~({\S}~{\ref{methods:Identifying the potential attacks and marking them on the AFDs}}).
For this step, using \cglspl{afd} becomes very useful for identifying vulnerabilities as well as visualizing the attacks.
First, we shall mark, in \textcolor{blue}{blue}, all known attacks, such as attacks that may already happen or that have been mentioned in the literature.
Next, we shall mark, in \textcolor{red}{red}, newly identified potential attacks that have been discovered as a result of using our \cglspl{afd}.\
Given the nature of cyber attacks' rapid development and constant change, this step will require additional updates from time to time (e.g., once new potential attacks are discovered).

\subsection[Mapping the discovered attacks into their relevant CAPECs]{Mapping the Discovered Attacks into Their Relevant \glsxtrshortpl{capec}}\label{methods:Mapping the Discovered Attacks into Their Relevant CAPECs}

This step includes the mapping of each attack, which was discovered in the previous step, into all relevant \cglspl{capec} that describe the attack or can potentially implement it.
Since \cgls{capec} provides standard definitions of many attack patterns, we plan that this mapping would be from many potential attacks into a significantly smaller number of \cglspl{capec}.\
Furthermore, since the domain of \cglspl{mid} is unique, \cgls{capec} may be missing a few patterns which are unique for cyber attacks on \cglspl{mid};\ in such cases, we shall suggest adding the missing patterns to \cgls{capec}.\

\subsection[Estimating the likelihood of the generic CAPECs into which the potential attacks are mapped]{Estimating the Likelihood of the Generic \glsxtrshortpl{capec} into Which the Potential Attacks Are Mapped}\label{methods:Estimating the likelihood of the generic CAPECs into which the potential attacks are mapped}

\cGls{capec} provides a rough estimate of the overall likelihood (see~{\S}~{\ref{background:CAPEC Mechanisms of Attack}}) to most of its attack patterns; however, these estimates are too general for the unique medical domain, and thus, may not be accurate enough for us.
To fine\=/tune \cgls{capec}'s likelihood estimates, we re\=/estimate these values with the assistance of a panel of four healthcare \cglspl{ise} who specialize in medical devices' security.
Users of the \cgls{tldr} methodology are welcome to fine\=/tune our estimates with the assistance of their own experts, to get even more accurate results, specific for their organizations.
Simply ask your experts to re\=/estimate the likelihood of each of the \cglspl{capec} and plug it into the tables we present in the results chapter~({\S}~{\ref{results:Results}}).
New attack incidents may provide new insights that would require to update the likelihood estimates from time to time.

\subsection[Computing the CAPEC-based likelihood estimates for each attack]{Computing the \glsxtrshort{capec}\=/based Likelihood Estimates for Each Attack}\label{methods:Computing the CAPEC-based likelihood estimates for each attack}

Finally, we compute the \cgls{capec}\=/based (i.e., overall) likelihood estimates for each attack as the \emph{mean} of the likelihood estimates, given in {\S}~{\ref{methods:Estimating the likelihood of the generic CAPECs into which the potential attacks are mapped}}, of the relevant \cglspl{capec} into which the potential attacks had been mapped in {\S}~{\ref{methods:Mapping the Discovered Attacks into Their Relevant CAPECs}}.
Note that just like in Fuzzy Logic, one could argue that the worst\=/case scenario should be considered by taking the \emph{maximum} likelihood of the relevant \cglspl{capec}, rather than the \emph{mean} likelihood, assuming that the attacker is aware of everything that we are, as Stine \etal{~\cite{Stine2017}} did in their risk scoring system (see~{\S}~{\ref{background:Information Security Risk Assessment Methodologies}}); %
however, we assume that an attacker choice of a potential \cgls{capec} is not based solely on the likelihood of the \cgls{capec}, since an attacker might have other reasons to choose a potential \cglspl{capec} (e.g., an attacker lacks knowledge or resources required to implement the \cgls{capec}).
Thus, we assume that all potential \cglspl{capec} mapped to an attack are equally likely, implying a \emph{mean} likelihood of the mapped \cglspl{capec} of the attack as the computed likelihood estimates of the attack.
Note that a constant shift \(c\) may be added (see~{\S}~{\ref{results:Validating the TLDR Methodology's Results}}).

\subsection[Decomposing each attack into several severity aspects and assigning them weights]{Decomposing Each Attack into Several Severity Aspects and Assigning Them Weights}\label{methods:Decomposing each attack into several severity aspects and assigning them weights}

The severity of attacks is usually decomposed into specific aspects that affect the organization (e.g., \cgls{cia} is used in traditional risk assessment methodologies, see~{\S}~{\ref{background:Information Security Risk Assessment Methodologies}}).
The decomposed severity aspects do not affect the organization equally; thus, we can assign predetermined relative importance weights for each decomposed severity aspect, based on the organization's policy.
The defined decomposed severity aspects and their weights will then be used as \emph{predefined default weights} throughout the \cgls{tldr} methodology.
New attack incidents may provide new insights that would require to update the weights from time to time.
Note that both the relative importance weights and the specific magnitudes of the impact of the severity aspects, represent measures of [dis]\emph{utility}, and are entirely different from the likelihood, which represents a measure of \emph{probability}.

In the current study, we are focusing on obtaining the \cgls{tldr} methodology's likelihood estimates for each attack. 
The severity decomposition steps of the \cgls{tldr} methodology and their validation are beyond the scope of the current study and are discussed in a separate paper that we are preparing.
One can view our use of the \cgls{tldr} methodology in the current study as using a trivial decomposition of the severity into a single\=/aspect severity --- i.e., the overall severity --- and assigning it a weight of one.

\subsection[Assessing the magnitude of the impact of each of the severity aspects for each attack]{Assessing the Magnitude of the Impact of Each of the Severity Aspects for Each Attack}\label{methods:Assessing the magnitude of the impact of each of the severity aspects for each attack}

One can interpret the severity as the expected \emph{magnitude of the impact}, denoting the assessment of the \emph{level}, or \emph{degree}, of damage assuming the attack was successful.
For example, one could envision the ordinal scale as being ``None, Very Low, Low, Moderate, High, Very High.'' 

We assess the expected magnitude of the impact of each of the decomposed severity aspects [in this particular study, we used only single\=/aspect severity --- the overall severity --- since this part is beyond the scope of the current study], for each attack individually on a scale of zero to five, with the assistance of a panel of \cglspl{me}.
Users of the \cgls{tldr} methodology can fine\=/tune our assessments with their own experts by re\=/assessing the severity with the assistance of their own experts, to get even more accurate results, specific for their organizations, and plug it into the tables we present in the results chapter~({\S}~{\ref{results:Results}}).
New attack incidents may provide new insights that would also require to update the expected magnitude of the impact of severity aspects.

\subsection[Computing the composite severity assessments for each attack]{Computing the Composite Severity Assessments for Each Attack}\label{methods:Computing the composite severity assessments for each attack}

We \emph{compute the composite severity assessments} for each attack as the \emph{weighted sum} of the magnitude of the impact of the decomposed severity aspects for each attack, weighted by the organization\=/specific relative importance weights of these aspects, as defined in \cref{eq:Severity Calculation}:
\begin{equation}\label{eq:Severity Calculation}
	{Severity}_j = \sum_{i=1}^{m} w_i\cdot s_{i,j} + c
\end{equation}
Where \(m\) is the number of decomposed severity aspects, \(w_i\) is the normalized \emph{relative importance weight} of the \(i\)\textsuperscript{th} \emph{decomposed} severity aspect, \(s_{i,j}\) is the assessment of the expected \emph{magnitude} (degree) \emph{of the impact} of the \(i\)\textsuperscript{th} decomposed severity aspect for the \(j\)\textsuperscript{th} specific attack, and \(c\) is a constant shift that may be added (see~{\S}~{\ref{results:Validating the TLDR Methodology's Results}}).

Note that the validation of this step is beyond the scope of the current study, in which we simply used the experts' the overall severity assessments for each attack.
This simplification can be viewed as using the \cgls{tldr} methodology with only single\=/aspect severity for each attack, with a weight of one.

\subsection[Integrating the likelihood and severity of each attack into its risk and prioritizing it]{Integrating the Likelihood and Severity of Each Attack into Its Risk and Prioritizing It}\label{methods:Integrating the likelihood and severity of each attack into its risk and prioritizing it}

We integrate the risk for each attack based on \cref{eq:Basic Risk Calculation} (see~{\S}~{\ref{background:Information Security Risk Assessment Methodologies}}), using the \cgls{capec}\=/based likelihood estimates~({\S}~{\ref{methods:Computing the CAPEC-based likelihood estimates for each attack}}) and the composite severity assessments~({\S}~{\ref{methods:Computing the composite severity assessments for each attack}}).
The attacks' risks can now be used to prioritize the attacks from the defenders' perspective.

\subsection[Validating the TLDR Methodology's Results]{Validating the \glsxtrshort{tldr} Methodology's Results}\label{methods:Validating the TLDR Methodology's Results}

To validate our results, we followed the same steps as the \cgls{tldr} methodology; however, we asked the \cglspl{ise} to estimate also the \emph{direct} overall likelihood for each attack individually, instead of using the \cgls{tldr} methodology.
We then defined two ``consensus'' vectors of the 23 potential attacks:
\begin{inparaenum}[(1)]
    \item the \glsxtrfull{mecble} (computed indirectly), and
    \item the \glsxtrfull{medle}.
\end{inparaenum}
To demonstrate the value and the validity of using the \cgls{tldr} methodology with respect to the relative ranking of the 23 potential threats, we calculated the pairwise Spearman's correlation between the \cgls{capec}\=/based (\cgls{mecble}) likelihood estimates and the direct estimates of each of our \cglspl{ise} and showed that the correlation is high, while the pairwise correlation between the mean of the direct estimates (the \cgls{medle}) and each of the \cglspl{ise}' overall direct estimates was lower.
To demonstrate that we maintained the validity of the risk assessment process, we calculated the paired \(T\)\=/test statistic between the \cgls{capec}\=/based (\cgls{mecble}) estimates and the direct \cglspl{ise}' estimates, and showed that the null hypothesis is accepted (i.e., there was no significant difference between them). % chktex 21
This implies that the \cgls{capec}\=/based (\cgls{mecble}) likelihood estimates, using the \cgls{tldr} methodology, are as valid as current direct risk assessment methodologies' likelihood estimates; however, the \cgls{capec}\=/based likelihood estimates, computed from the mapped \cglspl{capec}, are much easier to calculate.
Note that we asked the \cglspl{ise} to \emph{directly} estimate the likelihood of each attack individually \emph{before} we ask them to estimate it using the \cgls{tldr} methodology to eliminate a potential biased point of reference.

We shall also validate that the mapping of the discovered attacks into their relevant \cglspl{capec}, indeed results in a mapping of many attacks into a significantly smaller number of \cglspl{capec}.\
This, together with good correlations, will demonstrate, at least statistically, that instead of asking a panel of healthcare \cglspl{ise} to estimate the likelihood for each attack \emph{directly}, we can only ask them to estimate just a relatively small number of general \cgls{capec} likelihoods to achieve similar outcomes.
Also, we shall validate that using the \cglspl{afd}, which helps discover new attacks by measuring the number of newly discovered attacks.

%-------------------------------------------------------------------------------
\section{Results}\label{results:Results}
%-------------------------------------------------------------------------------

\begin{table*}[!htbp]
	\centering
	\caption[The application of the TLDR methodology to MIDs]{The application of the \glsxtrshort{tldr} methodology to \glsxtrshortpl{mid}.}\label{tab:The application of the TLDR methodology to MIDs}
	\begin{adjustbox}{width=.9\textwidth}
	\small
	\setlength\tabcolsep{1.5pt}
	\begin{tabular}{p{240pt}*{9}{P{6pt}}*{4}{>{\raggedright\arraybackslash}P{30pt}}}
		\cmidrule[\heavyrulewidth]{2-14}
		\multirow{2}{=}{%
		\textcolor{blue}{Blue}~represents known attacks.\newline
		\textcolor{red}{Red}~~represents new attacks that we have discovered.\newline
		\textsuperscript{*~}Shift explained in {\S}~{\ref{results:Validating the TLDR Methodology's Results}}.%
		} & \multicolumn{9}{M{81pt}}{\nameref{background:CAPEC Mechanisms of Attack}} &%
		\multicolumn{4}{M{120pt}}{Risk assessment} \\
		\cmidrule(lr){2-10}\cmidrule(lr){11-14}
		
		Defining attacks on the generic \glsxtrshort{mid} &%
		\rot{\Cref{CAPEC-75}} & \rot{\Cref{CAPEC-150}} &
		\rot{\Cref{CAPEC-165}} & \rot{\Cref{CAPEC-166}} & \rot{\Cref{CAPEC-542}} & \rot{\Cref{CAPEC-582}} & \rot{\Cref{CAPEC-601}} & \rot{\Cref{CAPEC-603}} & \rot{\hyperref[CAPEC-NEW]{CAPEC~NEW}} &%
		\multicolumn{1}{P{30pt}}{\rot{\emph{Severity}}} & \multicolumn{1}{P{30pt}}{\rot{\emph{Likelihood}}} & \multicolumn{1}{P{30pt}}{\rot{\emph{Likelihood (shifted)\textsuperscript{*~}}}} & \multicolumn{1}{P{30pt}}{\rot{\textbf{Risk}}} \\
		\midrule
		
		\attackref{blue}{attack:Ransomware} &%
		& & & & {\(\bullet \)} & & & & &%
		4.75 & 0.9 & 0.77 & \textbf{3.658} \\
		\attackref{blue}{attack:Disruption of patient-to-image linkage} &%
		& {\(\bullet \)} & {\(\bullet \)} & & {\(\bullet \)} & & & & &%
		4.75 & 0.75 & 0.62 & \textbf{2.945} \\
		\attackref{blue}{attack:Alteration of the imaging Exam's Results} &%
		& {\(\bullet \)} & {\(\bullet \)} & & {\(\bullet \)} & & & & &%
		4.5 & 0.75 & 0.62 & \textbf{2.79} \\
		\attackref{red}{attack:Contrast Material Over/Underdose} &%
		{\(\bullet \)} & & & & {\(\bullet \)} & & & & &%
		4.5 & 0.725 & 0.595 & \textbf{2.678} \\
		\attackref{red}{attack:Leakage of Patients' Private Information} &%
		& {\(\bullet \)} & & & & & & & &%
		3.25 & 0.75 & 0.62 & \textbf{2.015} \\
		\attackref{blue}{attack:Manipulation of Data Displayed on the Host's Monitor} &%
		& & & & {\(\bullet \)} & {\(\bullet \)} & {\(\bullet \)} & {\(\bullet \)} & {\(\bullet \)} &%
		4.25 & 0.6 & 0.47 & \textbf{1.998} \\		
		\attackref{blue}{attack:Mute Safety Alarms} &%
		& & & & {\(\bullet \)} & {\(\bullet \)} & {\(\bullet \)} & {\(\bullet \)} & {\(\bullet \)} &%
		3.5 & 0.6 & 0.47 & \textbf{1.645} \\
		\attackref{red}{attack:Disruption of the imaging Exam's Results} &%
		& {\(\bullet \)} & {\(\bullet \)} & & & {\(\bullet \)} & {\(\bullet \)} & & &%
		3.5 & 0.6 & 0.47 & \textbf{1.645} \\
		\attackref{blue}{attack:Activate False Safety Alarms} &%
		& & & & {\(\bullet \)} & {\(\bullet \)} & {\(\bullet \)} & {\(\bullet \)} & {\(\bullet \)} &%
		3.25 & 0.6 & 0.47 & \textbf{1.528} \\
		\attackref{red}{attack:Mechanical Disruption of MID's Motors} &%
		{\(\bullet \)} & & & & {\(\bullet \)} & & & & {\(\bullet \)} &%
		3 & 0.633 & 0.503 & \textbf{1.509} \\
		\attackref{blue}{attack:Restore System} &%
		& & & {\(\bullet \)} & & & & & &%
		2.5 & 0.55 & 0.42 & \textbf{1.05} \\
		
		\midrule
		\multicolumn{13}{l}{Defining attacks on the generic \glsxtrshort{ct}}\\
		\midrule
		
		\attackref{blue}{attack:Increase Milliamperage-Seconds} &%
		{\(\bullet \)} & & {\(\bullet \)} & & {\(\bullet \)} & & & & &%
		4.5 & 0.683 & 0.553 & \textbf{2.489} \\
		\attackref{blue}{attack:Increase Kilovoltage Peak} &%
		{\(\bullet \)} & & {\(\bullet \)} & & {\(\bullet \)} & & & & &%
		4.5 & 0.683 & 0.553 & \textbf{2.489} \\
		\attackref{blue}{attack:Radiation Overdose} &%
		{\(\bullet \)} & & {\(\bullet \)} & & {\(\bullet \)} & & & & &%
		4.5 & 0.683 & 0.553 & \textbf{2.489} \\
		\attackref{blue}{attack:Alteration of the IRS's Output Images} &%
		& {\(\bullet \)} & {\(\bullet \)} & & {\(\bullet \)} & & & & &%
		3.5 & 0.75 & 0.62 & \textbf{2.17} \\
		\attackref{blue}{attack:Configuration File Disruption} &%
		{\(\bullet \)} & & {\(\bullet \)} & & {\(\bullet \)} & & & & &%
		3.75 & 0.683 & 0.553 & \textbf{2.074} \\
        \attackref{red}{attack:Manipulation of CT Calibration} &%
		{\(\bullet \)} & & {\(\bullet \)} & {\(\bullet \)} & {\(\bullet \)} & & & & &%
		3.5 & 0.65 & 0.52 & \textbf{1.82} \\		
		\attackref{red}{attack:Disruption of the IRS's Output Images} &%
		& {\(\bullet \)} & {\(\bullet \)} & & {\(\bullet \)} & {\(\bullet \)} & & & &%
		2.25 & 0.7375 & 0.608 & \textbf{1.367} \\
		
		\midrule
		\multicolumn{13}{l}{Defining attacks on the generic \glsxtrshort{mri}}\\
		\midrule
		
		\attackref{blue}{attack:Overwhelm of the MRI's Receiving Coils with an Overpowered Magnetic Field} &%
		{\(\bullet \)} & & {\(\bullet \)} & & {\(\bullet \)} & & & & &%
		4.25 & 0.6833 & 0.553 & \textbf{2.352} \\
		\attackref{blue}{attack:Magnetic Field Disruption} &%
		{\(\bullet \)} & & {\(\bullet \)} & & {\(\bullet \)} & & {\(\bullet \)} & & &%
		3.5 & 0.6 & 0.47 & \textbf{1.645} \\
		\attackref{red}{attack:External RF Signal Disruption} &%
		& & & & & {\(\bullet \)} & {\(\bullet \)} & & & %
		4 & 0.525 & 0.395 & \textbf{1.58} \\		
		\attackref{blue}{attack:Activate Quenching of MRI} &%
		& & & & & & & & {\(\bullet \)} &%
		3.75 & 0.45 & 0.32 & \textbf{1.2} \\
		
		\midrule
		\multicolumn{13}{l}{Defining attacks on the generic ultrasound}\\
		\midrule
		
		\attackref{red}{attack:Disruption of MEMS Components} &%
		& & & & {\(\bullet \)} & & & & {\(\bullet \)} &%
		3 & 0.675 & 0.545 & \textbf{1.635} \\
		
		\bottomrule
	\end{tabular}
	\end{adjustbox}
\end{table*}

In this section, we present the results of the application of the \cgls{tldr} methodology (see~{\S}~{\ref{methods:Methods}}) to \cglspl{mid}.\
We first present the \textbf{T}hreat identification (\textbf{T}) part of the \cgls{tldr} methodology: creating the \cglspl{afd} of \cglspl{mid}, which helps us identify the potentially vulnerable components ({\S}~{\ref{results:identifying the potentially vulnerable components using AFDs}}) and identify 23 potential attacks on \cglspl{mid} ({\S}~{\ref{results:Identifying the potential attacks and marking them on the AFDs}}).
Following that, we present the ontology\=/based \textbf{L}ikelihood (\textbf{L}) part of the \cgls{tldr} methodology: %
mapping of the discovered attacks into their relevant \cglspl{capec} ({\S}~{\ref{results:Mapping the Discovered Attacks into Their Relevant CAPECs}}), estimating the likelihood of each \cgls{capec} with the assistance of the panel of healthcare \cglspl{ise}' ({\S}~{\ref{results:Estimating the likelihood of the generic CAPECs into which the potential attacks are mapped}}), and computing the \cgls{capec}\=/based likelihood estimates for each attack ({\S}~{\ref{results:Computing the CAPEC-based likelihood estimates for each attack}}).
As mentioned before, the severity \textbf{D}ecomposition (\textbf{D}) part of the \cgls{tldr} methodology ({\S}~{\ref{methods:Decomposing each attack into several severity aspects and assigning them weights}} to {\S}~{\ref{methods:Computing the composite severity assessments for each attack}}) is beyond the scope of the current study and will be discussed in a separate paper that we are preparing; thus, we replaced these steps with a single simple step of assessing the overall severity for each attack with the assistance of the panel of \cglspl{rme} ({\S}~{\ref{results:Assessing the magnitude of the impact of each of the severity aspects for each attack}}) and used it as its severity.
Following that, we present the \textbf{R}isk integration (\textbf{R}) part of the \cgls{tldr} methodology: integrating the risk for each attack and prioritizing it ({\S}~{\ref{results:Integrating the likelihood and severity of each attack into its risk and prioritizing it}}).
The results are summarized in \Cref{tab:The application of the TLDR methodology to MIDs}.
Finally, we validate the \cgls{tldr} methodology's results ({\S}~{\ref{results:Validating the TLDR Methodology's Results}}) using statistical metrics, such as the paired \(T\)\=/test and pairwise Spearman's correlation. % chktex 21

\subsection[Identifying the potentially vulnerable components using AFDs]{Identifying the Potentially Vulnerable Components Using \glsxtrshortpl{afd}}\label{results:identifying the potentially vulnerable components using AFDs}

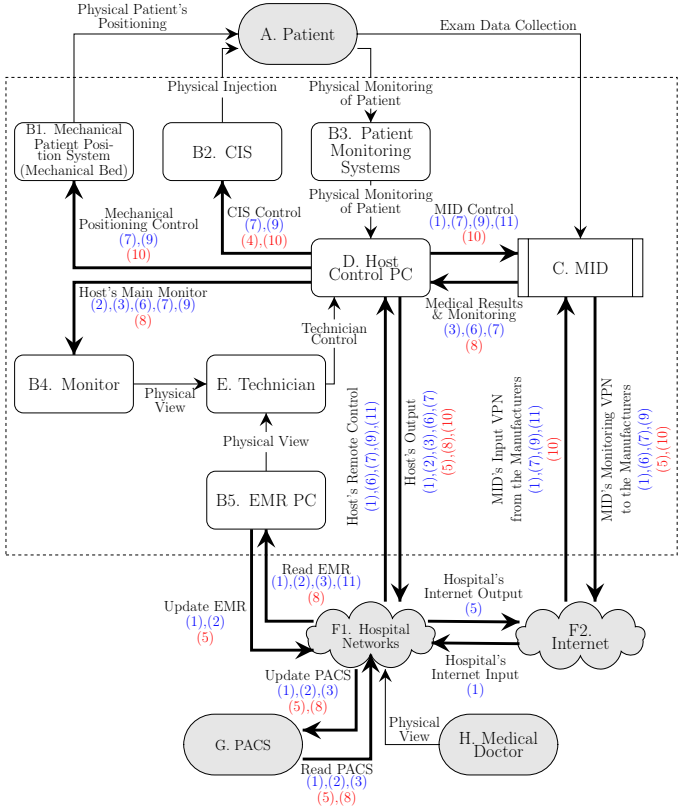
\begin{figure}[!htb]
	\centering
	\resizebox{\linewidth}{!}{\begin{tikzpicture}[
    align=center,
    node distance=2cm and 1cm,
]
    \draw[dashed] (-10,-1.5) rectangle (13,-18);
    
    \node (A) [terminator] {\afdref{itm:Patient}{Patient}};
    
    \node (B3) [component, below=of A, xshift=2.5cm] {\afdref{itm:Patient monitoring systems}{Patient Monitoring Systems}};
    \node (B2) [component, left=of B3] {\afdref{itm:CIS}{CIS}};
    \node (B1) [component, left=of B2, font=\Large] {\afdref{itm:Mechanical bed}{Mechanical Patient Position System (Mechanical Bed)}};
    
    \draw [directed] (B1) |- node[directedText, anchor=north, above right] {
        Physical Patient's\\
        Positioning
    } (A);
    \path (B2.north) |- node[directedText, anchor=south, below, yshift=-1.5cm] (B2A) {Physical Injection} (A.west);
    \draw [directed] (B2.north) -- (B2A) |- ([yshift=-.5cm]A.west);
    \path (A.east) -| node[directedText, anchor=south, below, yshift=-1.5cm] (AB3) {
        Physical Monitoring\\
        of Patient
    } (B3.north);
    \draw [directed] ([yshift=-.5cm]A.east) -| (AB3) -- (B3.north);
    
    \node (D) [component, below=of B3] {\afdref{itm:Host control PC}{Host Control PC}};
    \node (C) [subcomponent, right=of D, xshift=2cm] {\afdref{itm:The MID}{MID}};
    \node (B4top) [emptycomponent, below=of B1] {};
    
    \path (B3.south) -- node[directedText, anchor=center, yshift=.3cm] (B3D) {
        Physical Monitoring\\
        of Patient
    } (D.north);
    \draw [directed] (B3.south) -- (B3D) -- (D.north);
    \draw [directed] (A) -| node[directedText, anchor=north, above left] {Exam Data Collection} (C);
    
    \draw [bolded] ([yshift=.5cm]D.west) -| node[directedText, anchor=center, above right]{
        \Glsxtrshort{cis} Control\\
        \textcolor{blue}{(\ref{attack:Mute Safety Alarms}),(\ref{attack:Activate False Safety Alarms})}\\
        \textcolor{red}{(\ref{attack:Contrast Material Over/Underdose}),(\ref{attack:Mechanical Disruption of MID's Motors})}
    } (B2.south);
    \draw [bolded] (D.west) -| node[directedText, anchor=center, above right] {
        Mechanical\\
        Positioning Control\\
        \textcolor{blue}{(\ref{attack:Mute Safety Alarms}),(\ref{attack:Activate False Safety Alarms})}\\
        \textcolor{red}{(\ref{attack:Mechanical Disruption of MID's Motors})}
    } (B1.south);
    \draw [bolded] ([yshift=.5cm]D.east) -- node[directedText, anchor=center, above, yshift=.125cm] {
        \Glsxtrshort{mid} Control\\
        \textcolor{blue}{(\ref{attack:Ransomware}),(\ref{attack:Mute Safety Alarms}),(\ref{attack:Activate False Safety Alarms}),(\ref{attack:Restore System})}\\
        \textcolor{red}{(\ref{attack:Mechanical Disruption of MID's Motors})}
    } ([yshift=.5cm]C.west);
    \draw [bolded] ([yshift=-.5cm]C.west) -- node[directedText, anchor=center, below, yshift=-.375cm] {
        Medical Results\\
        \& Monitoring\\
        \textcolor{blue}{(\ref{attack:Alteration of the imaging Exam's Results}),(\ref{attack:Manipulation of Data Displayed on the Host's Monitor}),(\ref{attack:Mute Safety Alarms})}\\
        \textcolor{red}{(\ref{attack:Disruption of the imaging Exam's Results})}
    } ([yshift=-.5cm]D.east);
    
    \node (E) [component, below left=of D, xshift=1.5cm] {\afdref{itm:Technician}{Technician}};
    \node (B4) [component, left=of E, xshift=-1.5cm] {\afdref{itm:Monitor}{Monitor}};
    \node (F1toptop) [emptycomponent, below =of D]{};
    \node (F2toptop) [emptycomponent, below =of C]{};
    
    \draw [bolded] ([yshift=-.5cm]D.west) -| node[directedText, anchor=center, below right]{
        Host's Main Monitor\\
        \textcolor{blue}{(\ref{attack:Disruption of patient-to-image linkage}),(\ref{attack:Alteration of the imaging Exam's Results}),(\ref{attack:Manipulation of Data Displayed on the Host's Monitor}),(\ref{attack:Mute Safety Alarms}),(\ref{attack:Activate False Safety Alarms})}\\
        \textcolor{red}{(\ref{attack:Disruption of the imaging Exam's Results})}
    } (B4.north);
    \draw [directed] (B4) -- node[directedText, anchor=center, below] {
        Physical\\
        View
    } (E);
    \path (E.east) -| node[directedText, anchor=center, above, yshift=1.25cm] (ED) {
        Technician\\
        Control
    } ([xshift=-1.25cm]D.south);
    \draw [directed] (E.east) -| (ED) -- ([xshift=-1.25cm]D.south);
    
    \node (B5) [component, below=of E] {\afdref{itm:EMR}{EMR PC}};
    \node (F1top) [emptycomponent, below =of F1toptop]{};
    \node (F2top) [emptycomponent, below =of F2toptop]{};
    
    \path (B5) -- node[directedText, anchor=center] (B5E) {Physical View} (E);
    \draw [directed] (B5) -- (B5E) -- (E);
    
    \node (F1) [network, below=of F1top, yshift=-.5cm, font=\Large] {\afdref{itm:Hospital networks}{Hospital Networks}};
    \node (F2) [network, below=of F2top, yshift=-.5cm] {\afdref{itm:Internet}{Internet}};
    
    \draw [bolded] ([xshift=-.5cm]B5.south) |- node[directedText, anchor=center, above left]{
        Update \glsxtrshort{emr}\\
        \textcolor{blue}{(\ref{attack:Ransomware}),(\ref{attack:Disruption of patient-to-image linkage})}\\
        \textcolor{red}{(\ref{attack:Leakage of Patients' Private Information})}
    } ([yshift=-.5cm,xshift=.25cm]F1.west);
    \draw [bolded] ([yshift=.5cm,xshift=.25cm]F1.west) -| node[directedText, anchor=center, above right, yshift=.375cm]{
        Read \glsxtrshort{emr}\\
        \textcolor{blue}{(\ref{attack:Ransomware}),(\ref{attack:Disruption of patient-to-image linkage}),(\ref{attack:Alteration of the imaging Exam's Results}),(\ref{attack:Restore System})}\\
        \textcolor{red}{(\ref{attack:Disruption of the imaging Exam's Results})}
    } (B5.south);
    \draw [bolded] ([yshift=.5cm,xshift=-.25cm]F1.east) -- node[directedText, anchor=center, above]{
        Hospital's\\
        Internet Output\\
        \textcolor{blue}{(\ref{attack:Leakage of Patients' Private Information})}
    } ([yshift=.5cm,xshift=.125cm]F2.west);
    \draw [bolded] ([yshift=-.25cm,xshift=.125cm]F2.west) -- node[directedText, anchor=center, below, yshift=-.25cm]{
        Hospital's\\
        Internet Input\\
        \textcolor{blue}{(\ref{attack:Ransomware})}
    } ([yshift=-.25cm,xshift=-.125cm]F1.east);
    \draw [bolded] ([xshift=.5cm]F1.north) -- node[directedText, anchor=center, xshift=-.675cm, yshift=-.25cm]{
        \rotatebox{90}{\parbox{6cm}{\centering 
            Host's Remote Control\\
            \textcolor{blue}{(\ref{attack:Ransomware}),(\ref{attack:Manipulation of Data Displayed on the Host's Monitor}),(\ref{attack:Mute Safety Alarms}),(\ref{attack:Activate False Safety Alarms}),(\ref{attack:Restore System})}
        }}
    } ([xshift=.5cm]D.south);
    \draw [bolded] ([xshift=1cm]D.south) -- node[directedText, anchor=center, xshift=1.125cm, yshift=.25cm]{
    \rotatebox{90}{\parbox{4cm}{\centering
            Host's Output\\
            \textcolor{blue}{(\ref{attack:Ransomware}),(\ref{attack:Disruption of patient-to-image linkage}),(\ref{attack:Alteration of the imaging Exam's Results}),(\ref{attack:Manipulation of Data Displayed on the Host's Monitor}),(\ref{attack:Mute Safety Alarms})}\\
            \textcolor{red}{(\ref{attack:Leakage of Patients' Private Information}),(\ref{attack:Disruption of the imaging Exam's Results}),(\ref{attack:Mechanical Disruption of MID's Motors})}
    }}
    } ([xshift=1cm]F1.north);
    \draw [bolded] ([xshift=-.5cm]F2.north) -- node[directedText, anchor=center, xshift=-1.25cm]{
    \rotatebox{90}{\parbox{5cm}{\centering
        \Glsxtrshort{mid}'s Input \glsxtrshort{vpn}\\
        from the Manufacturers\\
        \textcolor{blue}{(\ref{attack:Ransomware}),(\ref{attack:Mute Safety Alarms}),(\ref{attack:Activate False Safety Alarms}),(\ref{attack:Restore System})}\\
        \textcolor{red}{(\ref{attack:Mechanical Disruption of MID's Motors})}
    }}
    } ([xshift=-.5cm]C.south);
    \draw [bolded] ([xshift=.5cm]C.south) -- node[directedText, anchor=center, xshift=1.5cm]{
    \rotatebox{90}{\parbox{5cm}{\centering
        \Glsxtrshort{mid}'s Monitoring \glsxtrshort{vpn}\\
        to the Manufacturers\\
        \textcolor{blue}{(\ref{attack:Ransomware}),(\ref{attack:Manipulation of Data Displayed on the Host's Monitor}),(\ref{attack:Mute Safety Alarms}),(\ref{attack:Activate False Safety Alarms})}\\
        \textcolor{red}{(\ref{attack:Leakage of Patients' Private Information}),(\ref{attack:Mechanical Disruption of MID's Motors})}
    }}
    } ([xshift=.5cm]F2.north);
    
    \node (G) [terminator, below left=of F1, font=\Large] {\afdref{itm:PACS}{PACS}};
    \node (H) [terminator, below right=of F1] {\afdref{itm:Medical doctor}{Medical Doctor}};
    
    \draw [bolded] ([yshift=-.5cm]G.east) -| node[directedText, anchor=center, below left, xshift=.25cm]{
        Read \glsxtrshort{pacs}\\
        \textcolor{blue}{(\ref{attack:Ransomware}),(\ref{attack:Disruption of patient-to-image linkage}),(\ref{attack:Alteration of the imaging Exam's Results})}\\
        \textcolor{red}{(\ref{attack:Leakage of Patients' Private Information}),(\ref{attack:Disruption of the imaging Exam's Results})} 
    } (F1.south);
    \draw [bolded] ([xshift=-.5cm, yshift=-.5cm]F1.south) |- node[directedText, anchor=center, above left, yshift=.375cm]{
        Update \glsxtrshort{pacs}\\
        \textcolor{blue}{(\ref{attack:Ransomware}),(\ref{attack:Disruption of patient-to-image linkage}),(\ref{attack:Alteration of the imaging Exam's Results})}\\
        \textcolor{red}{(\ref{attack:Leakage of Patients' Private Information}),(\ref{attack:Disruption of the imaging Exam's Results})}
    } ([yshift=.5cm]G.east);
    
    \draw [directed] (H.west) -| node[directedText, anchor=center, above right]{
        Physical\\
        View
    } ([xshift=.5cm, yshift=-.5cm]F1.south);

\end{tikzpicture}\unskip}
	\caption[The AFD of the generic MID.]{The \glsxtrshort{afd} of the generic \glsxtrshort{mid}, with the relevant \glsxtrshortpl{ifv} and attack vectors, that consists of components that typically appear in most \glsxtrshortpl{mid}.}\label{fig:The AFD of the generic MID3}
\end{figure}

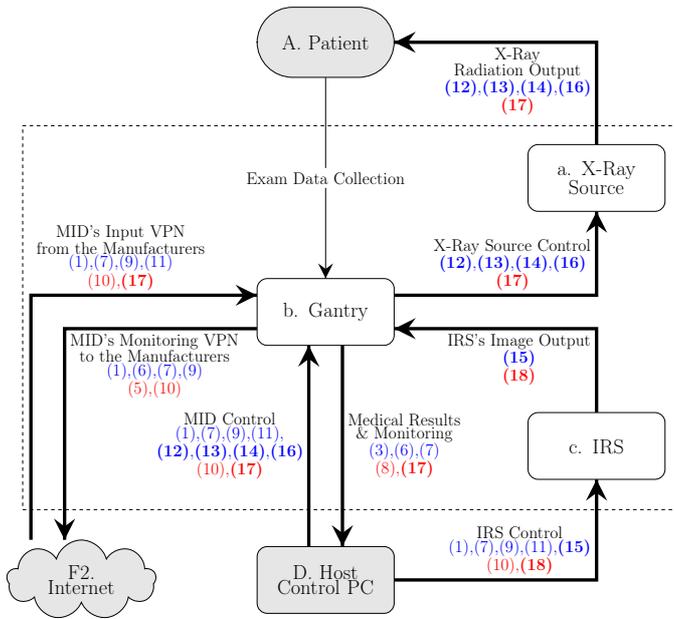
\begin{figure}[!htb]
	\centering
	\resizebox{\linewidth}{!}{\begin{tikzpicture}[
    align=center,
    node distance=2cm and 3cm,
]
    \draw[dashed] (-9,-2.5) rectangle (10.5,-14);
    
    \node (A) [terminator] {\afdref{itm:Patient}{Patient}};
    
    \node (btop) [emptycomponent, below=of A] {};
    \node (a) [component, right=of btop, xshift=1cm] {\afdref{itm:xray source}{X\=/Ray Source}};
    
    \draw [bolded] (a.north) |- node[directedText, anchor=center, below left] {
        X\=/Ray\\
        Radiation Output\\
        \textcolor{blue}{\textbf{(\ref{attack:Increase Milliamperage-Seconds})},\textbf{(\ref{attack:Increase Kilovoltage Peak})},\textbf{(\ref{attack:Radiation Overdose})},\textbf{(\ref{attack:Configuration File Disruption})}}\\
        \textcolor{red}{\textbf{(\ref{attack:Manipulation of CT Calibration})}}
    } (A.east);
    
    \node (b) [component, below=of btop] {\afdref{itm:Gantry}{Gantry}};
    
    \path (A) -- node[directedText, anchor=center] (Ab) {Exam Data Collection} (b);
    \draw [directed] (A) -- (Ab) -- (b);
    \draw [bolded] ([yshift=.5cm]b.east) -| node[directedText, anchor=center, above left] {
        X\=/Ray Source Control\\
        \textcolor{blue}{\textbf{(\ref{attack:Increase Milliamperage-Seconds})},\textbf{(\ref{attack:Increase Kilovoltage Peak})},\textbf{(\ref{attack:Radiation Overdose})},\textbf{(\ref{attack:Configuration File Disruption})}}\\
        \textcolor{red}{\textbf{(\ref{attack:Manipulation of CT Calibration})}}
    } (a.south);
    
    \node (Dtop) [emptycomponent, below=of b] {};
    \node (c) [component, right=of Dtop, xshift=1cm] {\afdref{itm:IRS}{IRS}};
    
    \draw [bolded] (c.north) |- node[directedText, anchor=center, below left] {
        \Glsxtrshort{irs}'s Image Output\\
        \textcolor{blue}{\textbf{(\ref{attack:Alteration of the IRS's Output Images})}}\\
        \textcolor{red}{\textbf{(\ref{attack:Disruption of the IRS's Output Images})}}
    } ([yshift=-.5cm]b.east);
    
    \node (D) [component, below=of Dtop, fill=gray!20!white] {\afdref{itm:Host control PC}{Host Control PC}};
    \node (F2) [network, left=of D] {\afdref{itm:Internet}{Internet}};
    
    \draw [bolded] ([xshift=-1.5cm]F2.north) |- node[directedText, anchor=center, above right]{
        \Glsxtrshort{mid}'s Input \glsxtrshort{vpn}\\
        from the Manufacturers\\
        \textcolor{blue}{(\ref{attack:Ransomware}),(\ref{attack:Mute Safety Alarms}),(\ref{attack:Activate False Safety Alarms}),(\ref{attack:Restore System})}\\
        \textcolor{red}{(\ref{attack:Mechanical Disruption of MID's Motors}),\textbf{(\ref{attack:Manipulation of CT Calibration})}}
    } ([yshift=.5cm]b.west);
    \draw [bolded] ([yshift=-.5cm]b.west) -| node[directedText, anchor=center, below right]{
        \Glsxtrshort{mid}'s Monitoring \glsxtrshort{vpn}\\
        to the Manufacturers\\
        \textcolor{blue}{(\ref{attack:Ransomware}),(\ref{attack:Manipulation of Data Displayed on the Host's Monitor}),(\ref{attack:Mute Safety Alarms}),(\ref{attack:Activate False Safety Alarms})}\\
        \textcolor{red}{(\ref{attack:Leakage of Patients' Private Information}),(\ref{attack:Mechanical Disruption of MID's Motors})}
    } ([xshift=-.5cm]F2.north);
    
    \draw [bolded] ([xshift=-.5cm]D.north) -- node[directedText, anchor=center, left]{
        \Glsxtrshort{mid} Control\\
        \textcolor{blue}{(\ref{attack:Ransomware}),(\ref{attack:Mute Safety Alarms}),(\ref{attack:Activate False Safety Alarms}),(\ref{attack:Restore System}),}\\
        \textcolor{blue}{\textbf{(\ref{attack:Increase Milliamperage-Seconds})},\textbf{(\ref{attack:Increase Kilovoltage Peak})},\textbf{(\ref{attack:Radiation Overdose})},\textbf{(\ref{attack:Configuration File Disruption})}}\\
        \textcolor{red}{(\ref{attack:Mechanical Disruption of MID's Motors}),\textbf{(\ref{attack:Manipulation of CT Calibration})}}
    } ([xshift=-.5cm]b.south);
    \draw [bolded] ([xshift=.5cm]b.south) -- node[directedText, anchor=center, right]{
        Medical Results\\
        \& Monitoring\\
        \textcolor{blue}{(\ref{attack:Alteration of the imaging Exam's Results}),(\ref{attack:Manipulation of Data Displayed on the Host's Monitor}),(\ref{attack:Mute Safety Alarms})}\\
        \textcolor{red}{(\ref{attack:Disruption of the imaging Exam's Results}),\textbf{(\ref{attack:Manipulation of CT Calibration})}}
    } ([xshift=.5cm]D.north);
    \draw [bolded] (D.east) -| node[directedText, anchor=center, above left] {
        \Glsxtrshort{irs} Control\\
        \textcolor{blue}{(\ref{attack:Ransomware}),(\ref{attack:Mute Safety Alarms}),(\ref{attack:Activate False Safety Alarms}),(\ref{attack:Restore System}),\textbf{(\ref{attack:Alteration of the IRS's Output Images})}}\\
        \textcolor{red}{(\ref{attack:Mechanical Disruption of MID's Motors}),\textbf{(\ref{attack:Disruption of the IRS's Output Images})}}
    } (c.south);
\end{tikzpicture}\unskip}
	\caption[The AFD of the generic CT]{The \glsxtrshort{afd} of the generic \glsxtrshort{ct} with the relevant \glsxtrshortpl{ifv} and attack vectors, based on~\cite{FitzGerald16, Suparta00}, which represents digital X\=/ray generators as well, due to their resemblance of \glsxtrshortpl{ct}.}\label{fig:The AFD of the generic CT3}
\end{figure}

\begin{figure}[!htb]
	\centering
	\resizebox{\linewidth}{!}{\begin{tikzpicture}[
    align=center,
    node distance=2cm and 1cm,
]
    \draw[dashed] (-8.5,-5.125) rectangle (9.5,-30.5);
    
    \draw[dashed] (-7.5,-14.25) rectangle (7.5,-25.5);
    
    \draw[dashed] (6.25,-21.5) -- (7.5,-21.5);
    \draw[dashed] (6.25,-21.5) -- (6.25,-25.5) node[pos=.5, text width=4cm, text centered, font=\LARGE, xshift=.5cm] {
        \rotatebox{90}{\parbox{4cm}{\centering
            \afdref{itm:RF screen}{RF Screen}
        }}
    };
    
    \node (D) [component, fill=gray!20!white] {\afdref{itm:Host control PC}{Host Control PC}};
    \node (F2) [network, left=of D, xshift=-2.5cm] {\afdref{itm:Internet}{Internet}};
    
    \node (a) [component, below=of D, yshift=-3cm] {\afdref{itm:Front-End controller}{Front\=/End Controller}};
    \node (d) [component, right=of a, xshift=2cm] {\afdref{itm:Synthesizer}{Synthesizer}};
    
    \draw [bolded] ([xshift=.5cm,yshift=-.5cm]F2.south) |- node[directedText, anchor=center, above right, yshift=3.5cm]{
        \Glsxtrshort{mri}'s Input \glsxtrshort{vpn}\\
        from the Manufacturers\\
        \textcolor{blue}{(\ref{attack:Ransomware}),(\ref{attack:Mute Safety Alarms}),(\ref{attack:Activate False Safety Alarms}),(\ref{attack:Restore System})}\\
        \textcolor{red}{(\ref{attack:Mechanical Disruption of MID's Motors})}
    } ([yshift=.5cm]a.west);
    \draw [bolded] (a.west) -| node[directedText, anchor=center, above left, yshift=3cm]{
        \Glsxtrshort{mri}'s Monitoring \glsxtrshort{vpn}\\
        to the Manufacturers\\
        \textcolor{blue}{(\ref{attack:Ransomware}),(\ref{attack:Manipulation of Data Displayed on the Host's Monitor}),(\ref{attack:Mute Safety Alarms}),(\ref{attack:Activate False Safety Alarms})}\\
        \textcolor{red}{(\ref{attack:Leakage of Patients' Private Information}),(\ref{attack:Mechanical Disruption of MID's Motors})}
    } ([xshift=-.5cm,yshift=-.5cm]F2.south);
    
    \draw [bolded] ([xshift=.5cm]D.south) -- node[directedText, anchor=center, right] {
        \Glsxtrshort{mri} Control\\
        \textcolor{blue}{(\ref{attack:Ransomware}),(\ref{attack:Mute Safety Alarms}),(\ref{attack:Activate False Safety Alarms}),(\ref{attack:Restore System}),}\textcolor{blue}{\textbf{(\ref{attack:Overwhelm of the MRI's Receiving Coils with an Overpowered Magnetic Field})},\textbf{(\ref{attack:Magnetic Field Disruption})},\textbf{(\ref{attack:Activate Quenching of MRI})}}\\
        \textcolor{red}{(\ref{attack:Mechanical Disruption of MID's Motors})}
    } ([xshift=.5cm]a.north);
    \draw [bolded] ([xshift=-.5cm]a.north) -- node[directedText, anchor=center, left, yshift=-.5cm] {
        Medical Results\\
        \& Monitoring\\
        \textcolor{blue}{(\ref{attack:Alteration of the imaging Exam's Results}),(\ref{attack:Manipulation of Data Displayed on the Host's Monitor}),(\ref{attack:Mute Safety Alarms})}\\
        \textcolor{red}{(\ref{attack:Disruption of the imaging Exam's Results})}
    } ([xshift=-.5cm]D.south);
    
    \draw [directed] ([yshift=.5cm]a.east) -- node[directedText, anchor=center, above]{
        Synthesizer\\
        Control
    } ([yshift=.5cm]d.west);
    
    \node (c) [component, below=of a, yshift=-1cm] {\afdref{itm:Gradient system}{Gradient System}};
    \node (b) [component, right=of c] {\afdref{itm:RF modulator and amplifier}{RF Modulator \& Amplifier}};
    
    \path ([yshift=-.5cm]a.east) -| node[directedText, anchor=center, yshift=-1cm] (ab) {
        \Glsxtrshort{rf} Transmission\\
        Control
    } ([xshift=-1cm]b.north);
    \draw [directed] (a) -| (ab) -- ([xshift=-1cm]b.north);
    \path (a) -- node[directedText, anchor=center] (ac) {
        Gradient Control\\
        \textcolor{blue}{\textbf{(\ref{attack:Magnetic Field Disruption})}}
    } (c);
    \draw [bolded] (a) -- (ac) -- (c);
    \path ([xshift=-.5cm]d.south) -- node[directedText, anchor=center, xshift=.25cm] (db) {
        Synthesizer\\
        Initialization
    } ([xshift=1cm]b.north);
    \draw [directed] ([xshift=-.5cm]d.south) -- (db) -- ([xshift=1.5cm]b.north);
    
    \node (e1) [component, below=of c] {\afdref{itm:Gradient coil}{Gradient Coil}};
    \node (e2) [component, left=of e1] {\afdref{itm:Superconducting magnet}{Superconducting Magnet}};
    
    \path (c) -- node[directedText, anchor=center, yshift=.25cm] (ce1) {
        Gradient Control\\
        \textcolor{blue}{\textbf{(\ref{attack:Magnetic Field Disruption})}}
    } (e1);
    \draw [bolded] (c) -- (ce1) -- (e1);
    \path ([yshift=-1cm]a) -| node[directedText, anchor=center, yshift=-4cm] (ae2) {
        Magnetic Field Control\\
        \textcolor{blue}{\textbf{(\ref{attack:Overwhelm of the MRI's Receiving Coils with an Overpowered Magnetic Field})},\textbf{(\ref{attack:Magnetic Field Disruption})},\textbf{(\ref{attack:Activate Quenching of MRI})}}
    } (e2);
    \draw [bolded] ([yshift=-1cm]a) -| (ae2) -- (e2);
    
    \node (A) [terminator, below=of e1] {\afdref{itm:Patient}{Patient}};
    \node (e3) [component, right=of A] {\afdref{itm:RF coil}{RF Coil}};
    
    \path (b.south) -- node[directedText, anchor=center] (be3) {\Glsxtrshort{rf} Coil Control} (e3.north);
    \draw [directed] (b.south) -- (be3) -- (e3.north);
    \path (e1.south) -- node[directedText, anchor=center, yshift=.5cm] (e1A) {Gradient Field Output} (A.north);
    \draw [directed] (e1.south) -- (e1A) -- (A.north);
    \path ([xshift=1cm]e2.south) -- node[directedText, anchor=center,xshift=-1cm] (e2A) {
        Magnetic Field\\
        \textcolor{blue}{\textbf{(\ref{attack:Overwhelm of the MRI's Receiving Coils with an Overpowered Magnetic Field})}}
    } (A.west);
    \draw [bolded] ([xshift=1cm]e2.south) -- (e2A) |- (A.west);
    \draw [bolded] (A.east) -- node[directedText, anchor=center, above, yshift=.5cm] {
        Physical \glsxtrshort{rf}\\
        \textcolor{red}{\textbf{(\ref{attack:External RF Signal Disruption})}}
    } (e3.west);
    
    \node (e4) [component, below=of A] {\afdref{itm:Surface coil}{Surface Coil}};
    \node (e5) [component, left=of e4] {\afdref{itm:Safety sensors}{Safety Sensors}};
   
    \path (A.south) -- node[directedText, anchor=center] (Ae4) {Physical Output} (e4.north);
    \draw [directed] (A.south) -- (Ae4) -- (e4.north);
    \path ([xshift=-1cm]e2.south) -- node[directedText, anchor=center, yshift=-1cm] (e2e5) {
        Monitor\\
        to Sensors\\
        \textcolor{blue}{(\ref{attack:Mute Safety Alarms}),(\ref{attack:Activate False Safety Alarms}),\textbf{(\ref{attack:Activate Quenching of MRI})}}
    } ([xshift=-1cm]e5.north);
    \draw [bolded] ([xshift=-1cm]e2.south) -- (e2e5) -- ([xshift=-1cm]e5.north);
    
    \node (g) [component, below=of e4] {\afdref{itm:Quadrature demodulator and IRS}{Quadrature Demodulator \& IRS}};
    \node (D2right) [emptycomponent, left=of g] {};
    \node (D2) [component, left=of D2right, fill=gray!20!white, xshift=-1cm] {\afdref{itm:Host control PC}{Host Control PC}};
    
    \path (e4.south) -- node[directedText, anchor=center] (e4g) {Exam Data Collection} (g.north);
    \draw [directed] (e4.south) -- (e4g) -- (g.north);
    \draw [bolded] (e5.west) -| node[directedText, anchor=center, above right, xshift=-1.25cm]{
        Sensors' Output\\
        \textcolor{blue}{(\ref{attack:Mute Safety Alarms}),(\ref{attack:Activate False Safety Alarms}),\textbf{(\ref{attack:Activate Quenching of MRI})}}
    } (D2.north);
    \draw [bolded] ([yshift=.25cm]g.west) -- node[directedText, anchor=center, above]{
        \Glsxtrshort{irs}'s Image Output\\
        \textcolor{blue}{(\ref{attack:Alteration of the IRS's Output Images})}\\
        \textcolor{red}{(\ref{attack:Disruption of the IRS's Output Images})}
    } ([yshift=.25cm]D2.east);
    \draw [bolded] ([yshift=-.25cm]D2.east) -- node[directedText, anchor=center, below]{
        \Glsxtrshort{irs} Control\\
        \textcolor{blue}{(\ref{attack:Ransomware}),(\ref{attack:Mute Safety Alarms}),(\ref{attack:Activate False Safety Alarms}),(\ref{attack:Restore System}),(\ref{attack:Alteration of the IRS's Output Images})}\\
        \textcolor{red}{(\ref{attack:Mechanical Disruption of MID's Motors}),(\ref{attack:Disruption of the IRS's Output Images})}
    } ([yshift=-.25cm]g.west);
    \path ([xshift=-1cm]e3.south) -- node[directedText, anchor=center,xshift=1cm] (e3g) {
        \Glsxtrshort{rf} Exam's\\
        Data Collection\\
        \textcolor{red}{\textbf{(\ref{attack:External RF Signal Disruption})}}
    } ([yshift=.5cm]g.east);
    \draw [bolded] ([xshift=-1cm]e3.south) -- (e3g) |- ([yshift=.5cm]g.east);
    
     \draw [bolded] ([xshift=1cm]d.south) |- node[directedText, anchor=center, right, yshift=10cm]{
        \rotatebox{90}{\parbox{5cm}{\centering 
            Synthesizer Correction
        }}
    } ([yshift=-.5cm]g.east);
\end{tikzpicture}\unskip}
	\caption[The AFD of the generic MRI]{The \glsxtrshort{afd} of the generic \glsxtrshort{mri} with the relevant \glsxtrshortpl{ifv} and attack vectors, based on~\cite{Higgins17,Brown14,McRobbie17,Elster01,Vlaardingerbroek03}.}\label{fig:The AFD of the generic MRI3}
\end{figure}
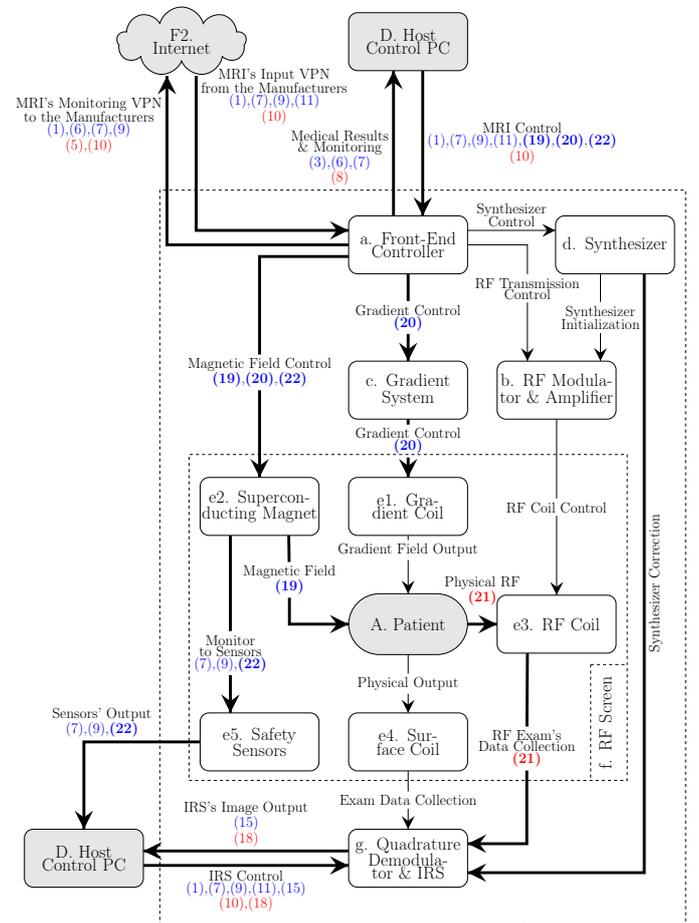

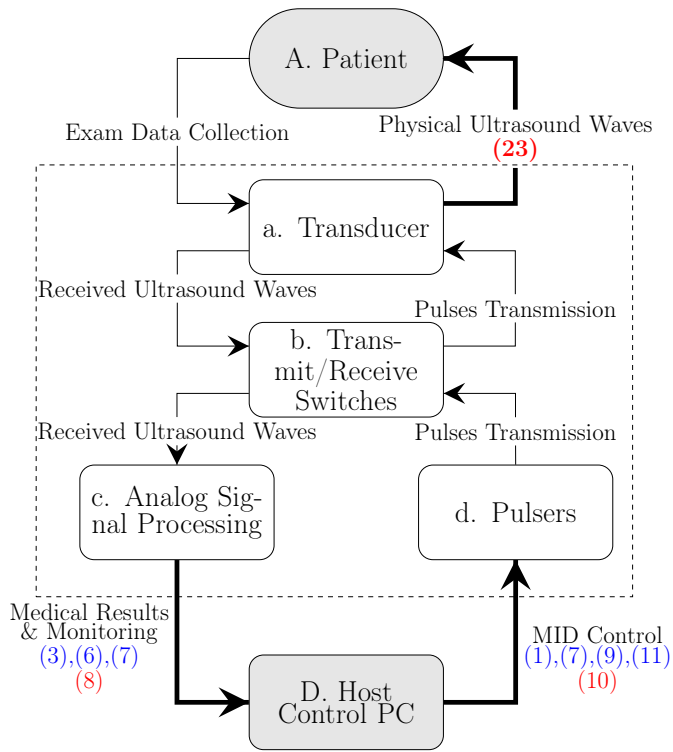
\begin{figure}[!htb]
	\centering
	\resizebox{\linewidth}{!}{\begin{tikzpicture}[
    align=center,
    node distance=1cm and 1cm,
]
    \draw[dashed] (-6.5,-2.25) rectangle (6,-11.375);
    
    \node (A) [terminator] {\afdref{itm:Patient}{Patient}};
    
    \node (a) [component, below=of A, yshift=-.5cm] {\afdref{itm:Transducer}{Transducer}};
    
    \path (A.west) -| node[directedText, anchor=center, below, xshift=-1.5cm, yshift=-1.25cm] (Aa) {Exam Data Collection} ([yshift=.5cm]a.west);
    \draw [directed] (A.west) -| (Aa) |- ([yshift=.5cm]a.west);
    \path ([yshift=.5cm]a.east) -| node[directedText, anchor=center, above, xshift=1.5cm, yshift=.75cm] (aA) {
        Physical Ultrasound Waves\\
        \textcolor{red}{\textbf{(23)}}
        } (A.east);
    \draw [bolded] ([yshift=.5cm]a.east) -| (aA) |- (A.east);
    
    \node (b) [component, below=of a] {\afdref{itm:Transmit/receive switches}{Transmit/Receive Switches}};
    
    \path ([yshift=-.5cm]a.west) -| node[directedText, anchor=center, below, xshift=-1.5cm, yshift=-.5cm] (ab) {Received Ultrasound Waves} ([yshift=.5cm]b.west);
    \draw [directed] ([yshift=-.5cm]a.west) -| (ab) |- ([yshift=.5cm]b.west);
    \path ([yshift=.5cm]b.east) -| node[directedText, anchor=center, above, xshift=1.5cm, yshift=.5cm] (ba) {Pulses Transmission} ([yshift=-.5cm]a.east);
    \draw [directed] ([yshift=.5cm]b.east) -| (ba) |- ([yshift=-.5cm]a.east);
    
    \node (Dtop) [emptycomponent, below=of b] {};
    \node (c) [component, left=of Dtop, xshift=1.5cm] {\afdref{itm:Analog signal processing}{Analog Signal Processing}};
    \node (d) [component, right=of Dtop, xshift=-1.5cm] {\afdref{itm:Pulsers}{Pulsers}};
    
    \path ([yshift=-.5cm]b.west) -| node[directedText, anchor=center, below, yshift=-.5cm] (bc) {Received Ultrasound Waves} (c.north);
    \draw [directed] ([yshift=-.5cm]b.west) -| (bc) -- (c.north);
    \path (d.north) -- node[directedText, anchor=center, below, xshift=.75cm, yshift=.25cm] (db) {Pulses Transmission} ([yshift=-.5cm]b.east);
    \draw [directed] (d.north) -- (db) |- ([yshift=-.5cm]b.east);
    
    \node (D) [component, below=of Dtop, fill=gray!20!white, yshift=-1cm] {\afdref{itm:Host control PC}{Host Control PC}};
    
    \draw [bolded] (D.east) -| node[directedText, anchor=center, above right]{
        \Glsxtrshort{mid} Control\\
        \textcolor{blue}{(\ref{attack:Ransomware}),(\ref{attack:Mute Safety Alarms}),(\ref{attack:Activate False Safety Alarms}),(\ref{attack:Restore System})}\\
        \textcolor{red}{(\ref{attack:Mechanical Disruption of MID's Motors})}
    } (d.south);
    \draw [bolded] (c.south) |- node[directedText, anchor=center, above left]{
        Medical Results\\
        \& Monitoring\\
        \textcolor{blue}{(\ref{attack:Alteration of the imaging Exam's Results}),(\ref{attack:Manipulation of Data Displayed on the Host's Monitor}),(\ref{attack:Mute Safety Alarms})}\\
        \textcolor{red}{(\ref{attack:Disruption of the imaging Exam's Results})}
    } (D.west);
\end{tikzpicture}\unskip}
	\caption[The AFD of the generic ultrasound]{The \glsxtrshort{afd} of a generic ultrasound with the relevant \glsxtrshortpl{ifv} and attack vectors, based on~\cite{Kutz09, TI17, element09}.}\label{fig:The AFD of the generic ultrasound3}
\end{figure}

Following the \cgls{tldr} methodology, we begin with identifying the potentially vulnerable components of \cglspl{mid} using \cglspl{afd}.\
In this section, we create the \cglspl{afd} of the generic \cglspl{mid} (see~\Cref{fig:The AFD of the generic MID3}), generic \cgls{ct} (see~\Cref{fig:The AFD of the generic CT3}), generic \cgls{mri} (see~\Cref{fig:The AFD of the generic MRI3}), and generic ultrasound (see~\Cref{fig:The AFD of the generic ultrasound3}).
Note that we already marked the potential attacks, which we identify in the following subsection.
For each \cgls{afd} of \cgls{mid}, we describe the presented components and the data flow inside the \cgls{mid}.\

\paragraph[The AFD of the generic MID]{The \glsxtrshort{afd} of the Generic \glsxtrshort{mid}}\label{results:The AFD of the generic MID}

The \cgls{afd} of the generic \cgls{mid} includes components, which typically appear in most \cglspl{mid}:\
\begin{enumerate}[noitemsep,nolistsep,label=\Alph*.,ref=\Alph*, leftmargin=*]
	\item\label{itm:Patient} The patient.
	\item\label{itm:Complementary devices} Complementary devices (i.e., devices that aid the \cgls{mid} during the scan):
	\begin{enumerate}[noitemsep,nolistsep,label=\Alph{enumi}\arabic*., ref=\Alph{enumi}\arabic*,leftmargin=*]
		\item\label{itm:Mechanical bed} Mechanical patient position system (mechanical bed): a bed on which the patient lies.
		It is controlled via the host control PC or the \cgls{mid}.\
		\item\label{itm:CIS} \cGls{cis}:\ a semi\=/automatic or manual mechanical system for injecting different contrast materials during the scan to enhance image results.\footnote{The \cgls{cis} is connected to the patient's blood system prior to the scan.} It is controlled via the host control PC or a special control unit.
		\item\label{itm:Patient monitoring systems} Patient monitoring systems: monitor the patient's physical conditions (e.g., \cgls{ecg} for monitoring heart rate).
		\item\label{itm:Monitor} Monitor: the technician's primary monitor in the control room, presenting essential information (e.g., scan's results, \cgls{mid}, and patient monitoring).
		\item\label{itm:EMR} \Glsxtrfull{emr} PC:\ presents the patient's \cgls{emr}, including the scan configurations specified by the physician.
		This functions as the technician's interface to the \cgls{ris}\footnote{The system that manages the workflow within the radiology department.} and the \glsxtrshort{pacs}.\
	\end{enumerate}
	\item\label{itm:The MID} The \glsxtrfull{mid}:\ performs the actual scan based on commands sent from the host control PC.\
	It is often connected to the manufacturers via a \cgls{vpn} over the Internet for technical monitoring.
	In the following sections, it is expanded into several \cglspl{afd} of unique \cglspl{mid} with specific attacks.
	\item\label{itm:Host control PC} Host control PC:\ the central control unit of the \cgls{mid}, operated by a certified technician, sending commands to the \cgls{mid} and other components.
	The commands are based on predefined scan protocol configurations, which are modified by the technician to fit the desired scan.
	It is connected to the hospital network for remote access and for sending scan results to the \glsxtrshort{pacs} database.
	\item\label{itm:Technician} The technician operates the \cgls{mid}.\
	\item\label{itm:Networks} Networks: hospitals usually consist of two networks:
	\begin{enumerate}[noitemsep,nolistsep,label=\Alph{enumi}\arabic*., ref=\Alph{enumi}\arabic*]
		\item\label{itm:Hospital networks} Hospital networks: internal networks, usually further divided into segments.
		Such networks are notorious for security and privacy issues.
		\item\label{itm:Internet} Internet: the \cgls{www}, including \cgls{vpn} connections for monitoring the \cglspl{mid}.\
	\end{enumerate}
	\item\label{itm:PACS} \cGls{pacs} database: archives all patients' images.
	The communication with it is done using the \cgls{dicom} protocol.
	\item\label{itm:Medical doctor} Medical doctor: specifies the required scan for the patient, evaluates the results, and can access the imaging results from any workstation in the hospital.
\end{enumerate}

The flow begins with the medical doctor (\ref{itm:Medical doctor}) requesting the scan from his office via a workstation in the hospital's network (\ref{itm:Networks}), which is then uploaded to the \cgls{emr} system.
Once the patient (\ref{itm:Patient}) arrives to have the scan, the technician (\ref{itm:Technician}) extracts the scan request manually (i.e., by physically viewing it) from the \cgls{emr} PC (\ref{itm:EMR}) and configures the host control PC (\ref{itm:Host control PC}) accordingly.
Then, he connects the patient (\ref{itm:Patient}) to complementary devices (\ref{itm:Complementary devices}) by placing the patient (\ref{itm:Patient}) on the mechanical bed (\ref{itm:Mechanical bed}) and connecting him to the \cgls{cis} (\ref{itm:CIS}), and the patient monitoring systems (\ref{itm:Patient monitoring systems}).
When the technician (\ref{itm:Technician}) initiates the scan, the host control PC (\ref{itm:Host control PC}) sends control signals to the \cgls{mid} (\ref{itm:The MID}), the mechanical bed (\ref{itm:Mechanical bed}), and the \cgls{cis} (\ref{itm:CIS}), and receives feedback from the \cgls{mid} (\ref{itm:The MID}) and the patient monitoring systems (\ref{itm:Patient monitoring systems}).
The received feedback is presented on the primary monitor (\ref{itm:Monitor}).

At the top of the \cgls{afd}, we present the interactions with the patient (\ref{itm:Patient}).
In case of problems or potential danger, the technician (\ref{itm:Technician}) can apply various safety mechanisms and stop the scanning process.
The scan results are collected via the host control PC (\ref{itm:Host control PC}) and sent to the \cgls{pacs} database (\ref{itm:PACS}), which is accessible to the medical doctor (\ref{itm:Medical doctor}) via the hospital's network (\ref{itm:Hospital networks}).
Often, the raw data (i.e., before reconstruction\footnote{Processing raw analog signal measurements into digital images.}) is stored internally by the \cgls{mid} (\ref{itm:The MID}) and is also accessible to the medical doctor (\ref{itm:Medical doctor}).

It is important to note that today, \cglspl{mid} are connected directly to the Internet (\ref{itm:Internet})~\cite{ISE16} in various ways, such as directly to the manufacturer for monitoring, troubleshooting, and more~\cite{Rotter16}, or via the \cgls{pacs} server (\ref{itm:PACS}) inside the hospital's internal network (\ref{itm:Hospital networks}), which is usually connected to the Internet (\ref{itm:Internet}).
Although these interactions are not part of the \cgls{mid} ecosystem, such connections make \cglspl{mid} very vulnerable since they enable potential access for attackers from anywhere in the world.

\paragraph[The AFD of the generic CT]{The \glsxtrshort{afd} of the Generic \glsxtrshort{ct}}\label{results:The AFD of the generic CT}

The \cgls{afd} of the generic \cgls{ct} includes the following components:
\begin{enumerate}[noitemsep,nolistsep,label=\alph*.,ref=\alph*, leftmargin=*]
	\item\label{itm:xray source} X\=/ray source emits ionizing radiation.
	\item\label{itm:Gantry} Gantry: the mechanical device that controls the X\=/ray source, including the X\=/ray detectors, which measure the output radiation.
	In \cgls{ct}, the patient lays inside the tube containing the X\=/ray sources and detectors.
	In some cases, manufacturers open a \cgls{vpn} connection from the Internet for monitoring purposes.
	\item\label{itm:IRS} \cGls{irs}: the computer that collects the raw data of the image, produced by the gantry's radiation measures during the scan, and reconstructs it into \cgls{dicom} images, which are sent to the host control PC.\
\end{enumerate}

Once the scan is initiated, the host control PC (\ref{itm:Host control PC}) sends commands to the gantry (\ref{itm:Gantry}), which controls the X\=/ray source (\ref{itm:xray source}) and measures the X\=/ray effects from the patient (\ref{itm:Patient}).
The gantry (\ref{itm:Gantry}) sends the raw results to the \cgls{irs} (\ref{itm:IRS}), via the host control PC (\ref{itm:Host control PC}) or directly in some devices, for reconstruction.\footnote{In some devices the gantry (\ref{itm:Gantry}) sends results directly to the \cgls{irs} (\ref{itm:IRS}).}
The reconstructed image is then sent to the host control PC (\ref{itm:Host control PC}), and the flow continues as it did in \Cref{fig:The AFD of the generic MID3}, including the interactions with the Internet (\ref{itm:Internet}).
The gantry's (\ref{itm:Gantry}) inner components breakdown is out of the scope of the current study.

\paragraph[The AFD of the generic MRI]{The \glsxtrshort{afd} of the Generic \glsxtrshort{mri}}\label{results:The AFD the generic MRI}

The \cgls{afd} of the generic \cgls{mri} includes the following components:
\begin{enumerate}[noitemsep,nolistsep,label=\alph*.,ref=\alph*,leftmargin=*]
	\item\label{itm:Front-End controller} Front\=/end controller: sometimes referred to as the \cgls{pp}, translates and forwards commands to control signals for different components (e.g., magnets, gradient systems, \cgls{rf} system, etc.).
	
	\item\label{itm:RF modulator and amplifier} \cGls{rf} modulator \& amplifier: generates \cgls{rf} pulses from the synthesizer and monitors \cgls{sar} limits,\footnote{The potential for heating the patient's tissue due to the application of the \cgls{rf} energy.} preventing an overpowered scan from taking place.
	This expensive component can amplify signals from \SI{10}{\kW} to an order of \SI{16}{\kW}.
	
	\item\label{itm:Gradient system} Gradient system: linearly varies the direction of the magnetic field to implement pulse sequences.
	
	\item\label{itm:Synthesizer} Synthesizer: generates an \cgls{rf} (i.e., the center frequency), which is used in the demodulation of the received signal to centralize it around zero.
	Without the synthesizer, the center frequency may differ from one patient to another, which may cause inaccurate results.
	
	\item\label{itm:MRI coils} \cGls{mri} coils:%
	\begin{enumerate}[noitemsep,nolistsep,label=\alph{enumi}\arabic*., ref=\alph{enumi}\arabic*, leftmargin=*]
		\item\label{itm:Gradient coil} Gradient coil: generates three orthogonal linear magnetic field gradients, varying the magnetic field for spatially encoding of the \cgls{mr} signal's location.
		It is mounted on a cylindrical former inside the bore of the magnet.
		The higher the maximum gradient amplitude, the smaller the \cgls{fov}, and the thinner the slice widths.
		Several specialized scanning techniques (e.g., diffusion imaging) require high peak gradient amplitudes.
		\item\label{itm:Superconducting magnet} Superconducting magnet: the main component of the \cgls{mri}, with magnetic field strength (i.e., flux density) of \SIrange{1.5}{3}{\tesla} for clinical use, and over \SI{3}{\tesla} for research purposes.
		The higher the magnetic field strength, the better the \cgls{snr} (i.e., the better the image quality) is; therefore, the \cgls{snr} is the most useful image quality parameter.
		The homogeneity of the magnetic field is another crucial factor, which is measured in \cgls{ppm} within a given spherical volume, defined as the \cgls{dsv}.\
		There are four main types of magnets used in \cglspl{mri}: air\=/cored, iron\=/cored (i.e., electromagnet), permanent, and superconducting, which is the most popular today.
		Superconducting magnets are held at temperatures that approach absolute zero (\SI{-273.13}{\degreeCelsius} or \SI{0}{\kelvin}) to allow zero electrical resistance.
		To achieve that, these magnets are held in a cryogenic bath (i.e., liquid helium) at less than \SI{4.2}{\kelvin}.
		When a problem occurs, and the magnet needs to be turned off immediately, the magnet can be put in quench state, and the liquid helium, keeping the magnet cold, is released, causing the magnet to stop functioning altogether.
		\item\label{itm:RF coil} \cGls{rf} coil: used to transmit and receive \cgls{rf} signals, where the transmission must be a homogeneous \cgls{rf} magnetic field.
		\item\label{itm:Surface coil} Surface coil: a physical coil that is placed directly over the anatomical region of interest for better signal reception.
		\item\label{itm:Safety sensors} Safety sensors: alerts about dangerous situations.
		They include thermal and oxygen level sensors, which alert about abnormal temperature and dangerous depletion of oxygen, which may happen when there is a helium gas leak from the quench pipe.
	\end{enumerate}
	
	\item\label{itm:RF screen} \cGls{rf} screen: acts as a Faraday cage to reduce the extended fringe field from the magnet and the interference from external transmitters.
	
	\item\label{itm:Quadrature demodulator and IRS} Quadrature demodulator \& \cgls{irs}:\ boosts signals detected from the surface coils, mixes them with the \cgls{rf} from the synthesizer, converts the analog raw data into digital data, and reconstructs the image.
\end{enumerate}

The flow begins when the \cgls{mri} enters the initialization stage, which includes tuning the components.
When a scan configuration is entered, the host control PC (\ref{itm:Host control PC}) verifies that the values are within the hardware limits (e.g., maximum gradient strength) or software limits (e.g., the maximum number of slices).
Afterward, the commands are sent to the front\=/end controller (\ref{itm:Front-End controller}), which translates and forwards the commands to the gradient system (\ref{itm:Gradient system}), the synthesizer (\ref{itm:Synthesizer}), the magnet (\ref{itm:Superconducting magnet}), and the \cgls{rf} coil (\ref{itm:RF coil}).
The \cgls{rf} modulator (\ref{itm:RF modulator and amplifier}) verifies that the \cgls{rf} signal is within a safe range, and the gradient system (\ref{itm:Gradient system}) creates the gradient field.

The internal flow inside the \cgls{mri}'s gantry (which is presented as a logical encapsulation) begins when the gradient coil (\ref{itm:Gradient coil}) generates the magnetic field gradients for spatially encoding the location of the \cgls{mr} signal.
The magnet (\ref{itm:Superconducting magnet}) produces the magnetic field, and the \cgls{rf} coil (\ref{itm:RF coil}) measures the \cgls{rf} radiation emitted from the patient (\ref{itm:Patient}) and sends raw results to the \cgls{irs} (\ref{itm:Quadrature demodulator and IRS}) for reconstruction.
Note that the synthesizer (\ref{itm:Synthesizer}) is also connected to the \cgls{irs} (\ref{itm:Quadrature demodulator and IRS}) for the initial tuning of the signals.
The surface coil (\ref{itm:Surface coil}) is used to increase signal reception.
The \cgls{rf} screen (\ref{itm:RF screen}) protects the coils from the outside \cgls{rf} signals, which may interfere with the results.
The reconstructed images are sent back to the host control PC (\ref{itm:Host control PC}), and the flows continue as it did in \Cref{fig:The AFD of the generic MID3}.

\paragraph[The AFD of the generic ultrasound]{The \glsxtrshort{afd} of the Generic Ultrasound}\label{results:The AFD of the generic ultrasound}

The \cgls{afd} of the generic ultrasound includes the following components:
\begin{enumerate}[noitemsep,nolistsep,label=\alph*.,ref=\alph*,leftmargin=*]
	\item\label{itm:Transducer} Transducer: a physical component that transmits ultrasonic signals and measures the time between the transmission and reception.
	This component can be divided into transmit and receive transducers.
	\item\label{itm:Transmit/receive switches} Transmit/receive switches: control when the transducer will transmit and when it will receive since the transducer performs both functions.
	\item\label{itm:Analog signal processing} Analog signal processing: reconstructs the image using the analog ultrasonic signals detected (similar to the \cgls{irs}).
	\item\label{itm:Pulsers} Pulsers: creates the electric pulses for the ultrasonic transmission.
\end{enumerate}

The technician configures the scan using the host control PC (\ref{itm:Host control PC}), which sends commands to the pulsers (\ref{itm:Pulsers}) that create the ultrasonic pulses.
The switches (\ref{itm:Transmit/receive switches}) manage the incoming and outgoing pulses so that they do not interfere with one another.
Once the pulses reach the transducer (\ref{itm:Transducer}), the operator presses the transducer (\ref{itm:Transducer}) against the skin of the patient (\ref{itm:Patient}) and measures the returning pulses.
The measurements are received in an analog form and sent to the analog signal processing (\ref{itm:Analog signal processing}) for reconstruction.
The final results are sent to the host control PC (\ref{itm:Host control PC}), and the flows continue as in \Cref{fig:The AFD of the generic MID3}.

\subsection[Identifying the potential attacks and marking them on the AFDs]{Identifying the Potential Attacks and Marking Them on the \glsxtrshortpl{afd}}\label{results:Identifying the potential attacks and marking them on the AFDs}

In this section, we provide a detailed description of each of the 23 potential attacks that we have found.
We list the attacks based on the relevant \cgls{mid} that it affects.

\paragraph[Generic MID attacks]{Generic \glsxtrshort{mid} Attacks}\label{results:Generic MID Attacks}

\begin{attack}[Ransomware~\cite{Larson17, Liptak17, Millar17, Brenner17, BBC17, UngoedThomas17, Hern17, CarrieWong2017, Foxx17, Skinner16, Strickland16, Wong16, Winton16, Mahler18}]\label{attack:Ransomware}
	Since \cglspl{mid} rely on computers to operate, ransomware attacks (\Cref{CAPEC-542}) such as the \nameref{appendix:WannaCry} cyber attack (see~{\S}~{\ref{appendix:WannaCry}}) mentioned earlier, can lock the host control PC and make the \cgls{mid} nonoperational, damaging the availability on a large scale.
	Also, it can encrypt the scan results or essential system files in critical situations and may put patients at risk.
	New ransomware attacks may also involve malicious firmware updates of various components of the \cgls{mid}.\
	In those cases, cleaning the host PC will not restore the system.
\end{attack}

\begin{attack}[Disruption of patient\=/to\=/image linkage~\cite{Ayala16, Mahler18}]\label{attack:Disruption of patient-to-image linkage}
	The scan results are linked to the specific \cgls{emr} of the patient being scanned.
	Linking different results and patient information to specific patients is done inside the host control PC, and the linked results are sent to the \cgls{pacs} server using the \cgls{dicom} protocol.
	Targeted malware (\Cref{CAPEC-542}) resident on the host control PC can disrupt these linkages (\Cref{CAPEC-165}), by either linking the result to the wrong patient or relating the wrong results to the patient being scanned.
	This attack is sophisticated and requires specific insider knowledge (\Cref{CAPEC-150}).
	This attack can affect many patients (i.e., scale), causing incorrect diagnosis of patients, improper treatment, and privacy issues.
	This attack is technically similar to \Cref{attack:Alteration of the imaging Exam's Results} (\nameref{attack:Alteration of the imaging Exam's Results}).
\end{attack}

\begin{attack}[Alteration of the imaging exam's results~\cite{Mahler18,Becker2018,Mahler19}]\label{attack:Alteration of the imaging Exam's Results}
	The attacker attempts to alter the output image of the imaging exam in order either to hide a medical condition (e.g., remove a dangerous tumor) or to deceive (e.g., inject a nonexistent tumor), leading to improper treatment of patients.
	Since patients often receive treatment based on the imaging results, the outcome of such an attack can be very dangerous, including direct physical harm to patients, which could be fatal in some cases.
	Furthermore, if performed well, it is very challenging to detect the attack and notice that something is wrong; thus, an attacker could create a small change in the image for all patients, which may have no significant influence on the patient in the short run but could damage the health of many patients (i.e., scale) in the long run.
\end{attack}

\begin{attack}[Contrast material over/underdose]\label{attack:Contrast Material Over/Underdose}
	The \cgls{cis} is often used in \cgls{mid} exams in order to inject different contrast materials and enhance the scan's results.
	The \cgls{cis} is a mechanical system that is connected to the patient's blood system and remotely operated via a special control unit or the host control PC.\
	The \cgls{cis} is semi\=/automatically activated by the technician and the host control PC during the exam.
	Targeted malware (\Cref{CAPEC-542}) resident on the host control PC can manipulate the amount of contrast material injected.
	The severity of this attack may result in disrupted exam results due to incorrect contrast material dose and potential adverse side effects to the patient due to over\=/injection of the contrast material (e.g., overdose with potential toxicity).
	In addition, misuse of this system may damage it.
\end{attack}

\begin{attack}[Leakage of patients' private information]\label{attack:Leakage of Patients' Private Information}
	\cGlspl{mid} have access to sensitive private information that enables patient identification, as well as the patient's \cgls{emr}, in order to link the exam results to the patient's \cgls{emr}.\
	In addition, \cglspl{mid} are connected to the hospital's network, and in many cases, are connected directly to the manufacturer's monitoring systems via the Internet.
	This makes \cglspl{mid} a potential gateway for leakage of patients' sensitive private information such as exam results or other private information that the \cgls{mid} has access to.
	Also, this attack may leak private information of many patients (i.e., scale).
\end{attack}

\begin{attack}[Manipulation of data displayed on the host's monitor~\cite{Ayala16}]\label{attack:Manipulation of Data Displayed on the Host's Monitor}
	The host's monitor is the primary monitor that the technician uses during the scan.
	This attack is similar to \Cref{attack:Activate False Safety Alarms} (\nameref{attack:Activate False Safety Alarms}), however, in addition, it displays all of the information related to the scan, the patient, and the \cgls{mid}, such as the scan's configurations and results, the patient's physical condition, and the control and configuration panel of the \cgls{mid}.\
	The severity of this attack is only related to the patient.
	A targeted malware may present false information to the monitor, which may be completely different from the real information (e.g., a normal heart rate while the actual heart rate is high, or \viceversa), in order to influence the scan, affecting the physical condition of the patient as well as the diagnosis.
	Such malware can easily evade detection; hence, it may affect many patients before being discovered.
\end{attack}

\begin{attack}[Mute safety alarms~\cite{Ayala16}]\label{attack:Mute Safety Alarms}
	Technically similar to \Cref{attack:Activate False Safety Alarms} (\nameref{attack:Activate False Safety Alarms}); however, in this attack, instead of activating false alarms, the attacker tries to mute safety alarms.
	This may prevent the \cgls{mid} from being automatically or manually stopped if a dangerous situation occurs, which could be fatal.
\end{attack}

\begin{attack}[Disruption of the imaging exam's results~\cite{Mahler18}]\label{attack:Disruption of the imaging Exam's Results}
	Technically similar to \Cref{attack:Alteration of the imaging Exam's Results} (\nameref{attack:Alteration of the imaging Exam's Results}); however, this attack focuses merely on disrupting the exam's results, possibly requiring repeating the exam.
	The attacker can do so by manipulating files (\Cref{CAPEC-165}) and common resources (\Cref{CAPEC-150}) or by jamming (\Cref{CAPEC-601}) and disabling the route of the output image (\Cref{CAPEC-582}), which interferes with patient diagnosis.
	While this attack may not always be severe, in cases that require immediate care to the patient, even the briefest delay may be fatal.
\end{attack}

\begin{attack}[Activate false safety alarms~\cite{Ayala16}]\label{attack:Activate False Safety Alarms}
	Most \cglspl{mid} include various safety alarms and monitors, such as fire alarms, oxygen alarms, heart rate monitors, blood pressure monitors, etc.
	These alarms include sensors that are configured to automatically stop the \cgls{mid}'s operation in case of a problem or issue and alert the technician to turn off the \cgls{mid} manually.
	Some alarm systems include standalone alarms (e.g., a red light or speaker), which are more challenging to activate.
	However, some alarms are connected to a centralized computer, such as the host control PC, or other computers in the \cgls{mid}'s ecosystem.
	A targeted malware (\Cref{CAPEC-542}) on such a computer can activate false alarms to cause the \cgls{mid} to automatically or manually (by the technician) stop during treatment (e.g., interventional radiology), which may put the patient at increased risk.
	In some cases, such operation could also damage the device itself, affecting its availability.
	This attack pattern uses environment hardware and sensors (our new attack pattern: \hyperref[CAPEC-NEW]{CAPEC~NEW} \nameref{CAPEC-NEW}), as well as additional obstruction (\hrefblue{https://capec.mitre.org/data/definitions/607.html}{CAPEC\=/607: Obstruction})\footnote{\hrefblue{https://capec.mitre.org/data/definitions/607.html}{https://capec.mitre.org/data/definitions/607.html}} attack patterns: disabling the route to/from the alarm or the sensor (\Cref{CAPEC-582}), jamming alarms that use \cgls{rf} communication (\Cref{CAPEC-601}), and blockage of alarm or sensor (\Cref{CAPEC-603}).
	The severity of such an attack range from damage to the \cgls{mid} to physical harm to the patient.
\end{attack}

\begin{attack}[Mechanical disruption of \glsxtrshort{mid}'s motors~\cite{Mahler18}]\label{attack:Mechanical Disruption of MID's Motors}
	\cGls{mid} ecosystems include several components with mechanical motors: mechanical bed motors, scanner motors, rotation motors, etc.
	The motors receive instructions from a control unit (e.g., the host control PC).
	Targeted malware (\Cref{CAPEC-542}) can change the commands controlling these motors (i.e., \hyperref[CAPEC-NEW]{CAPEC~NEW}: manipulate environment hardware or sensors), causing an undesired change in the movement of these motors.
	By misusing the motors, this attack can directly damage the device, possibly wrecking the motors and damaging their availability and the availability of the \cgls{mid}.\
	Also, since these motors physically interact with the patient, this attack may potentially harm the patient by making the motors move in a way that hurts the patient.
\end{attack}

\begin{attack}[Restore system~\cite{Ayala16}]\label{attack:Restore System}
	This attack attempts to use the system restore option (\Cref{CAPEC-166}) found in most computers in \cgls{mid} ecosystems (e.g., the host control PC) or externally restore the system to factory default configurations.
	This attack may cause a loss of data (e.g., scan results) or a loss of custom configurations set by the technician.
	This may affect the availability of the device, resulting in delayed treatment for patients.
\end{attack}

\paragraph[Generic CT Attacks]{Generic \glsxtrshort{ct} Attacks}\label{results:Generic CT Attacks}

\begin{attack}[Increase milliamperage\=/seconds \si{[\mAs]}~\cite{Ayala16, Mahler18}]\label{attack:Increase Milliamperage-Seconds}
	Similar to \Cref{attack:Increase Kilovoltage Peak} (\nameref{attack:Increase Kilovoltage Peak}); however, this attack targets another specific parameter of the radiation.
	This parameter measures the radiation produced \si{[\mA]} over a set of time \si{[\s]}~\cite{Bushong2013a}; increasing the \si{\mA} results in an increase in the quantity of radiation.
	Malicious manipulation of this parameter by targeted malware is very dangerous and may cause distorted or incorrect image results and radiation overdose to the patient, which may put the patient at risk.
\end{attack}

\begin{attack}[Increase kilovoltage peak \si{[\kVp]}~\cite{Ayala16, Mahler18}]\label{attack:Increase Kilovoltage Peak}
	Similar to \Cref{attack:Radiation Overdose} (\nameref{attack:Radiation Overdose}); however, this attack targets a specific parameter of the radiation.
	The \si{\kVp} parameter determines the highest energy of X\=/ray photon, and the radiation dose to the patient is directly proportional to the square of \si{\kVp}.
	Increasing the \si{\kVp} will decrease the contrast seen between soft tissue and bone~\cite{Fuchs71}.
	Malicious manipulation of this parameter by targeted malware is very dangerous and may cause distorted or incorrect image results as well as radiation overdose to the patient, which may put the patient at risk.
\end{attack}

\begin{attack}[Radiation overdose~\cite{Ayala16, Mahler18}]\label{attack:Radiation Overdose}
	Similar to \Cref{attack:Configuration File Disruption} (\nameref{attack:Configuration File Disruption}); however, this attack targets the radiation configurations directly.
	The implications of this attack are very dangerous, as it may cause radiation overdose to the patient, which may put the patient at risk.
	Furthermore, the attacker may increase the radiation dosage to amounts that are not immediately noticeable but could potentially increase the risk of the development of cancer in the future; this attack may affect many patients (i.e., scale) before it is discovered.
\end{attack}

\begin{attack}[Alteration of the \glsxtrshort{irs}'s output images~\cite{Mahler18,Becker2018,Mahler19}]\label{attack:Alteration of the IRS's Output Images}
	Similar to \Cref{attack:Alteration of the imaging Exam's Results} (\nameref{attack:Alteration of the imaging Exam's Results}); however, this attack targets a different component (\Cref{CAPEC-150}): the \cgls{irs} inside the \cgls{ct}, affecting the diagnosis of patients.
	In addition, this attack may affect many patients whose images were not yet sent to the \cgls{pacs}.\
	It is also vulnerable to the same attacks as \Cref{attack:Disruption of the IRS's Output Images}.
\end{attack}

\begin{attack}[Configuration file disruption~\cite{Ayala16, Mahler18}]\label{attack:Configuration File Disruption}
	The entire \cgls{ct} scan operation is defined in the scan configuration file, inside the host control PC, making it extremely important.
	By manipulating the file, an attacker can change the \cgls{ct}'s behavior.
	Moreover, the attacker can change the output commands directly for a similar effect.
	Targeted malware (\Cref{CAPEC-542}) resident on the host control PC can manipulate this file (\Cref{CAPEC-165}, \Cref{CAPEC-75}) to control the entire \cgls{ct} operation.
	Since the \cgls{ct} operation physically affects the patient, this attack also affects the patient in this way, as well as affecting the patient's diagnosis.
	In addition, misuse of the configurations may damage the \cgls{mid}.\
	Since the attacker can manipulate configurations, which may not be noticeable immediately, this attack may affect many patients (i.e., scale) before it is discovered.
\end{attack}

\begin{attack}[Manipulation of \glsxtrshort{ct} calibration]\label{attack:Manipulation of CT Calibration}
	The \cgls{ct} is pre\=/calibrated so that it will work correctly.
	This attack attempts to manipulate this calibration and cause the device to use parameters with an offset (\Cref{CAPEC-166}).
	Incorrect calibration may damage the device, possibly destroying it and damaging the availability of the \cgls{mid}.\
	Many people may use the \cgls{ct} with the wrong calibration before the problem is discovered.
	In addition, incorrect calibration of the device may result in misdiagnosis of the patient or radiation overdose (e.g., calibrating the radiation parameters to include a large offset in each scan).
\end{attack}

\begin{attack}[Disruption of the \glsxtrshort{irs}'s output images~\cite{Mahler18}]\label{attack:Disruption of the IRS's Output Images}
	Similar to \Cref{attack:Disruption of the imaging Exam's Results} (\nameref{attack:Disruption of the imaging Exam's Results}); however, this attack targets a different component (\Cref{CAPEC-150}): the \cgls{irs} inside the \cgls{ct} modality.
	This component is usually a standard Windows PC, and it is difficult to update it regularly since it runs the specific software of the manufacturer.
	This makes it very vulnerable to attacks, with a severity similar to \Cref{attack:Disruption of the imaging Exam's Results} (\nameref{attack:Disruption of the imaging Exam's Results}).
\end{attack}

\paragraph[Generic MRI Attacks]{Generic \glsxtrshort{mri} Attacks}\label{results:Generic MRI Attacks}

\begin{attack}[Overwhelm of the \glsxtrshort{mri}'s receiving coils with an overpowered magnetic field~\cite{Ayala16}]\label{attack:Overwhelm of the MRI's Receiving Coils with an Overpowered Magnetic Field}
	Similar to \Cref{attack:Configuration File Disruption} (\nameref{attack:Configuration File Disruption}); however, targeting the \cgls{mri} receiving coils.
	A targeted malware (\Cref{CAPEC-542}) can create a powerful magnetic field by manipulating the scan configuration files (\Cref{CAPEC-75} and \Cref{CAPEC-165}), which can damage the receiving coils and possibly cause damage to other electronics nearby that are sensitive to magnetic fields.
\end{attack}

\begin{attack}[Magnetic field disruption~\cite{Ayala16}]\label{attack:Magnetic Field Disruption}
	Technically similar to \Cref{attack:Overwhelm of the MRI's Receiving Coils with an Overpowered Magnetic Field} (\nameref{attack:Overwhelm of the MRI's Receiving Coils with an Overpowered Magnetic Field}); however, this attack focus on disrupting output images.
	In addition to the attack patterns described in \Cref{attack:Overwhelm of the MRI's Receiving Coils with an Overpowered Magnetic Field}, this attack may also use jamming of the \cgls{rf} signal (\Cref{CAPEC-601}).
	This attack may result in high \cgls{rf} radiation capable of causing heat burns to the patient and possible damage to the machine.
\end{attack}

\begin{attack}[External \glsxtrshort{rf} signal disruption]\label{attack:External RF Signal Disruption}
	An attacker can position a strong \cgls{rf} antenna inside or close to the \cgls{mri}'s Faraday cage (e.g., position the antenna inside a truck and park the truck very close to the \cgls{mri} room), which can interfere with the \cgls{mri}'s \cgls{rf} measures (\Cref{CAPEC-601} and \Cref{CAPEC-582}).
	This can corrupt the exam's results and cause increased \cgls{rf} radiation, which may put the patient at risk.
	If such an antenna is covertly placed, this attack may affect many patients before it is discovered.
\end{attack}

\begin{attack}[Activate quenching of \glsxtrshort{mri}~\cite{Ayala16}]\label{attack:Activate Quenching of MRI}
	Similar to \Cref{attack:Activate False Safety Alarms} (\nameref{attack:Activate False Safety Alarms}); however, specific to the \cgls{mri}.\
	\cGls{mri} usually uses superconducting magnets, which use liquid hydrogen to maintain a low temperature.
	Quenching refers to a situation in which the temperature of the magnet suddenly rises (e.g., due to a fire or sudden release of the liquid hydrogen), which may be triggered by our new attack pattern \hyperref[CAPEC-NEW]{CAPEC~NEW}: \nameref{CAPEC-NEW}.
	In such cases, there is a danger of asphyxia, hypothermia, and ruptured eardrums to the patient.
	In addition, quenching may take weeks to fix, affecting the availability of the \cgls{mri} to many patients whose treatment may be delayed.
	This may also damage the device itself.
\end{attack}

\paragraph{Generic Ultrasound Attacks}\label{results:Generic ultrasound attacks}

\begin{attack}[Disruption of \glsxtrshort{mems} components~\cite{Pan17}]\label{attack:Disruption of MEMS Components}
	\cGls{mems} are components consisting of microscopic devices with mechanical moving parts, widely used in a variety of devices and applications, including medical devices such as pacemakers and insulin pumps~\cite{Aram16}.
	Since \cgls{mems} consists of mechanical moving parts, they are sensitive to interference caused by sound vibrations.
	At the Black Hat USA conference held in 2017 in Las Vegas, Pan \etal~\cite{Pan17} demonstrated that ultrasound effects on devices with \cgls{mems} components could disrupt \cgls{vr} devices and self\=/balancing vehicles (e.g., Segway).
	A targeted malware (\Cref{CAPEC-542}) can affect the ultrasound hardware directly (our new attack pattern \hyperref[CAPEC-NEW]{CAPEC~NEW}: \nameref{CAPEC-NEW}) so that it acts similarly to that presented by Pan \etal~\cite{Pan17} and disrupts \cgls{mems} components near or inside the patient, such as \cglspl{icd} and insulin pumps, making these devices malfunction; this may be fatal; however, no such experiment has been performed to assess this threat.
\end{attack}

\subsection[Mapping the discovered attacks into their relevant CAPECs]{Mapping the Discovered Attacks into Their Relevant \glsxtrshortpl{capec}}\label{results:Mapping the Discovered Attacks into Their Relevant CAPECs}

In \Cref{tab:The application of the TLDR methodology to MIDs}, we present the results of our mapping of the discovered attacks into their relevant \cglspl{capec} (see~{\S}{\ref{methods:Mapping the Discovered Attacks into Their Relevant CAPECs}}).
As we can see, we were able to map the 23 potential attacks into one or more of the following eight \emph{general} \cgls{capec} standard attack patterns and to an additional new attack pattern that we suggest adding:

\setcounter{CAPEC}{74}
\begin{CAPEC}[\textcolor{blue}{\href{https://capec.mitre.org/data/definitions/75.html}{Manipulating Writeable Configuration Files}}]\label{CAPEC-75}
	An adversary manipulates configuration files that define how the system operates (e.g., the operation of the device during an imaging scan).
	This attack differs from \Cref{CAPEC-165} as it does \emph{not} attempt to cause \emph{improper} behavior, but instead, \emph{it attempts to cause proper behavior for improper configurations}.
\end{CAPEC}

\setcounter{CAPEC}{149}
\begin{CAPEC}[\textcolor{blue}{\href{https://capec.mitre.org/data/definitions/150.html}{Collect Data from Common Resource Locations}}]\label{CAPEC-150}
	An attacker easily accesses sensitive data stored in predefined or common resource locations (e.g., imaging results are saved in a default directory).
\end{CAPEC}

\setcounter{CAPEC}{164}
\begin{CAPEC}[\textcolor{blue}{\href{https://capec.mitre.org/data/definitions/165.html}{File Manipulation}}]\label{CAPEC-165}
	An adversary attempts to cause the improper behavior of the system by modifying the content or attributes of files that the system processes (e.g., manipulate the image result of the imaging scan).
	This attack is commonly combined with \Cref{CAPEC-150}, which focuses on accessing the location of the files.
\end{CAPEC}

\setcounter{CAPEC}{165}
\begin{CAPEC}[\textcolor{blue}{\href{https://capec.mitre.org/data/definitions/166.html}{Force the System to Reset Values}}]\label{CAPEC-166}
	An attacker reconfigures the system to an initial state, possibly by using the system's existing reset functions (e.g., Windows restore), potentially allowing the attacker to bypass security (e.g., use default passwords) and disable services.
\end{CAPEC}

\setcounter{CAPEC}{541}
\begin{CAPEC}[\textcolor{blue}{\href{https://capec.mitre.org/data/definitions/542.html}{Targeted Malware}}]\label{CAPEC-542}
	An attacker develops a targeted malware, usually exploiting a vulnerability in the targeted system or the \cgls{it} infrastructure (e.g., the \nameref{appendix:WannaCry} attack).
\end{CAPEC}

\setcounter{CAPEC}{581}
\begin{CAPEC}[\textcolor{blue}{\href{https://capec.mitre.org/data/definitions/582.html}{Route Disabling}}]\label{CAPEC-582}
	An attacker disrupts the network communication route (i.e., channel) between targets.
\end{CAPEC}

\setcounter{CAPEC}{600}
\begin{CAPEC}[\textcolor{blue}{\href{https://capec.mitre.org/data/definitions/601.html}{Jamming}}]\label{CAPEC-601}
	An attacker disrupts the communication using \cgls{rf} signals or illicit traffic, causing possible \cgls{dos}.\
\end{CAPEC}

\setcounter{CAPEC}{602}
\begin{CAPEC}[\textcolor{blue}{\href{https://capec.mitre.org/data/definitions/603.html}{Blockage}}]\label{CAPEC-603}
	An attacker denies the system access or delivery of critical resources, causing a possible \cgls{dos} attack.
\end{CAPEC}

In addition to the eight standard \cgls{capec} attack patterns, we define and use a new attack pattern in the \cgls{tldr} methodology, following the \cgls{capec} format:

\begin{CAPEC-NEW}[Manipulate Environment Hardware or Sensors]\label{CAPEC-NEW}
	Child of \hrefblue{https://capec.mitre.org/data/definitions/176.html}{CAPEC~176}\footnote{\hrefblue{https://capec.mitre.org/data/definitions/176.html}{https://capec.mitre.org/data/definitions/176.html}}.
	An adversary interferes with various sensors of the system in an attempt to affect the system's behavior.
	E.g., an adversary may increase the temperature in the environment of the system in order to cause the temperature sensor to activate, affecting the system.
\end{CAPEC-NEW}

We suggest including \hyperref[CAPEC-NEW]{CAPEC~NEW} in the official \cgls{capec} ontology since we were unable to identify an existing attack pattern in the taxonomy that corresponds to this pattern.

\subsection[Estimating the likelihood of the generic CAPECs into which the potential attacks are mapped]{Estimating the Likelihood of the Generic \glsxtrshortpl{capec} into Which the Potential Attacks Are Mapped}\label{results:Estimating the likelihood of the generic CAPECs into which the potential attacks are mapped}

In \Cref{tab:Estimating the likelihood of the generic CAPECs into which the potential attacks are mapped provided by our panel of healthcare ISEs}, we present the results, in descending order, of our panel of four healthcare \cglspl{ise}, each individually, \cglspl{capec} likelihood estimates (see~{\S}~{\ref{methods:Estimating the likelihood of the generic CAPECs into which the potential attacks are mapped}}), which we use throughout the rest of the \cgls{tldr} methodology.
All experts had a similar level of expertise, each individually with over ten years of experience in the relevant field, \cglspl{ciso} of major \cglspl{hmo}; therefore, to obtain each value, we computed a simple mean of their answers.
Note that the experts were not asked to map each attack into the relevant \cglspl{capec}; this was done by the authors.

\begin{table}[!htb]
    \centering
    \caption[Estimating the likelihood of the generic CAPECs into which the potential attacks are mapped, provided by our panel of healthcare ISEs]{Estimating the likelihood of the generic \glsxtrshortpl{capec} into which the potential attacks are mapped, provided by our panel of healthcare \glsxtrshortpl{ise}.}\label{tab:Estimating the likelihood of the generic CAPECs into which the potential attacks are mapped provided by our panel of healthcare ISEs}
    \begin{tabular}{ll}
		\toprule
		\Glsxtrshort{capec} ID & Estimation \\%
		\midrule
		\Cref{CAPEC-542} & 4.5 \\%
		\Cref{CAPEC-150} & 3.75 \\%
		\Cref{CAPEC-582} & 3.5 \\%
		\Cref{CAPEC-165} & 3 \\%
		\Cref{CAPEC-603} & 3 \\%
		\Cref{CAPEC-75} & 2.75 \\%
		\Cref{CAPEC-166} & 2.75 \\%
		\hyperref[CAPEC-NEW]{CAPEC~NEW} & 2.25 \\%
		\Cref{CAPEC-601} & 1.75 \\%
		\bottomrule
	\end{tabular}%
\end{table}

\subsection[Computing the CAPEC-based likelihood estimates for each attack]{Computing the \glsxtrshort{capec}\=/Based Likelihood Estimates for Each Attack}\label{results:Computing the CAPEC-based likelihood estimates for each attack}

In \Cref{tab:The application of the TLDR methodology to MIDs}, we present the computed \cgls{capec}\=/based likelihood estimates for each attack, using the \cgls{capec} likelihood estimates from \Cref{tab:Estimating the likelihood of the generic CAPECs into which the potential attacks are mapped provided by our panel of healthcare ISEs}.
The resulting likelihood estimates are higher by an average of 0.13 compared with the direct estimates (see~{\S}~{\ref{results:Validating the TLDR Methodology's Results}}); thus, we have shifted the likelihood estimates by a constant of \(c = -0.13\).

\subsection[Decomposing each attack into several severity aspects and assigning them weights]{Decomposing Each Attack into Several Severity Aspects and Assigning Them Weights}\label{results:Decomposing each attack into several severity aspects and assigning them weights}

This step is beyond the scope of the current study (see~{\S}~{\ref{methods:Decomposing each attack into several severity aspects and assigning them weights}}) and will be discussed in a separate paper that we are preparing.

\subsection[Assessing the magnitude of the impact of each of the severity aspects for each attack]{Assessing the Magnitude of the Impact of Each of the Severity Aspects for Each Attack}\label{results:Assessing the magnitude of the impact of each of the severity aspects for each attack}

In \Cref{tab:The application of the TLDR methodology to MIDs}, we present the results of the single\=/aspect severity assessments for each attack by our panel of four \cglspl{rme} (see~{\S}~{\ref{methods:Assessing the magnitude of the impact of each of the severity aspects for each attack}}).
All experts had a similar level of expertise, each individually with over ten years of clinical experience in their relevant fields, heads of large hospitals' radiology departments; therefore, to obtain each value, we computed a simple mean of their answers.
Since the current study focuses on the ontology\=/based \textbf{L}ikelihood (\textbf{L}) part of the \cgls{tldr} methodology, we simplified this process by using just the mean of the single\=/aspect severity assessments of our \cglspl{rme} for each attack.
In a separate paper, we discuss the severity \textbf{D}ecomposition (\textbf{D}) part of the \cgls{tldr} methodology and its validation.

\subsection[Computing the composite severity assessments for each attack]{Computing the Composite Severity Assessments for Each Attack}\label{results:Computing the composite severity assessments for each attack}

In the scope of the current study, this step is trivial (see~{\S}~{\ref{methods:Computing the composite severity assessments for each attack}}), since we only have single\=/aspect severity; thus, the composite severity assessments are simply the single\=/aspect severity assessments from {\S}~{\ref{results:Assessing the magnitude of the impact of each of the severity aspects for each attack}}.

\subsection[Integrating the likelihood and severity of each attack into its risk and prioritizing it]{Integrating the Likelihood and Severity of Each Attack into Its Risk and Prioritizing It}\label{results:Integrating the likelihood and severity of each attack into its risk and prioritizing it}

Finally, in \Cref{tab:The application of the TLDR methodology to MIDs}, we integrated the risk for each attack by plugging into \Cref{eq:Basic Risk Calculation} the \cgls{capec}\=/based likelihood estimates (see~{\S}~{\ref{results:Computing the CAPEC-based likelihood estimates for each attack}}) and the composite severity assessments (see~{\S}~{\ref{results:Computing the composite severity assessments for each attack}}).

\subsection[Validating the TLDR Methodology's Results]{Validating the \glsxtrshort{tldr} Methodology's Results}\label{results:Validating the TLDR Methodology's Results}

We validated our results, following {\S}~{\ref{methods:Validating the TLDR Methodology's Results}}, by following the same steps as the \cgls{tldr} methodology; however, we asked the panel of healthcare \cglspl{ise} to estimate also the \emph{direct} overall likelihood for each attack individually, instead of using the \cgls{tldr} methodology.
We then defined two ``consensus'' vectors of the 23 potential attacks:
\begin{inparaenum}[(1)]
    \item the \glsxtrfull{mecble} (computed indirectly), and
    \item the \glsxtrfull{medle}.
\end{inparaenum}
We demonstrated that we maintained the validity of the risk assessment process, by calculating the paired \(T\)\=/test statistic between the \cgls{capec}\=/based (\cgls{mecble}) estimates and the direct \cglspl{ise}' estimates, and showing that the null hypothesis is accepted (i.e., there was no significant difference between them). % chktex 21
In \Cref{tab:paried t-test}, we show that when we apply a constant shift of the \cgls{capec}\=/based likelihood estimates, the results turn out to be significantly similar (the second\=/row~(\ref{itm:shifted})).
This is expected due to the different scales that the two methods create.

\begin{table}[!htb]
    \small%
    \centering
    \caption[The paired T-test]{The paired \(T\)\=/test (\(P\)\=/value) results of the direct likelihood estimates vs. % chktex 21
    \begin{inparaenum}[(a)]
        \item\label{itm:noshift} the \cgls{capec}\=/based likelihood estimates; and
        \item\label{itm:shifted} the \cgls{capec}\=/based likelihood estimates shifted by -0.13.
    \end{inparaenum}
    }\label{tab:paried t-test}%
    \setlength\tabcolsep{1.5pt}
    \begin{tabular}{lll}
        \toprule
        & Likelihood &~Risk \\
        \midrule
        (\ref{itm:noshift}) & 5.756 (\SI{8.645e-6}) &~6.026 (\SI{4.585e-6}) \\
		(\ref{itm:shifted}) & 0.043 (0.966) & -0.008 (0.994) \\
        \bottomrule
    \end{tabular}%
\end{table}%

\begin{table*}[!ht]
    \small%
    \centering
    \caption[Spearman's pairwise correlation of the MECBLE and MEDLE with each of the ISEs' estimates]{Spearman's pairwise correlation of the \glsxtrshort{mecble} and \glsxtrshort{medle} with each of the \glsxtrshortpl{ise}' estimates.}\label{tab:Spearman's pairwise correlation of the MECBLE and MEDLE with each of the ISEs' estimates}%
    \setlength\tabcolsep{1.5pt}
    \begin{tabular}{lllll}
        \toprule
        Spearman (\(P\)\=/value) & \Glsxtrshort{mecble}\=/\Glsxtrshort{ise}1 & \Glsxtrshort{mecble}\=/\Glsxtrshort{ise}2 & \Glsxtrshort{mecble}\=/\Glsxtrshort{ise}3 &~\Glsxtrshort{mecble}\=/\Glsxtrshort{ise}4 \\ % chktex 21
        \midrule
		Computed Likelihood & 0.414 (0.050) & 0.365 (0.087) & 0.368 (0.084) & -0.371 (0.082) \\
        \bottomrule%

		\addlinespace%
		\addlinespace%
		
		\toprule
        & \Glsxtrshort{medle}\=/\Glsxtrshort{ise}1 & \Glsxtrshort{medle}\=/\Glsxtrshort{ise}2 & \Glsxtrshort{medle}\=/\Glsxtrshort{ise}3 &~\Glsxtrshort{medle}\=/\Glsxtrshort{ise}4 \\ % chktex 21
        \midrule
        Direct Likelihood & 0.271 (0.211) & 0.837 (0.) & 0.141 (0.521) &~0.445 (0.033) \\
        \bottomrule
    \end{tabular}%
\end{table*}%

We demonstrated the value and the validity of using the \cgls{tldr} methodology with respect to the relative ranking of the 23 potential threats, we calculated the pairwise Spearman's correlation between the \cgls{capec}\=/based (\cgls{mecble}) likelihood estimates and the direct estimates of each of our \cglspl{ise} and showed that the correlation is high, while the pairwise correlation between the mean of the direct estimates (the \cgls{medle}) and each of the \cglspl{ise}' overall direct estimates was lower.
Note that we asked the \cglspl{ise} to \emph{directly} estimate the likelihood of each attack individually \emph{before} we ask them to estimate it using the \cgls{tldr} methodology to eliminate a potential biased point of reference.
In \Cref{tab:Spearman's pairwise correlation of the MECBLE and MEDLE with each of the ISEs' estimates}, we show that the healthcare \cglspl{ise} seem to agree with the \cgls{mecble} on the \cgls{capec}\=/based likelihood estimates, while two out of four \cglspl{ise} did not agree with the \cgls{medle} on the estimates of the direct likelihood (implying high variance between the correlations).
Thus, we can imply that the \cgls{capec}\=/based (\cgls{mecble}) likelihood estimates, using the \cgls{tldr} methodology, are as valid as current direct risk assessment methodologies' likelihood estimates; however, the \cgls{capec}\=/based likelihood estimates, computed from the mapped \cglspl{capec}, are much easier to calculate.

We also validated that the mapping of the discovered attacks into their relevant \cglspl{capec} indeed results in a mapping of many attacks into a significantly smaller number of just nine \cglspl{capec}.\
Also, we validated that using the \cglspl{afd} helps discover new attacks, since we discovered \emph{eight} new attacks after using this method.

%-------------------------------------------------------------------------------
\section{Summary}\label{sec:Summary}
%-------------------------------------------------------------------------------

In this study, we have presented the \cgls{tldr} methodology for information security risk assessment for medical devices, its application to \cglspl{mid} and the validation of its results correctness with the assistance of a panel of four senior healthcare \cglspl{ise}, \cglspl{ciso} of major \cglspl{hmo}, and a panel of four senior \cglspl{rme}, the heads of several large medical centers' imaging departments.
We first presented the essential background on medical device security, information security risk assessment methodologies, \cgls{capec} mechanisms of attacks, and \cglspl{mid}.\
We have shown that nowadays, medical devices are facing many security challenges; however, it is difficult to accurately assess the risk of these challenges using current risk assessment methodologies, which are too general and do not seem to fit the unique medical domain.
The application of the \cgls{tldr} methodology to \cglspl{mid} included:
\begin{inparaenum}[(1)]
	\item identifying the potentially vulnerable components of \cglspl{mid} by creating four \cglspl{afd} for the generic \cgls{mid}, a generic \cgls{ct}, a generic \cgls{mri}, and a generic ultrasound;
	\item identifying a total of 23 potential attacks on \cglspl{mid}:\ 15 known attacks and eight new attacks that we discovered using the \cglspl{afd};\
	\item mapping of all 23 discovered attacks to just eight out of 517 existing \cgls{capec}, and a new, \nth{518}, \cgls{capec} attack pattern that we suggest adding;
	\item estimating the likelihood of the nine generic \cglspl{capec} with the assistance of our panel of four senior healthcare \cglspl{ise} and using these estimates to compute the \cgls{capec}\=/based likelihood estimates of each attack;
	\item computing, for all medical device attacks, the composite severity assessments of each attack with the assistance of a panel of four senior \cglspl{rme};\ and
	\item integrating the likelihood and severity of each attack into its risk and prioritizing it.
\end{inparaenum}

We have shown that by using \cglspl{afd}, we were able to discover eight new potential attacks on \cglspl{mid}, in addition to the 15 known attacks.
This implies that \cglspl{afd} helps discover new attacks and shows the importance of initially identifying potentially vulnerable components.

We also validated that the mapping of the discovered attacks into their relevant \cglspl{capec} indeed results in a mapping of many attacks into a significantly smaller number of just nine \cglspl{capec}.\
Thus, it saves significant time and effort and leads to greater consistency in risk assessment.

We have suggested adding \hyperref[CAPEC-NEW]{CAPEC~NEW}, the new \cgls{capec} that we have defined, to the official \cgls{capec} taxonomy since it may be relevant to other domains.
It involves the manipulation of environment hardware or sensors as part of the attack pattern.
Other potentially relevant domains might include operational technology or critical infrastructure domains that use devices that rely on such environment hardware or sensors.

%-------------------------------------------------------------------------------
\section{Discussion}\label{sec:Discussion}
%-------------------------------------------------------------------------------

The analysis of the pairwise Spearman's correlation between the \cgls{capec}\=/based (\cgls{mecble}) likelihood estimates and the direct estimates of each of our \cglspl{ise}, had shown that the \cglspl{ise} strongly agree with the \cgls{mecble} using the \cgls{tldr} methodology while disagreeing with the mean of the direct estimates (the \cgls{medle}).
This shows that it is much easier for \cglspl{ise} to reach a consensus when using the \cgls{tldr} methodology.
In fact, when the \cglspl{ise} were asked to estimate the likelihood directly, they could not reach a consensus.

Moreover, we demonstrate, at least statistically, that by systematically decomposing and mapping the attacks into their relevant \cgls{capec} standard attack patterns and estimating the likelihood of just these \cglspl{capec} with a panel of healthcare \cglspl{ise}, we can reasonably accurately compute the likelihood estimates of all attacks; the \cgls{capec}\=/based likelihood and risk correlated well with our panel of healthcare \cglspl{ise}' \emph{direct} individual attack estimates of the likelihood.

We have also shown that such ontology\=/based mapping of attacks is beneficial to the analysis of attacks on \cglspl{mid}.\
The mapping of discovered attacks into a set of existing \cglspl{capec} enables us to exploit repeatedly the one\=/time likelihood estimates of these \cglspl{capec} in the medical (here, radiology) domain, and in some cases, even the default \cglspl{capec} likelihood estimates, to compute the \cgls{capec}\=/based likelihood estimates of the new potential attacks.
Unfortunately, not all of the mapped \cglspl{capec} that we found include a likelihood estimate in the official \cgls{capec} ontology, thus we can use for now our \cglspl{capec} likelihood estimates instead, and perhaps update these estimates in the future, once the \cgls{capec} ontology is updated.

The overall enumeration and charting, mapping into \cglspl{capec}, likelihood estimates, severity assessments, and overall risk integration, as defined in the \cgls{tldr} methodology~({\S}~{\ref{methods:Methods}}), is potentially relevant to many other medical domains and devices, its application to \cglspl{mid} in this study was given as a use case example.
Such mapping could likely be beneficial to other medical devices as well since we did not base it on any \cgls{mid}\=/specific aspects.

Using the \cgls{tldr} methodology implies that \cglspl{ise} are used only for the qualitative phase of mapping potential attacks to \cglspl{capec}.\
Note that in most cases, mapping attacks to \cglspl{capec} is straightforward, and may be performed by a less experienced \cgls{ise}.\
Once this relatively simple task is performed, we use the predefined likelihoods of the respective \cglspl{capec}.\
There is no need to ask the panel of \cglspl{ise} for the quantitative estimates again.

%-------------------------------------------------------------------------------
\section{Limitations}\label{sec:Limitations}
%-------------------------------------------------------------------------------

We used only four senior \cglspl{ise} to directly estimate the likelihood of  each of the 23 potential attacks, as well as the likelihood of the nine \cglspl{capec} into which these attacks were mapped.
It is possible that using even more \cglspl{ise} might modify the results, although the fact that even with four \cglspl{ise}, it was possible to reach a reasonable [mean] consensus, seems to indicate that using a higher number is likely to further strengthen the results.

We applied the full \cgls{tldr} methodology only to the radiology domain and to \cglspl{mid}.\
Applying the methodology to other medical domains and additional medical devices in a future study is quite possible as well, to show that the \cgls{tldr} methodology is useful in other medical domains and devices.

%-------------------------------------------------------------------------------
\section{Conclusions}\label{sec:Conclusions}
%-------------------------------------------------------------------------------

From the \cgls{afd} of the generic \cgls{mid}, we can conclude that many high\=/risk attacks could be targeting it; thus, potentially affecting many \cglspl{mid}.\
The host control PC, connected to the \cglspl{mid} and the Internet (via hospital networks), is particularly vulnerable, poses many threats (e.g., sabotage, physical harm to patients), and is the most apparent initial entry point for attackers.
The \cgls{ct} device itself is also very vulnerable and is potentially exposed to additional attacks that may cause direct harm to patients due to the use of ionizing X\=/ray radiation.

From our prioritization of risks, we can conclude that a ransomware attack (attack~\ref{attack:Ransomware}) poses the highest risk for a generic \cgls{mid} and the overall \emph{highest risk attack of all \cglspl{mid}}, followed by a disruption of patient\=/to\=/image linkage attack (attack~\ref{attack:Disruption of patient-to-image linkage}) and an alteration of the imaging exam's results attack (attack~\ref{attack:Alteration of the imaging Exam's Results}).
Furthermore, a radiation overdose attack (attack~\ref{attack:Radiation Overdose}) poses the highest risk for a generic \cgls{ct} and an overwhelm of the \cgls{mri}'s receiving coils with an overpowered magnetic field attack (attack~\ref{attack:Overwhelm of the MRI's Receiving Coils with an Overpowered Magnetic Field}) poses the highest risk for a generic \cgls{mri}.\

This work reinforces our understanding that in order to better protect medical devices, it is necessary to start by correctly prioritizing the protection efforts through a consistent and accurate information security risk assessment of the potential threats, such as by using the \cgls{tldr} methodology that we had proposed in this paper.
Then, further research and development of detection and prevention techniques against attacks must be performed.
Such techniques should be implemented both outside (e.g., through hospital networks) and inside medical devices ecosystems since each device can consist of an entire ecosystem of components, all of which must be better secured.

We hope that our work contributes to a better risk assessment of the risk to medical devices posed by cyber security threats. We consider it potentially valuable for researchers, healthcare organizations, hospitals, medical device manufacturers, and \cglspl{ise}, leading to a prioritized assessment of the potential risks, followed by research and development of new defensive mechanisms.
Indeed, we have already begun investigating several such means for mitigation of the potential attacks that we had discovered.

%-------------------------------------------------------------------------------
\section*{Acknowledgments}
%-------------------------------------------------------------------------------
We want to thank Erez~Shalom,~PhD, for helping us in this research.
We want to thank the experts that participated in the panel of \glsxtrfullpl{rme}, Arnon~Makori,~MD,~MHA, Ilan~Shelef,~MD, Gad~Levy,~MD, and the other experts.
We want to thank the experts that participated in the panel of healthcare \glsxtrfullpl{ise}, Russ~Neff,~CISSP, Itzik~Kochav, Israel~Goldenberg~,\glsxtrshort{ciso}, and the other experts.

%-------------------------------------------------------------------------------
\bibliography{tldr}

\begin{thebibliography}{10}

\bibitem{Gibson2017}
William Gibson.
\newblock {\em {Burning chrome}}.
\newblock 1982.

\bibitem{Larson17}
Selena Larson.
\newblock {Massive Cyberattack Targeting 99 Countries Causes Sweeping Havoc}.
\newblock {\em CNN}, 2017.

\bibitem{Liptak17}
Andrew Liptak.
\newblock {The WannaCry ransomware attack has spread to 150 countries}.
\newblock {\em The Verge}, 2017.

\bibitem{Millar17}
Sheila~A. Millar and Tracy~P. Marshall.
\newblock {WannaCry: Are Your Security Tools Up to Date?}
\newblock {\em The National Law Review}, 2017.

\bibitem{Brenner17}
Bill Brenner.
\newblock {WannaCry: the ransomware worm that didn’t arrive on a phishing
  hook}.
\newblock {\em Naked Security by Sophos}, 2017.

\bibitem{BBC17}
{Cyber-attack: Europol says it was unprecedented in scale}.
\newblock {\em British Broadcasting Corporation (BBC)}, pages 1--7, 2017.

\bibitem{UngoedThomas17}
Jon Ungoed-Thomas, Robin Henry, and Dipesh Gadher.
\newblock {Cyber-attack guides promoted on YouTube}.
\newblock {\em The Sunday Times}, 2017.

\bibitem{Hern17}
Alex Hern and Samuel Gibbs.
\newblock {What is WannaCry ransomware and why is it attacking global
  computers?}
\newblock {\em The Guardian}, 2017.

\bibitem{CarrieWong2017}
Julia Carrie~Wong and Olivia Solon.
\newblock {Massive ransomware cyber-attack hits nearly 100 countries around the
  world}, 2017.

\bibitem{Foxx17}
{NHS cyber-attack: GPs and hospitals hit by ransomware}.
\newblock {\em British Broadcasting Corporation (BBC)}, 2017.

\bibitem{Skinner16}
Curtis Skinner.
\newblock {Los Angeles hospital paid hackers {\$}17,000 ransom in bitcoins}.
\newblock {\em Aol.}, 2016.

\bibitem{Strickland16}
Eliza Strickland.
\newblock {5 Major Hospital Hacks: Horror Stories from the Cybersecurity
  Frontlines}.
\newblock {\em IEEE Spectrum}, 2016.

\bibitem{Davis16}
Jessica Davis.
\newblock {Latest cybersecurity threat, 'Locky,' spreads faster than any other
  virus}, 2016.

\bibitem{Wong16}
Julia Wong.
\newblock {Los Angeles hospital returns to faxes and paper charts after
  cyberattack}.
\newblock {\em The Guardian}, 2016.

\bibitem{Winton16}
Richard Winton.
\newblock {Hollywood hospital pays {\$}17,000 in bitcoin to hackers; FBI
  investigating}.
\newblock {\em The Los Angeles Times}, 2016.

\bibitem{Leveson93}
N.G. Leveson and C.S. Turner.
\newblock {An investigation of the Therac-25 accidents}.
\newblock {\em Computer}, 26(7):18--41, 7 1993.

\bibitem{Kramer12}
Daniel~B. Kramer, Matthew Baker, Benjamin Ransford, Andres Molina-Markham,
  Quinn Stewart, Kevin Fu, and Matthew~R. Reynolds.
\newblock {Security and Privacy Qualities of Medical Devices: An Analysis of
  FDA Postmarket Surveillance}.
\newblock {\em PLoS ONE}, 7(7):e40200, 7 2012.

\bibitem{Ayala16}
Luis Ayala.
\newblock {\em {Cybersecurity for Hospitals and Healthcare Facilities}}.
\newblock Apress, Berkeley, CA, 2016.

\bibitem{ISO73:2009}
{ISO}.
\newblock {ISO 73:2009, Risk management - Vocabulary}, 2009.

\bibitem{Popov2016}
Georgi Popov, Bruce~K Lyon, and Bruce Hollcroft.
\newblock {\em {Risk assessment: A practical guide to assessing operational
  risks}}.
\newblock John Wiley {\&} Sons, 2016.

\bibitem{Rausand2013}
Marvin Rausand.
\newblock {\em {Risk assessment: theory, methods, and applications}}, volume
  115.
\newblock John Wiley {\&} Sons, 2013.

\bibitem{HHS06a}
{HHS}.
\newblock {HIPAA Security Standards for Covered Entities}, 2006.

\bibitem{Camara15}
Carmen Camara, Pedro Peris-Lopez, and Juan~E. Tapiador.
\newblock {Security and privacy issues in implantable medical devices: A
  comprehensive survey}.
\newblock {\em Journal of Biomedical Informatics}, 55:272--289, 6 2015.

\bibitem{Panescu08}
Dorin Panescu.
\newblock {Emerging Technologies [wireless communication systems for
  implantable medical devices]}.
\newblock {\em IEEE Engineering in Medicine and Biology Magazine},
  27(2):96--101, 3 2008.

\bibitem{Li11}
{Chunxiao Li}, Anand Raghunathan, Niraj~K. Jha, Chunxiao Li, Anand Raghunathan,
  and Niraj~K. Jha.
\newblock {Hijacking an insulin pump: Security attacks and defenses for a
  diabetes therapy system}.
\newblock In {\em 2011 IEEE 13th International Conference on e-Health
  Networking, Applications and Services}, pages 150--156. IEEE, 6 2011.

\bibitem{Fu09}
Kevin Fu.
\newblock {Inside risksReducing risks of implantable medical devices}.
\newblock {\em Communications of the ACM}, 52(6):25, 6 2009.

\bibitem{Hei10}
Xiali Hei, Xiaojiang Du, Jie Wu, and Fei Hu.
\newblock {Defending Resource Depletion Attacks on Implantable Medical
  Devices}.
\newblock In {\em 2010 IEEE Global Telecommunications Conference GLOBECOM
  2010}, pages 1--5, Miami, FL, USA, 12 2010. IEEE.

\bibitem{Franzen13}
Carl Franzen.
\newblock {Dick Cheney had the wireless disabled on his pacemaker to avoid risk
  of terrorist tampering}, 2013.

\bibitem{FBICyberDivision2014a}
{FBI Cyber Division} and {Federal Bureau of Investigation Cyber Division}.
\newblock {Health Care Systems and Medical Devices at Risk for Increased Cyber
  Intrusions for Financial Gain}.
\newblock Technical report, FBI, 2014.

\bibitem{USIS16}
Va~Office~of Information~Security.
\newblock {Information Security Monthly Incident Report - January 2016}.
\newblock Technical report, DEPARTMENT OF VETERANS AFFAIRS, 2016.

\bibitem{USVA16}
{U.S. Department of Veterans Affairs}.
\newblock {VetPop2016 Dataset}, 2016.

\bibitem{Ferrara2019}
Adi Ferrara.
\newblock {Cybersecurity in Medical Imaging.}
\newblock {\em Radiologic technology}, 90(6):563--575, 7 2019.

\bibitem{Fenz14}
Stefan Fenz, Johannes Heurix, Thomas Neubauer, and Fabian Pechstein.
\newblock {Current challenges in information security risk management}.
\newblock {\em Information Management {\&} Computer Security}, 22(5):410--430,
  11 2014.

\bibitem{NISTSP12}
{National Institute of Standards and Technology (NIST)}.
\newblock {Guide for conducting risk assessments}.
\newblock Technical report, National Institute of Standards and Technology,
  Gaithersburg, MD, 2012.

\bibitem{isoiec2700508}
{BS ISO/IEC}.
\newblock {ISO/IEC 27005:2008, Information Technology --- Security techniques -
  Information Security Risk Management}.
\newblock 3, 2008.

\bibitem{OCTAVE03}
Christopher~J. Alberts, Audrey~J. Dorofee, James~F. Stevens, and Carol Woody.
\newblock {Introduction to the OCTAVE Approach}.
\newblock Technical report, 2003.

\bibitem{Farquhar1991}
Bill Farquhar.
\newblock {One approach to risk assessment}.
\newblock {\em Computers {\&} Security}, 10(1):21--23, 2 1991.

\bibitem{Yazar02}
Zeki Yazar.
\newblock {A Qualitative Risk Analysis and Management Tool - CRAMM}, 2002.

\bibitem{Freund2015}
Jack Freund and Jack Jones.
\newblock {\em {Measuring and managing information risk a FAIR approach}}.
\newblock 2015.

\bibitem{ISAMM07}
Carlo Harpes, André~Andre Adelsbach, Stefano Zatti, and Nestor Peccia.
\newblock {Quantitative Risk Assessment With ISAMM on ESA's Operations Data
  System}.
\newblock {\em Proceedings of TTC}, pages 173--176, 2007.

\bibitem{Stine2017}
Ian Stine, Mason Rice, Stephen Dunlap, and John Pecarina.
\newblock {A cyber risk scoring system for medical devices}.
\newblock {\em International Journal of Critical Infrastructure Protection}, 4
  2017.

\bibitem{FDA2016}
{U.S. Food and Drug Administration} and {FDA}.
\newblock {Postmarket Management of Cybersecurity in Medical Devices Guidance
  for Industry and Food and Drug Administration Staff Additional Copies}.
\newblock 2016.

\bibitem{Yasqoob2019IntegratedDevices}
Tahreem Yasqoob, Haider Abbas, and Narmeen Shafqat.
\newblock {Integrated Security, Safety, and Privacy Risk Assessment Framework
  for Medical Devices}.
\newblock {\em IEEE Journal of Biomedical and Health Informatics}, pages 1--1,
  2019.

\bibitem{CAPEC}
{the MITRE Corporation}.
\newblock {Common Attack Pattern Enumeration and Classification (CAPEC)}.

\bibitem{Hendee2010b}
William Hendee, Ehsan Samei, Elizabeth Krupinski, and William Hendee.
\newblock {The Handbook of Medical Image Perception and Techniques}.
\newblock {\em Medical Physics}, 37(11):6112--6112, 11 2010.

\bibitem{FDA17MedicalImaging}
{U.S. Food and Drug Administration (FDA)}.
\newblock {Medical Imaging}, 2017.

\bibitem{FDA17CT}
{U.S. Food and Drug Administration (FDA)}, {U.S. Food {\&} Drug
  Administration}, and {U.S. Food and Drug Administration (FDA)}.
\newblock {Medical X-ray Imaging - Computed Tomography (CT)}, 2017.

\bibitem{FDA17Xray}
{U.S. Food and Drug Administration (FDA)} and {U.S. Food {\&} Drug
  Administration (FDA)}.
\newblock {Medical X-ray Imaging - Radiography}, 2017.

\bibitem{hartwig09}
Valentina Hartwig, Giulio Giovannetti, Nicola Vanello, Massimo Lombardi, Luigi
  Landini, and Silvana Simi.
\newblock {Biological effects and safety in magnetic resonance imaging: a
  review.}
\newblock {\em International Journal of Environmental Research and Public
  Health}, 6(12):1778--1798, 6 2009.

\bibitem{Mahesh13}
Mahadevappa Mahesh.
\newblock {The Essential Physics of Medical Imaging, Third Edition.}
\newblock {\em Medical Physics}, 40(7):077301, 6 2013.

\bibitem{Benseler06}
J.~S. Benseler.
\newblock {\em {The Radiology Handbook : A Pocket Guide To Medical Imaging}}.
\newblock Ohio University Press, 2006.

\bibitem{RSNA17}
Inc Radiological Society~of North~America and {Radiological Society of North
  America}.
\newblock {Radiation Dose in X-Ray and CT Exams}, 2017.

\bibitem{Brown14}
Robert~W. Brown, Yu-Chung~N. Cheng, E.~Mark. Haacke, Michael~R. Thompson, and
  Ramesh. Venkatesan.
\newblock {\em {Magnetic Resonance Imaging: Physical Principles and Sequence
  Design, Second Edition}}.
\newblock John Wiley {\&} Sons Ltd, Chichester, UK, 2 edition, 4 2014.

\bibitem{Beyer11}
Thomas Beyer, Lutz~S Freudenberg, David~W Townsend, and Johannes Czernin.
\newblock {The future of hybrid imaging-part 1: hybrid imaging technologies and
  SPECT/CT}.
\newblock {\em Insights into Imaging}, 2(2):161--169, 4 2011.

\bibitem{Townsend08}
David~W Townsend.
\newblock {Combined Positron Emission Tomography-Computed Tomography: The
  Historical Perspective}.
\newblock {\em Seminars in Ultrasound, CT and MRI}, 29(4):232--235, 8 2008.

\bibitem{AHA16}
{American Heart Association}.
\newblock {Single Photon Emission Computed Tomography (SPECT)}, 2016.

\bibitem{Bruza93}
P~D Bruza and Th~P Van Der~Weide.
\newblock {The Semantics of Data Flow Diagrams}.
\newblock In {\em In Proceedings of the International Conference on Management
  of Data}, pages 66--78. McGraw-Hill Publishing Company, 1993.

\bibitem{Mahler18}
Tom Mahler, Nir Nissim, Erez Shalom, Israel Goldenberg, Guy Hassman, Arnon
  Makori, Itzik Kochav, Yuval Elovici, and Yuval Shahar.
\newblock {Know Your Enemy: Characteristics of Cyber-Attacks on Medical Imaging
  Devices}.
\newblock {\em arXiv}, 1801.05583, 1 2018.

\bibitem{Becker2018}
Anton~S. Becker, Lukas Jendele, Ondrej Skopek, Nicole Berger, Soleen Ghafoor,
  Magda Marcon, and Ender Konukoglu.
\newblock {Injecting and removing malignant features in mammography with
  CycleGAN: Investigation of an automated adversarial attack using neural
  networks}.
\newblock {\em arXiv}, 1811.07767, 11 2018.

\bibitem{Mahler19}
Yisroel Mirsky, Tom Mahler, Ilan Shelef, and Yuval Elovici.
\newblock {CT-GAN: Malicious Tampering of 3D Medical Imagery using Deep
  Learning}.
\newblock In {\em 28th USENIX Security Symposium (USENIX Security 19)}, pages
  461--478, Santa Clara, CA, 2019. USENIX Association.

\bibitem{Pan17}
Aimin Pan, Bo~Yang, Shangyuan LI, Wang Kang, and Wang Zhengbo.
\newblock {Sonic Gun to Smart Devices: Your Devices Lose Control Under
  Ultrasound/Sound}.
\newblock In {\em Black Hat USA}, 2017.

\bibitem{ISE16}
{Independent Security Evaluators}.
\newblock {Securing Hospitals - A research study and blueprint}.
\newblock Technical report, 2016.

\bibitem{Rotter16}
Joanna Rotter.
\newblock {8 Ways Medical Device Manufacturers Can Use Strategic IoT Solutions
  to Set Their Services Apart}, 2015.

\bibitem{FitzGerald16}
Paul FitzGerald, James Bennett, Jeffrey Carr, Peter~M. Edic, Daniel Entrikin,
  Hewei Gao, Maria Iatrou, Yannan Jin, Baodong Liu, Ge~Wang, Jiao Wang, Zhye
  Yin, Hengyong Yu, Kai Zeng, and Bruno De~Man.
\newblock {Cardiac CT: A system architecture study}.
\newblock {\em Journal of X-Ray Science and Technology}, 24(1):43--65, 3 2016.

\bibitem{Suparta00}
{Gede B. Suparta}.
\newblock {Focusing Computed Tomography}, 2000.

\bibitem{Higgins17}
D~M Higgins.
\newblock {ReviseMRI.com : System architecture}, 2017.

\bibitem{McRobbie17}
Donald~W. McRobbie, Elizabeth~A. Moore, and Martin~J. Graves.
\newblock {\em {MRI from Picture to Proton}}.
\newblock Cambridge University Press, Cambridge, 2017.

\bibitem{Elster01}
Allen~D. Elster and Jonathan~H. Burdette.
\newblock {\em {Questions and Answers in Magnetic Resonance Imaging}}.
\newblock Mosby, 2 edition, 8 2001.

\bibitem{Vlaardingerbroek03}
Marinus~T. Vlaardingerbroek and Jacques~A. Boer.
\newblock {\em {Magnetic Resonance Imaging, Theory and Practice}}.
\newblock Springer-Verlag Berlin Heidelberg, Berlin, Heidelberg, 3 edition, 7
  2003.

\bibitem{Kutz09}
Myer Kutz.
\newblock {\em {Biomedical Engineering and Design Handbook, Volume 2}}.
\newblock McGraw-Hill, New York, Chicago, San Francisco, Lisbon, London,
  Madrid, Mexico City, Milan, New Delhi, San Juan, Seoul, Singapore, Sydney,
  Toronto, 2009.

\bibitem{TI17}
{Texas Instruments}.
\newblock {Ultrasound System Design Resources and Block Diagram}.

\bibitem{element09}
{tech3}.
\newblock {Block Diagrams- Texas Instruments- Ultrasound System Design
  Resources and Block Diagram}, 2009.

\bibitem{Bushong2013a}
Stewart~C. Bushong.
\newblock {\em {Radiologic Science for Technologists : Physics, Biology, and
  Protection.}}
\newblock Elsevier Health Sciences, 2013.

\bibitem{Fuchs71}
Arthur~W. Fuchs.
\newblock {\em {Principles of Radiographic Exposure and Processing}}.
\newblock Charles C. Thomas, second edition, 1971.

\bibitem{Aram16}
Siamak Aram, Rouzbeh~A. Shirvani, Eros~G. Pasero, and Mohamd~F. Chouikha.
\newblock {Implantable Medical Devices; Networking Security Survey}.
\newblock {\em Journal of Internet Services and Information Security (JISIS)},
  6(3):40--60, 2016.

\bibitem{CVE-2017-0144}
{Microsoft Corporation}.
\newblock {CVE-2017-0144}.
\newblock Technical report, 2017.

\bibitem{Newman17}
Lily~H. Newman.
\newblock {How an Accidental 'Kill Switch' Slowed Friday's Massive Ransomware
  Attack}, 2017.

\bibitem{MalwareTech17}
{MalwareTech}.
\newblock {How to Accidentally Stop a Global Cyber Attacks}, 2017.

\bibitem{Greenberg17}
Andy Greenberg.
\newblock {Hackers are Trying to Reignite WannaCry with Nonstop Botnet
  Attacks}, 2017.

\end{thebibliography}
%-------------------------------------------------------------------------------

\appendix

%-------------------------------------------------------------------------------
\section{WannaCry}\label{appendix:WannaCry}
%-------------------------------------------------------------------------------

The \nameref{appendix:WannaCry} crypto\=/worm~\cite{Larson17, Liptak17, Millar17}, which emerged in May 2017, is a worldwide ransomware cyber attack that infected over 200,000 devices in more than 150 nations~\cite{Brenner17}, with a significant impact on the FedEx delivery company's services in the USA, Spanish telecom and gas companies, Renault car production factories in France, Russia's interior ministry, and the \cgls{nhs}~\cite{BBC17} of the UK.\
This was the first, large\=/scale cyber attack that affected the healthcare industry directly by infecting tens of thousands of \cgls{nhs} hospitals' devices, including \cglspl{mid} such as \cglspl{mri}, according to the \textit{Sunday Times}~\cite{UngoedThomas17} reports, causing them to be non\=/operational by encrypting them.
This attack caused several hospitals to turn away patients~\cite{Hern17, CarrieWong2017} and divert ambulance routes~\cite{Foxx17}.
	
The \nameref{appendix:WannaCry} attack was based on the EternalBlue exploit, a Windows \cgls{smb} protocol exploit tool (CVE\=/2017\=/0144%
\footnote{\hrefblue{https://cve.mitre.org/cgi-bin/cvename.cgi?name=CVE-2017-0144}{https://cve.mitre.org/cgi-bin/cvename.cgi?name=CVE-2017-0144}}% chktex 8
~\cite{CVE-2017-0144}), which affected many Windows versions such as Windows Vista, Windows Server 2008, Windows 7, Windows 8.1, Windows Server 2012, Windows RT 8.1, Windows 10, and Windows Server 2016.
It was leaked in April 2017 by the \textit{Shadow Brokers} hacker group, among several other exploits believed to belong to the \cgls{nsa}.\
The exploit allows arbitrary code execution by remote attackers via crafted packets.
Although in March 2017 (before the exploit was leaked) Microsoft issued a critical security patch fixing the exploit~\cite{CVE-2017-0144}, many computers were still vulnerable, because they had not been regularly updated for various reasons (e.g., technical issues, strict regulations, or insufficient awareness), as was the case with the \cgls{nhs} medical devices and computers.

The \nameref{appendix:WannaCry} crypto\=/worm was eventually contained due to the discovery of a software kill switch~\cite{Newman17} (i.e., a unique code that was inserted by the attacker to allow the malware execution to be stopped).
In this case, the kill switch was a particular \cgls{url} used by the malware to track activity from infected machines.
This domain was found to act as a kill switch that shuts down the software before it executed its payload, stopping the spread of the ransomware.
Some believe that this was inserted by the attackers to prevent the malware from running on quarantined machines and sandbox environments used by \cgls{av} researchers~\cite{MalwareTech17}.
Such environments, respond to all queries from the machine with some traffic in order to trick the suspicious software into ``thinking'' that it is still connected to the Internet; hence, when the \nameref{appendix:WannaCry} attempts to contact the discovered \cgls{url}, which does not actually exist and receives an answer, it means that the ransomware is running in a sandbox environment (i.e., receiving traffic from a nonexistent \cgls{url}) and therefore should do nothing, to evade detection by \cgls{av} researchers.
When discovered, some hackers even tried using a \textit{Mirai} botnet variant to create a distributed attack on this domain (i.e., the \nameref{appendix:WannaCry}'s kill switch), intending to shut it down and reignite the attack~\cite{Greenberg17}.

%%% TO CREATE ABBRIV NEED TO MANUALLY CALL makeglossaries -d "out" "tldr"
\printglossary[type=\glsxtrabbrvtype]\label{appendix:Abbreviations}

% \tikzexternalize[prefix=tikzCache/]
\end{document}